\documentclass[letter]{aa}
\usepackage[varg]{txfonts}
\usepackage{graphicx}
\usepackage{natbib,twoopt}
\usepackage[breaklinks=true]{hyperref} 
\usepackage{lscape}
\bibpunct{(}{)}{;}{a}{}{,} 
\makeatletter
 \newcommandtwoopt{\citeads}[3][][]{\href{https://ui.adsabs.harvard.edu/abs/#3/abstract}%
 {\def\hyper@linkstart##1##2{}%
 \let\hyper@linkend\@empty\citealp[#1][#2]{#3}}}
 \newcommandtwoopt{\citepads}[3][][]{\href{https://ui.adsabs.harvard.edu/abs/#3/abstract}%
 {\def\hyper@linkstart##1##2{}%
 \let\hyper@linkend\@empty\citep[#1][#2]{#3}}}
 \newcommandtwoopt{\citetads}[3][][]{\href{https://ui.adsabs.harvard.edu/abs/#3/abstract}%
 {\def\hyper@linkstart##1##2{}%
 \let\hyper@linkend\@empty\citet[#1][#2]{#3}}}
 \newcommandtwoopt{\citeyearads}[3][][]%
 {\href{https://ui.adsabs.harvard.edu/abs/#3/abstract}
 {\def\hyper@linkstart##1##2{}%
 \let\hyper@linkend\@empty\citeyear[#1][#2]{#3}}}
\makeatother

\begin{document}

   \title{Memories of past close encounters in extreme trans-Neptunian space: 
          Finding unseen planets using pure random searches}
   \author{C. de la Fuente Marcos$^{1}$
           \and
           R. de la Fuente Marcos$^{2}$}
   \authorrunning{C. de la Fuente Marcos \and R. de la Fuente Marcos}
   \titlerunning{Close encounters in extreme trans-Neptunian space}
   \offprints{C. de la Fuente Marcos, \email{nbplanet@ucm.es}}
   \institute{$^{1}$ Universidad Complutense de Madrid,
              Ciudad Universitaria, E-28040 Madrid, Spain \\
              $^{2}$AEGORA Research Group,
              Facultad de Ciencias Matem\'aticas,
              Universidad Complutense de Madrid,
              Ciudad Universitaria, E-28040 Madrid, Spain}
   \date{Received 11 January 2021 / Accepted 3 February 2021}

   \abstract
      {The paths followed by the known extreme trans-Neptunian objects (ETNOs) 
       effectively avoid direct gravitational perturbations from the four giant 
       planets, yet their orbital eccentricities are in the range between 
       0.69--0.97. Solar system dynamics studies show that such high values of 
       the eccentricity can be produced via close encounters or secular 
       perturbations. In both cases, the presence of yet-to-be-discovered 
       trans-Plutonian planets is required. Recent observational evidence cannot 
       exclude the existence, at 600~AU from the Sun, of a planet of 
       five~Earth~masses.
       }
      {If the high eccentricities of the known ETNOs are the result of 
       relatively recent close encounters with putative planets, the mutual 
       nodal distances of sizeable groups of ETNOs with their assumed perturber 
       may still be small enough to be identifiable geometrically. In order to 
       confirm or reject this possibility, we used Monte Carlo random search 
       techniques.  
       }
      {Two arbitrary orbits may lead to close encounters when their mutual nodal
       distance is sufficiently small. We  generated billions of random 
       planetary orbits with parameters within the relevant ranges and computed 
       the mutual nodal distances with a set of randomly generated orbits with 
       parameters consistent with those of the known ETNOs and their 
       uncertainties. We monitored which planetary orbits had the maximum 
       number of potential close encounters with synthetic ETNOs and we studied 
       the resulting distributions. 
       }
      {We provide narrow ranges for the orbital parameters of putative planets 
       that may have experienced orbit-changing encounters with known ETNOs. 
       Some sections of the available orbital parameter space are strongly 
       disfavored by our analysis.
       }
      {Our calculations suggest that more than one perturber is required if 
       scattering is the main source of orbital modification for the known 
       ETNOs. Perturbers might not be located farther than 600~AU and they  
       have to follow moderately eccentric and inclined orbits to be capable of
       experiencing close encounters with multiple known ETNOs.
       }


   \maketitle

   \section{Introduction\label{Intro}}
      Extreme trans-Neptunian objects (ETNOs) serve as unique probes into the gravity perturbations shaping the outer solar system beyond 
      the classical trans-Neptunian or Kuiper belt (see e.g., \citealt{2019AJ....158...43K}). The trajectories followed by the known ETNOs 
      effectively avoid direct gravitational perturbations from the four giant planets, yet their orbital eccentricities are in the range 
      0.69--0.97. Efficient drivers for the eccentricity excitation of small bodies include close encounters with planets (see e.g., 
      \citealt{1990CeMDA..49..111C}) and the von~Zeipel-Lidov-Kozai mechanism \citep{1910AN....183..345V,1962AJ.....67..591K,
      1962P&SS....9..719L,2019MEEP....7....1I}. In both cases above and in the case of ETNOs, the presence of yet-to-be-discovered 
      trans-Plutonian planets is required. 

      \citet{2020A&A...640A...6F} used the INPOP19a planetary ephemerides that include Jupiter-updated positions by the Juno mission and a 
      reanalysis of Cassini observations to show that there is no clear evidence for the existence of the so-called Planet~9 predicted by 
      \citet{2016AJ....151...22B} as an explanation for the orbital architecture of the known ETNOs. However, \citet{2020A&A...640A...6F} 
      concluded that if Planet~9 exists, it cannot be closer than 500~AU, if it has a mass of 5~$M_{\oplus}$, and no closer than 650~AU, if 
      it has a mass of 10~$M_{\oplus}$. The latest version of the Planet~9 hypothesis \citep{2019PhR...805....1B} predicts the existence of 
      a planet with a mass in the range 5--10~$M_{\oplus}$, following an orbit with a value of the semi-major axis in the range of 
      400--800~AU, eccentricity in the range of 0.2--0.5, and inclination in the interval between (15{\degr}, 25{\degr}). A number of 
      exoplanets have already been observed orbiting at hundreds of AUs from their host stars (see e.g., \citealt{2014ApJ...780L...4B,
      2014ApJ...787....5N,2021AJ....161...22N}) and theoretical calculations have confirmed plausible pathways for their formation (see 
      e.g., \citealt{2015ApJ...806...42K,2016ApJ...825...33K}).

      If the high eccentricities of known ETNOs are the result of relatively recent close encounters with putative planets, the mutual 
      nodal distances of sizeable groups of ETNOs with their assumed perturber may still be small enough to be identifiable geometrically.
      Here, we use Monte Carlo random search techniques to identify orbits that may lead to the maximum number of potential close encounters 
      with synthetic ETNOs whose orbital parameters are consistent with those of the real ETNOs and their uncertainties. This letter is organized 
      as follows. In Sect.~\ref{Data}, we review our methodology and present the data used in our analyses. In Sect.~\ref{Results}, we apply 
      our methodology and discuss its results. Our conclusions are summarized in Sect.~\ref{Conclusions}.

   \section{Methods and data description\label{Data}}
      This work explores a "what if" scenario in which the starting hypothesis states that a sizeable number of known ETNOs have experienced 
      relatively recent close encounters with putative planets; the timescale comes constrained by their orbital periods that are in the 
      range 1867--50116~yr, so encounters may have taken place during the last 10$^{3}$--10$^{5}$~yr. In this work, therefore, we are testing the 
      hypothesis statistically. If this hypothesis is plausible, a statistically significant number of compatible planetary perturber orbits 
      should emerge from the analysis of a very large sample of orbits. If the distributions of one or more of the orbital parameters of the 
      candidate are flat, the starting hypothesis must be rejected as this would show that there is no favored orbital solution for the perturber; conversely, if 
      all the distributions produce consistent intervals that are statistically significant, the plausibility of the starting hypothesis can be 
      considered as confirmed. Plausibility concerns the likelihood of acceptance, not the likelihood of being true or better than competing 
      scenarios. The problem under investigation here is equivalent to a nondifferenciable optimization that is well-suited for a uniform Monte 
      Carlo random search \citep{MU49}. Our methodology brings together geometry and statistics in our attempt to find the confocal ellipse 
      that passes the closest to the maximum number of known orbits of a certain dynamical class; this approach is fundamentally different 
      from those involving $N$-body calculations and statistics (see e.g., \citealt{2016MNRAS.460L.123D,2017MNRAS.467L..66D,
      2017Ap&SS.362..198D}). \citet{2020AstBu..75..459K} recently applied a geometric approach within the context of the Planet~9 
      hypothesis.  

      \subsection{Methodology}
         Two arbitrary orbits may experience close encounters when their mutual nodal distance is sufficiently small. Recurrent (or even 
         single) encounters within 1~Hill radius \citep{1992Icar...96...43H} of a massive body may change the orbit of a small body 
         significantly. The mutual nodal distance between the orbits of a small body (an ETNO in our case) and an arbitrary planet can be 
         computed as described in Appendix~\ref{nodaleqns}. Orbits are defined by the values of semi-major axis, $a$ (that controls size), 
         eccentricity, $e$ (that controls shape), and those of the angular elements --- inclination, $i$, longitude of the ascending node, 
         $\Omega$, and argument of perihelion, $\omega  $ ---  that control the orientation in space of the orbit; the perihelion distance or
         pericenter, $q$, is given by the expression $q = a\,(1 - e)$. The actual position of an object in its orbit is controled by a 
         fourth angle, the mean anomaly, $M$. Our geometric approach leaves this angle out of the analysis and, therefore, it is not capable of 
         predicting the current location of the perturber, if it is, in fact, real. 

         We generated 2$\times$10$^{10}$ random planetary orbits with uniformly distributed relevant parameters: $\Omega_{\rm p}$ and 
         $\omega_{\rm p}$ $\in(0\degr, 360\degr)$, $i_{\rm p}\in(0\degr, 80\degr)$, $e_{\rm p}\in(0, 0.9),$ and $q_{\rm p}\in(x, 1000 - x)$~AU, 
         with $x=300, 400, 500,$ and 600~AU so $a_{\rm p}=q_{\rm p}/(1-e_{\rm p})$. For each random planetary orbit, we computed the 
         mutual nodal distances between a set of synthetic ETNOs and the planet. Each set of synthetic ETNOs was generated using the 
         mean values and standard deviations of the orbit determinations of the known ETNOs as pointed out in Appendix~\ref{ETNOelements}. 

         For each combination of random planetary orbit and synthetic (but compatible with the observations) set of ETNOs, we have counted 
         how many synthetic ETNOs had at least one mutual nodal distance with the planet under 5~AU (for $x=300$~AU), 7.5~AU (for 
         $x=400$~AU), and 10~AU (for $x=500$ and 600~AU). We then proceeded to record the random planetary orbit if the count was $\geq5$, 
         in order to maximize the number of potential close approaches between planet and set of ETNOs. We then studied the resulting 
         distributions. In order to analyze the results, we produced histograms using the Matplotlib library \citep{2007CSE.....9...90H} 
         with sets of bins computed using NumPy \citep{2011CSE....13b..22V} by applying the Freedman and Diaconis rule \citep{FD81}. 
         Instead of using frequency-based histograms, we considered counts to form a probability density so the area under the histogram 
         will sum to one.

         The nodal distance separation criteria for selective counting are not arbitrary but motivated by the results in 
         \citet{2020A&A...640A...6F}, a 2~$M_{\oplus}$ has a Hill radius of 4.5~AU (if $a_{\rm p}=400$~AU and $e_{\rm p}=0.1$), a 
         5~$M_{\oplus}$ has a Hill radius of 9.2~AU (if $a_{\rm p}=542$~AU and $e_{\rm p}=0.01$). Therefore, we implicitly assume that the 
         farther away the pericenter of the planet is, the more massive it may be.

      \subsection{Data sources}
         Here, we work with publicly available data from Jet Propulsion Laboratory's (JPL) Small-Body Database 
         (SBDB)\footnote{\href{https://ssd.jpl.nasa.gov/sbdb.cgi}{https://ssd.jpl.nasa.gov/sbdb.cgi}} and HORIZONS on-line solar system data 
         and ephemeris computation service,\footnote{\href{https://ssd.jpl.nasa.gov/?horizons}{https://ssd.jpl.nasa.gov/?horizons}} both 
         provided by the Solar System Dynamics Group \citep{2015IAUGA..2256293G}. Assuming the definition in \citet{2014Natur.507..471T}, 
         that the ETNOs have $q>30$~AU and $a>150$~AU, the sample of known ETNOs now includes 39 objects with reliable orbits (see 
         Appendix~\ref{ETNOelements}) whose data have been retrieved from JPL's SBDB and HORIZONS using tools provided by the Python package 
         Astroquery \citep{2019AJ....157...98G}. In the following section, we use barycentric elements because within the context of the 
         ETNOs, barycentric orbit determinations better account  for their changing nature as Jupiter follows its 12~yr orbit around the Sun; 
         Appendix~\ref{heliocentric} shows results based on heliocentric orbits that are consistent with those of the barycentric ones. 

   \section{Results and discussion\label{Results}} 
\begin{figure*}
        \centering
         \includegraphics[width=0.245\linewidth]{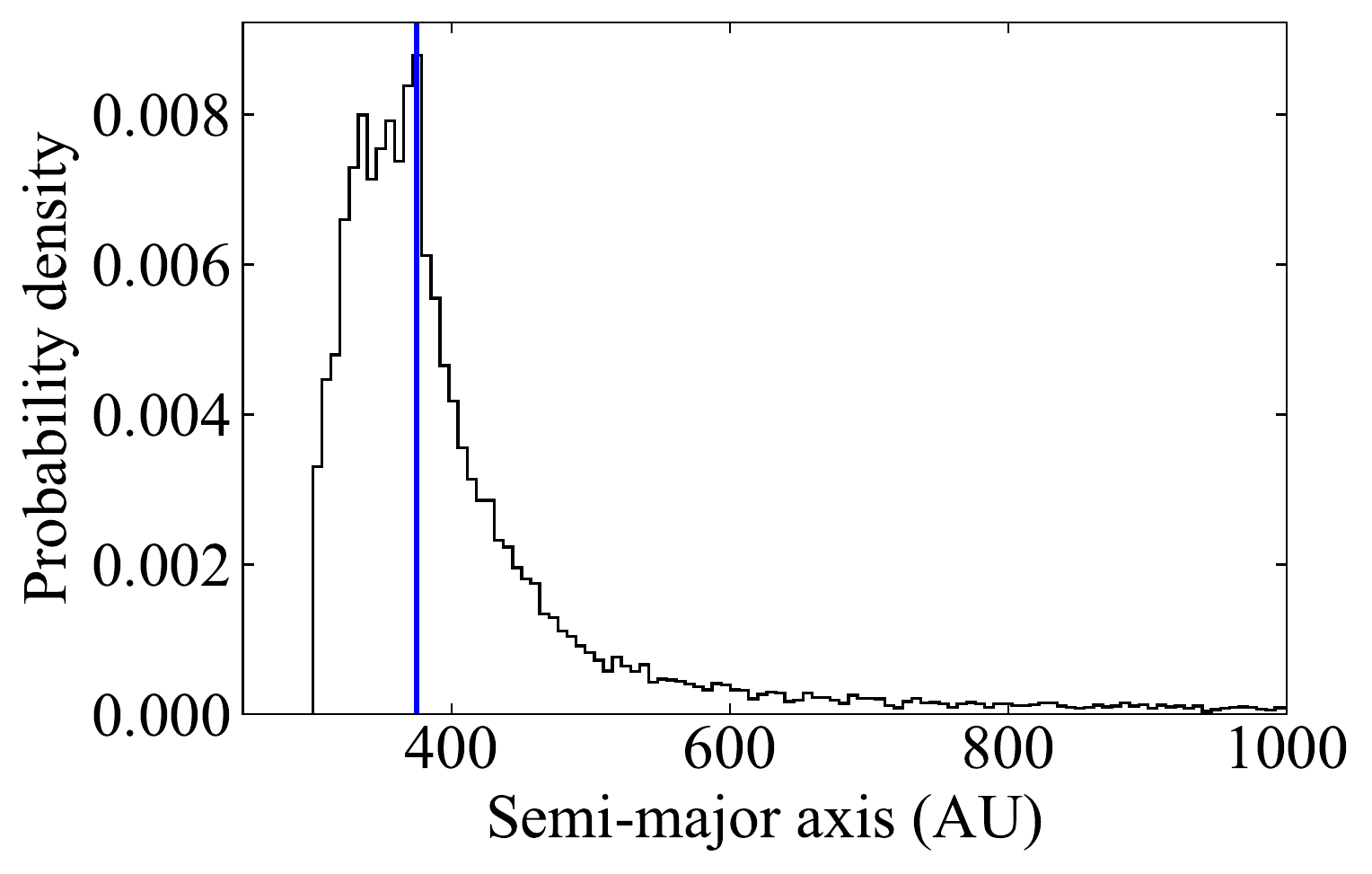}
         \includegraphics[width=0.245\linewidth]{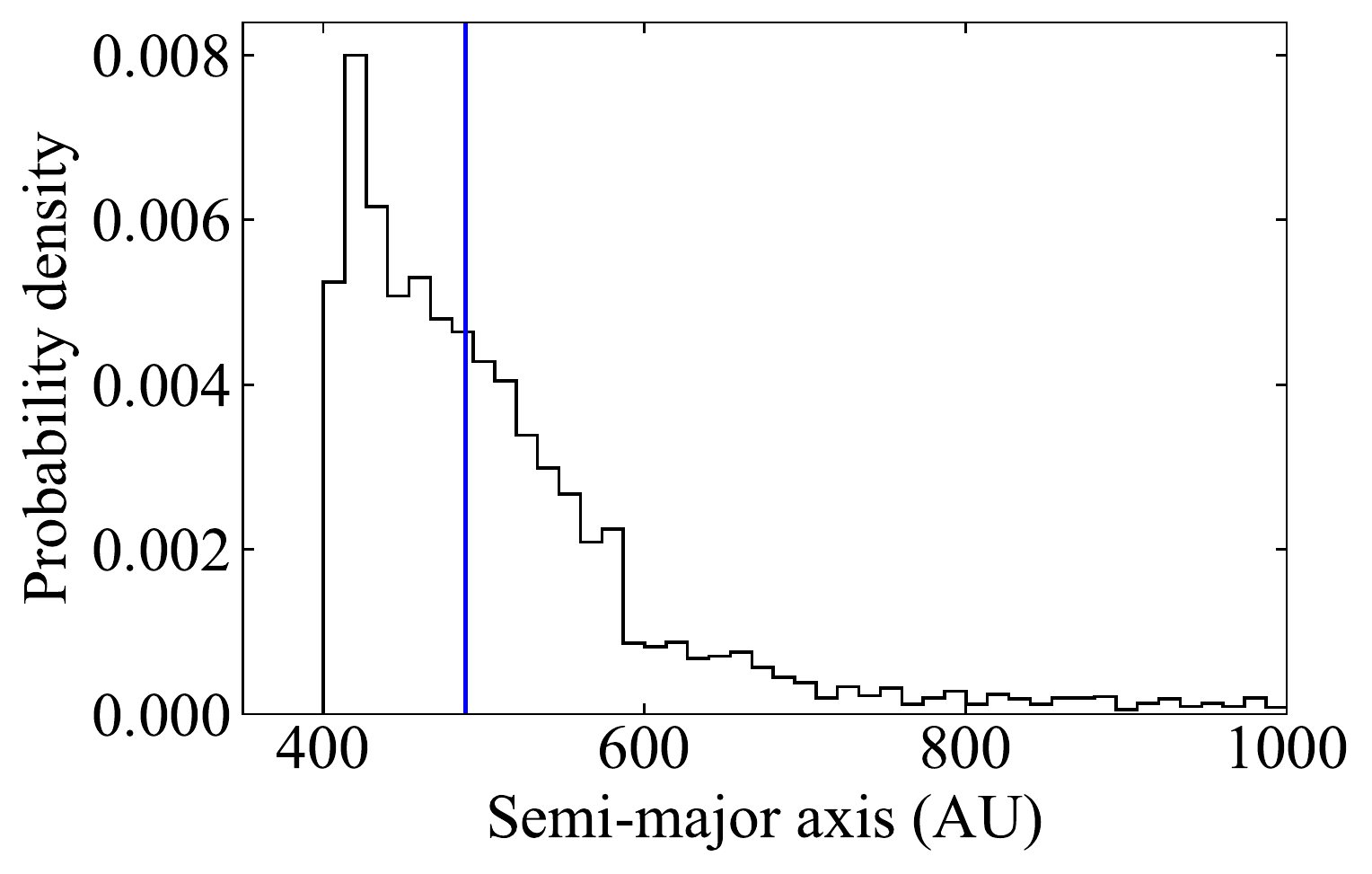}
         \includegraphics[width=0.245\linewidth]{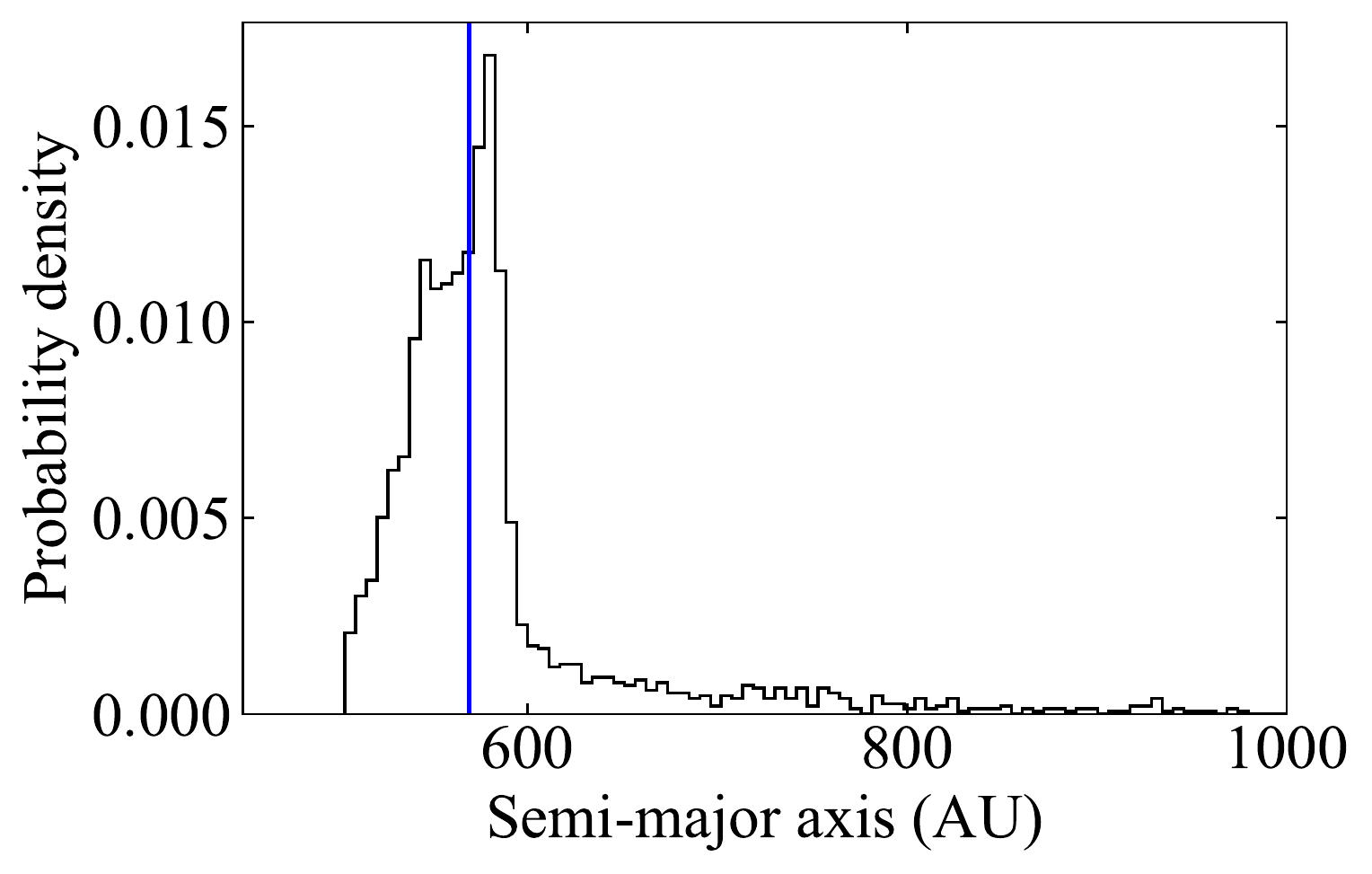}
         \includegraphics[width=0.245\linewidth]{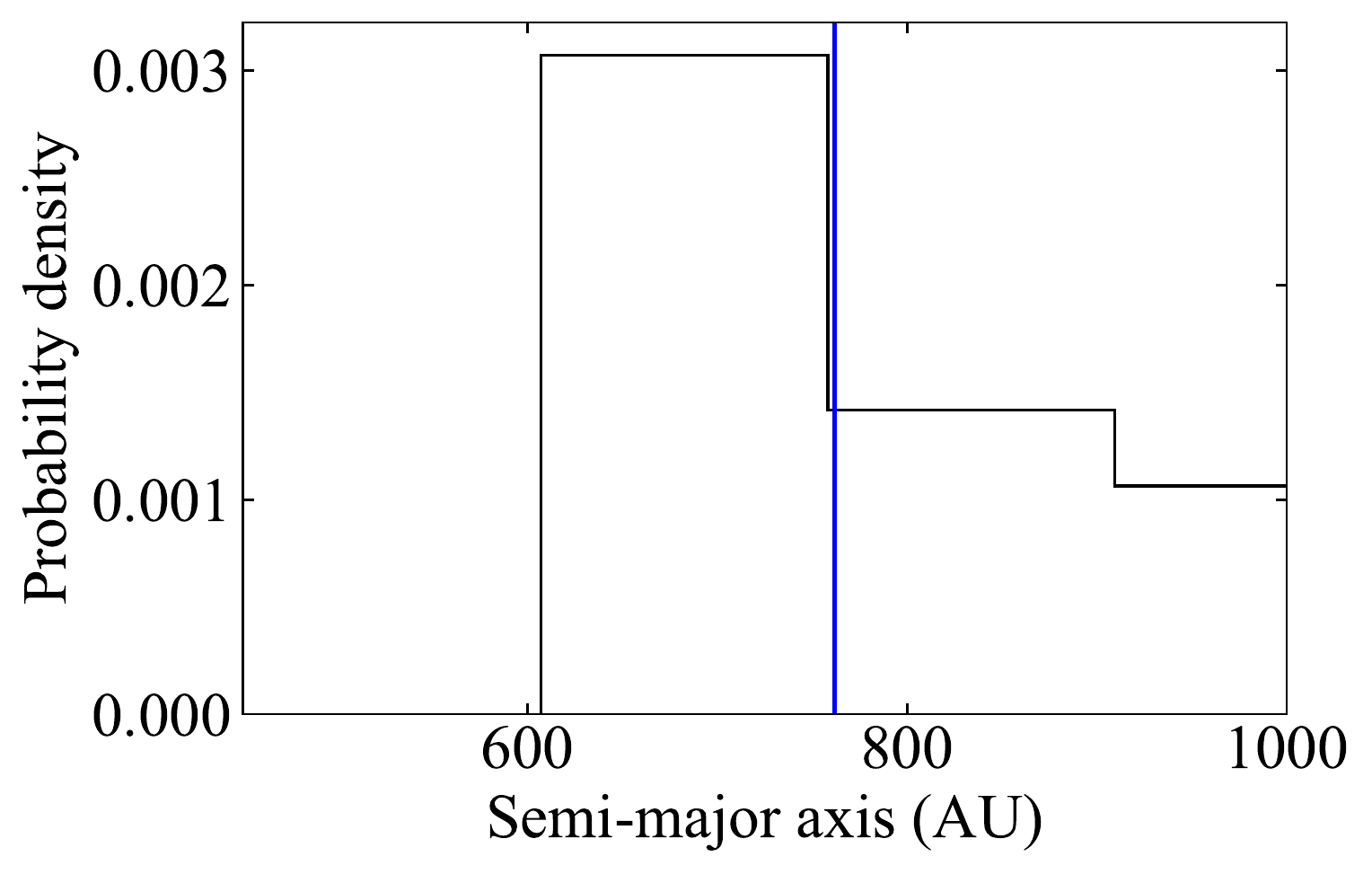}
         \includegraphics[width=0.245\linewidth]{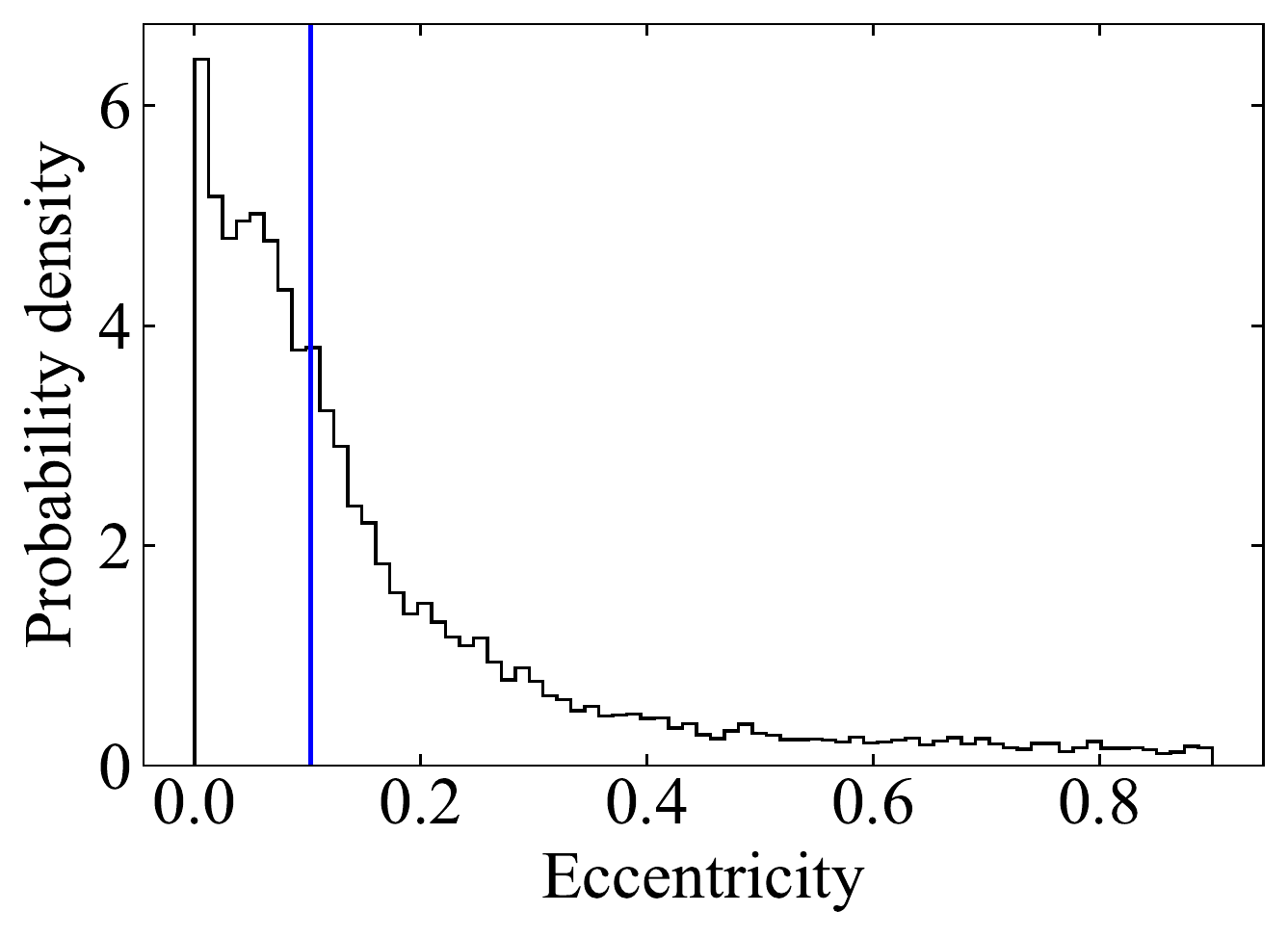}
         \includegraphics[width=0.245\linewidth]{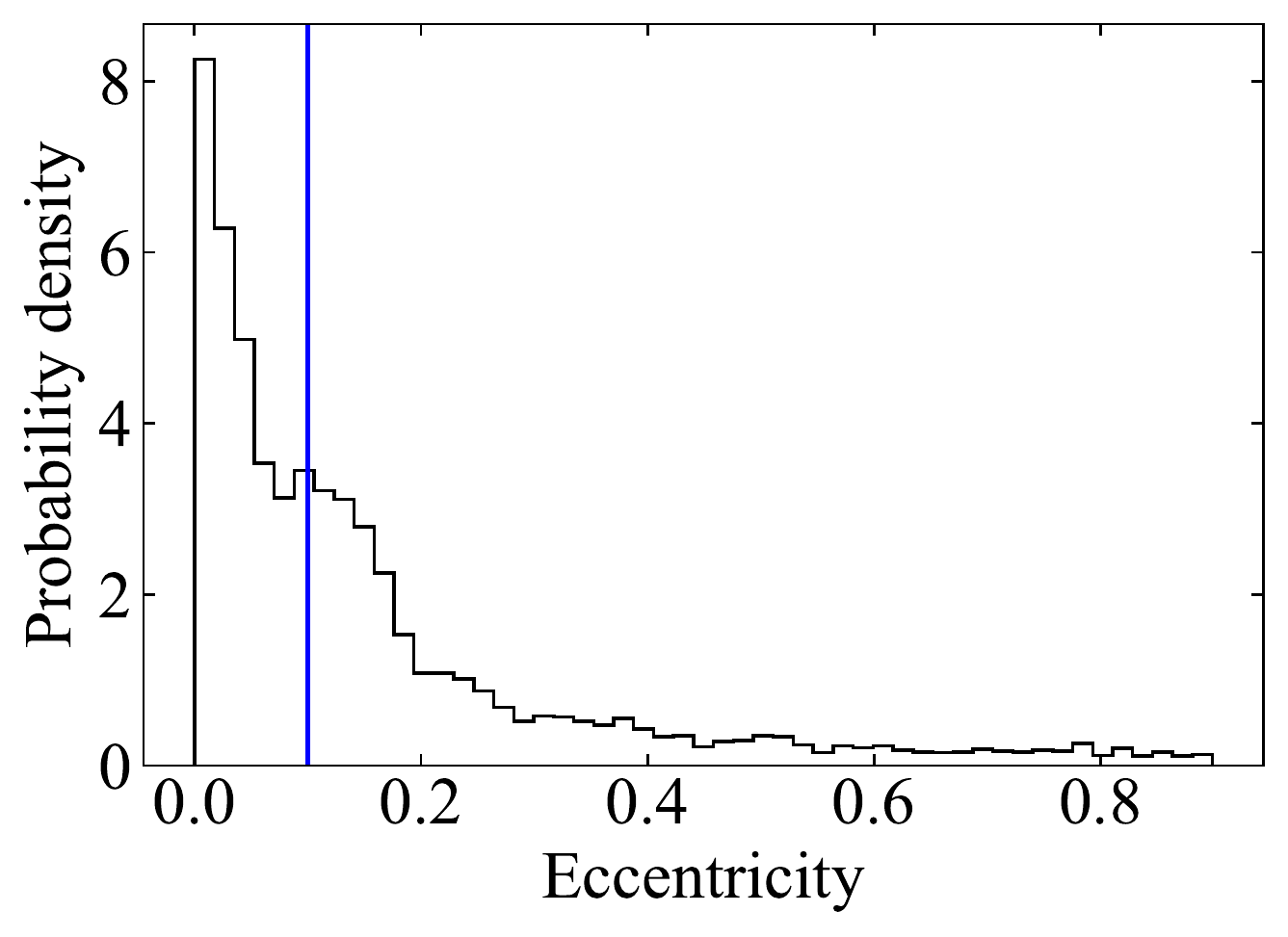}
         \includegraphics[width=0.245\linewidth]{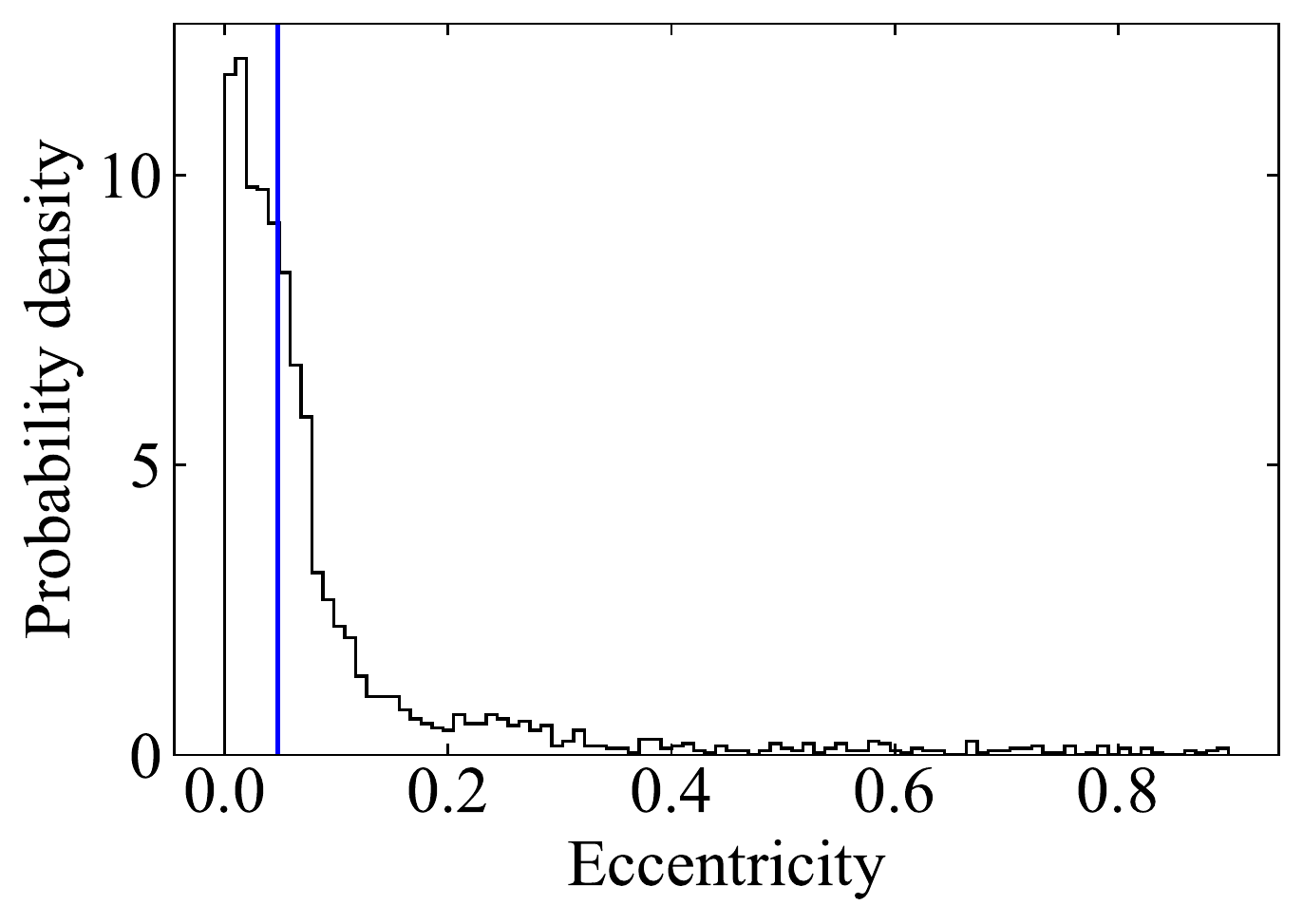}
         \includegraphics[width=0.245\linewidth]{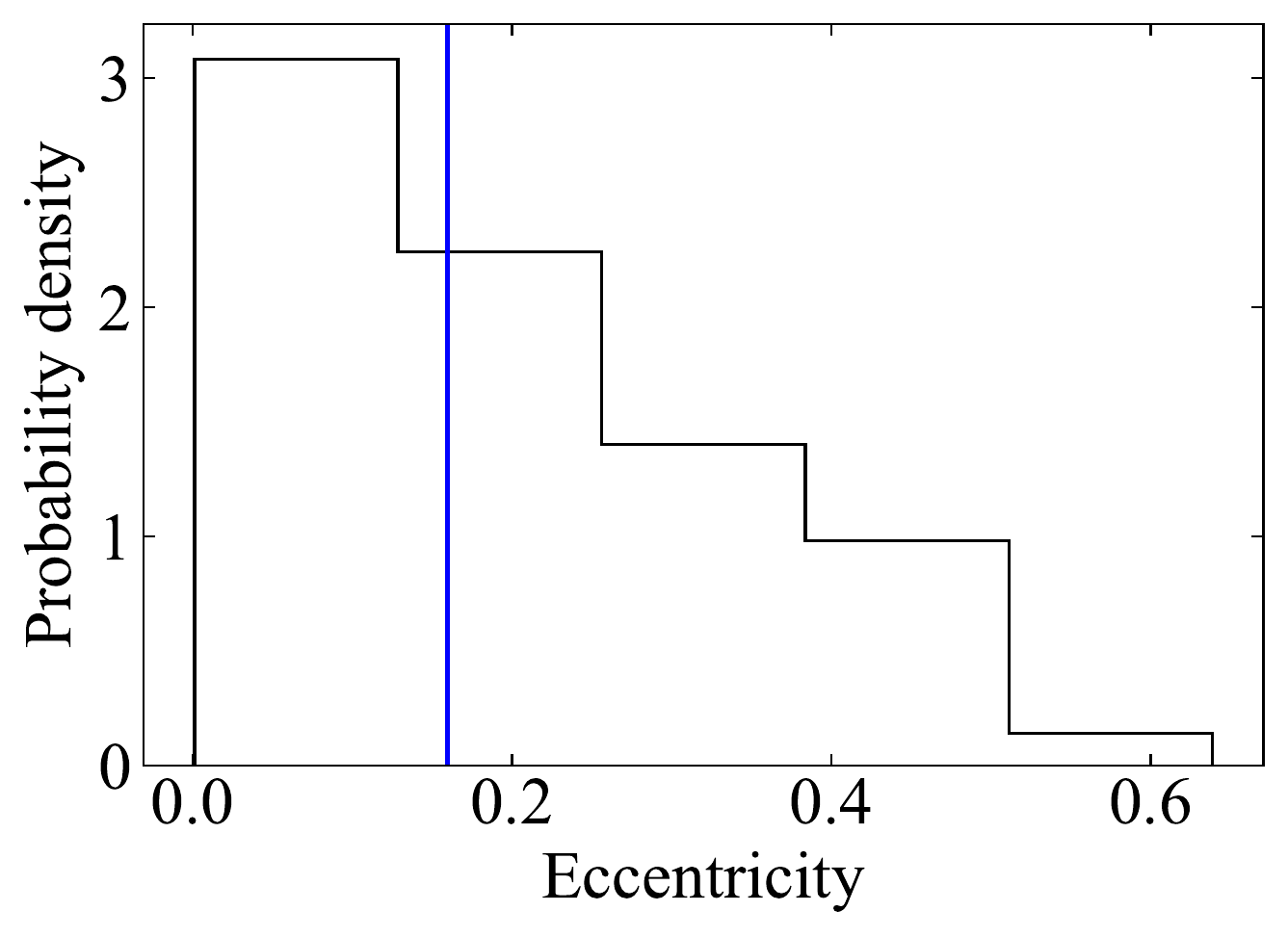}
         \includegraphics[width=0.245\linewidth]{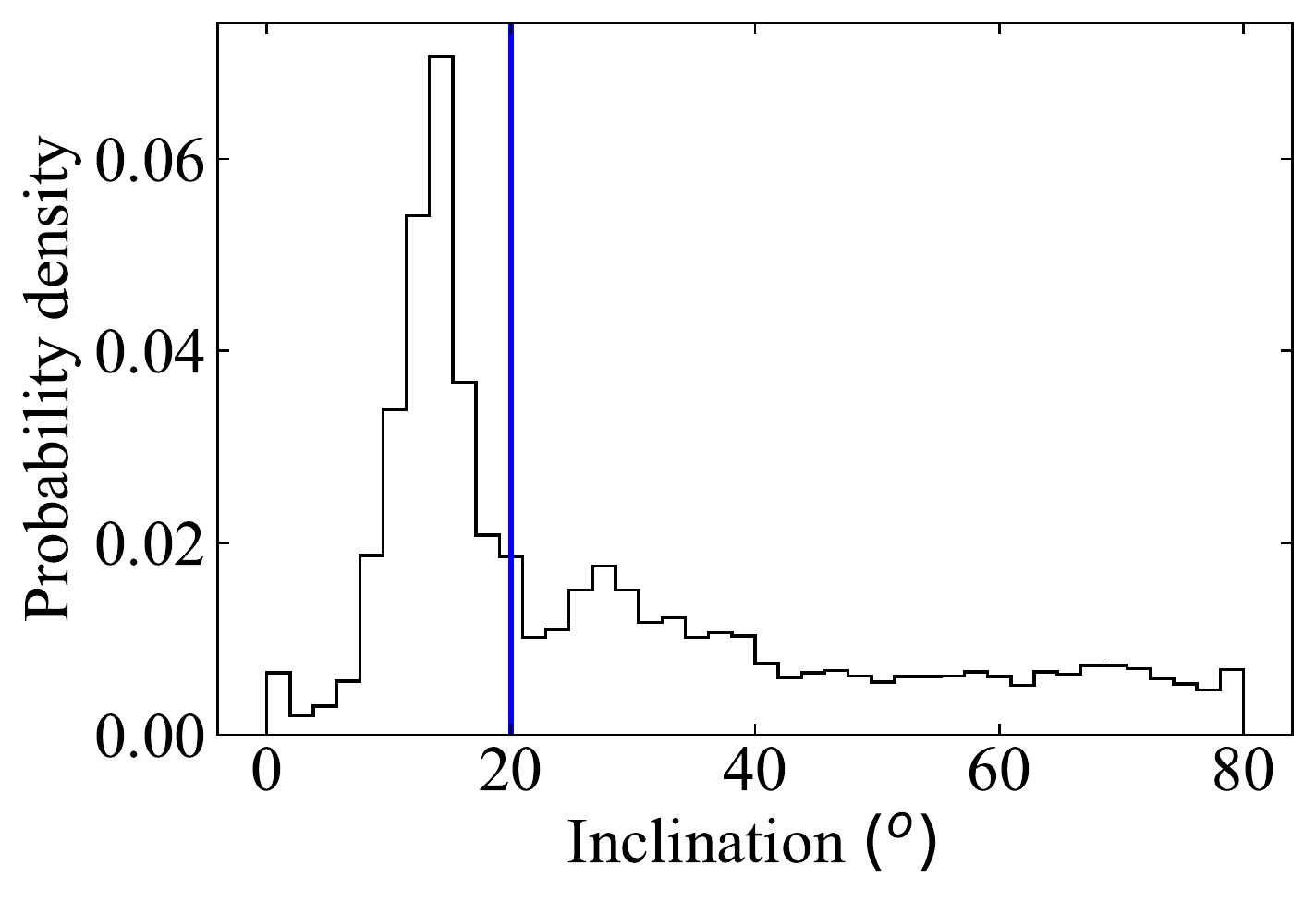}
         \includegraphics[width=0.245\linewidth]{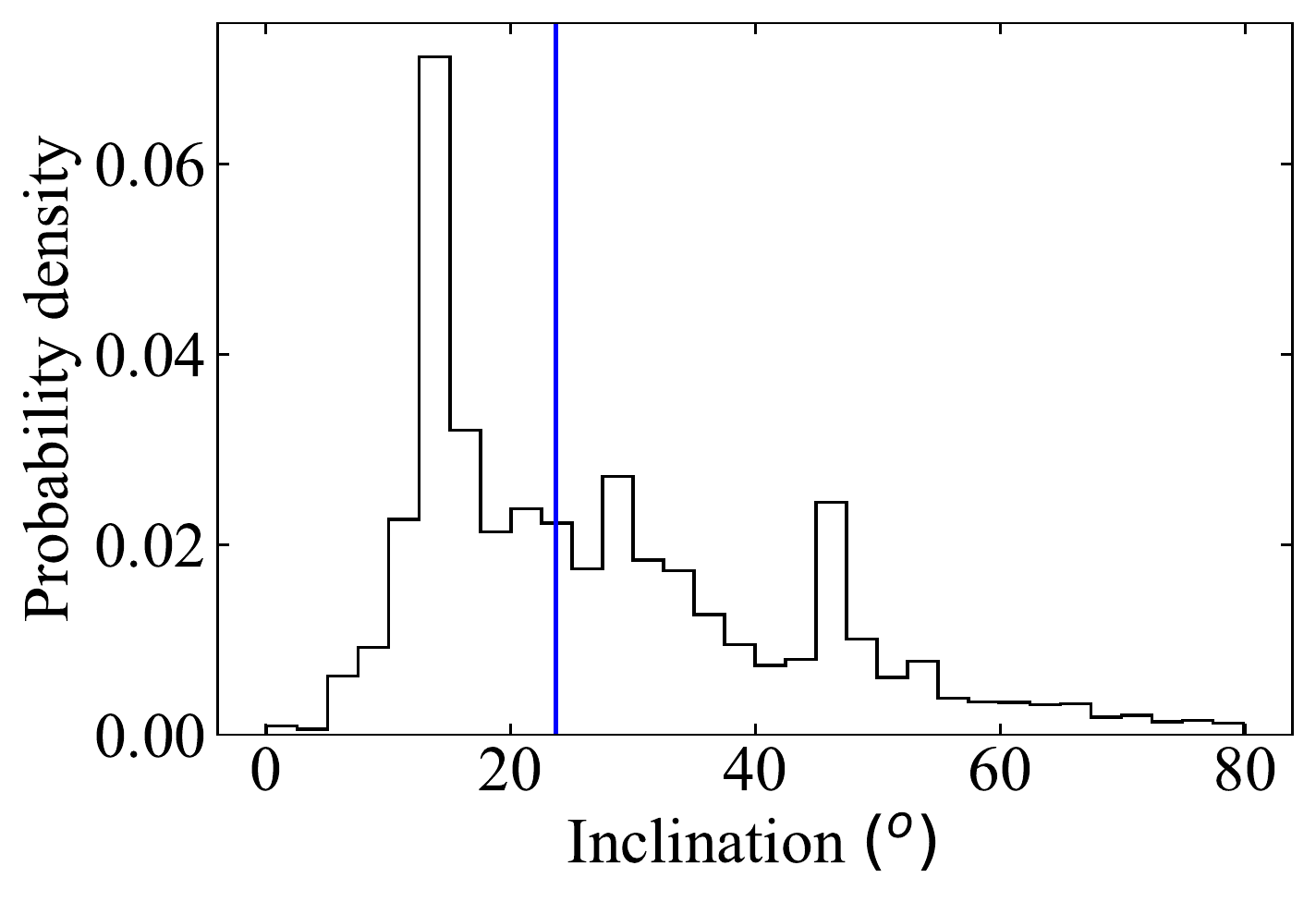}
         \includegraphics[width=0.245\linewidth]{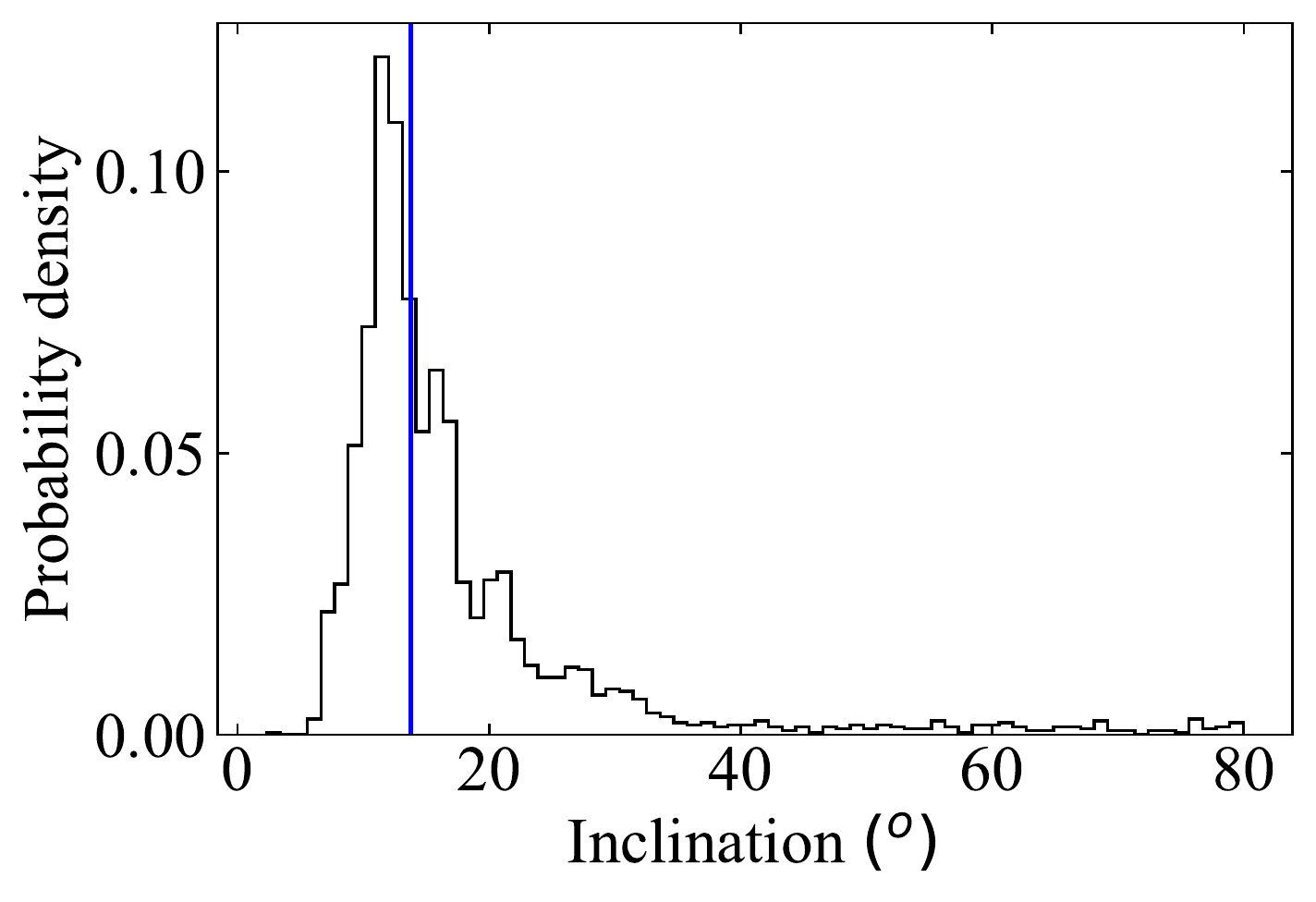}
         \includegraphics[width=0.245\linewidth]{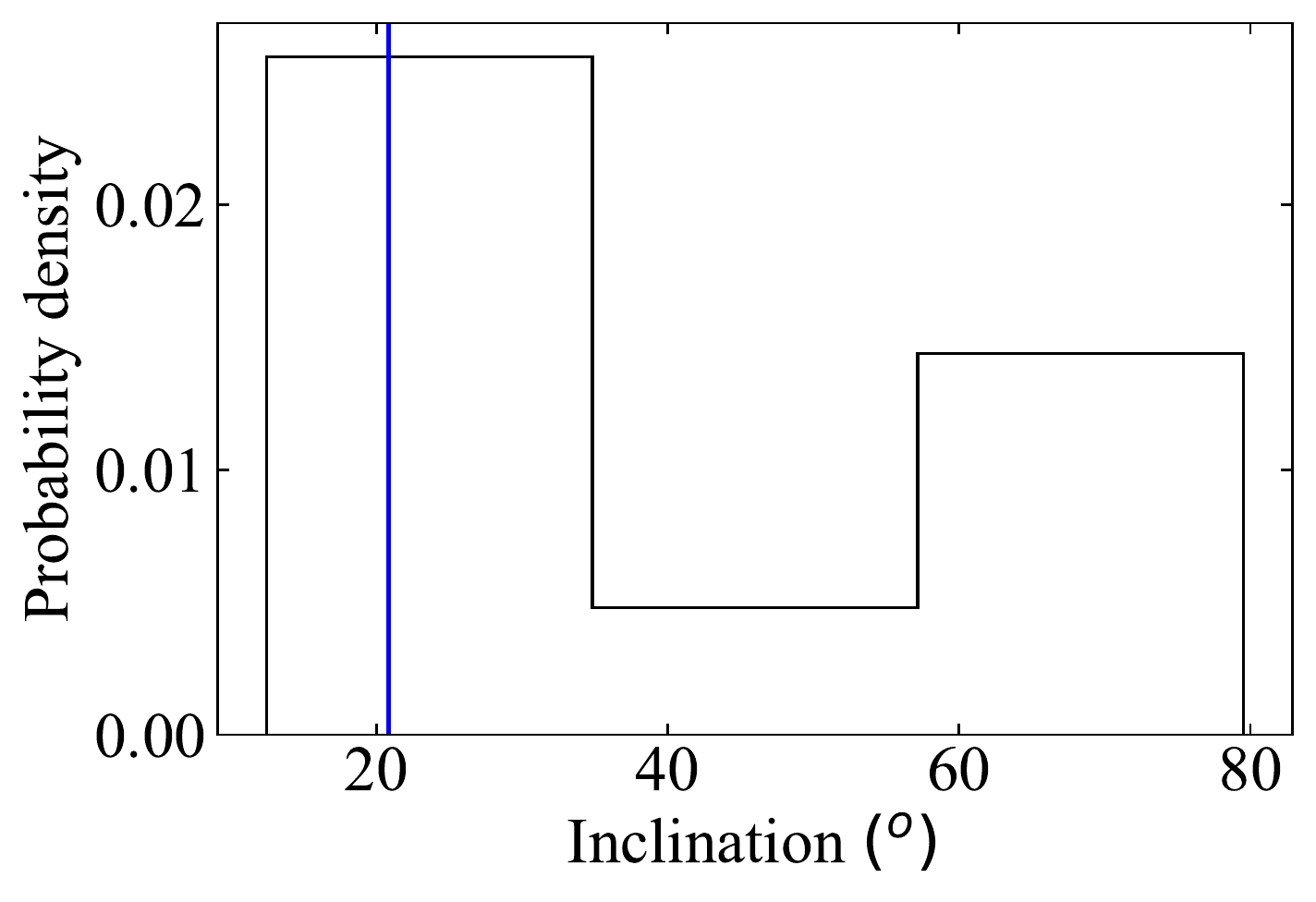}
         \includegraphics[width=0.245\linewidth]{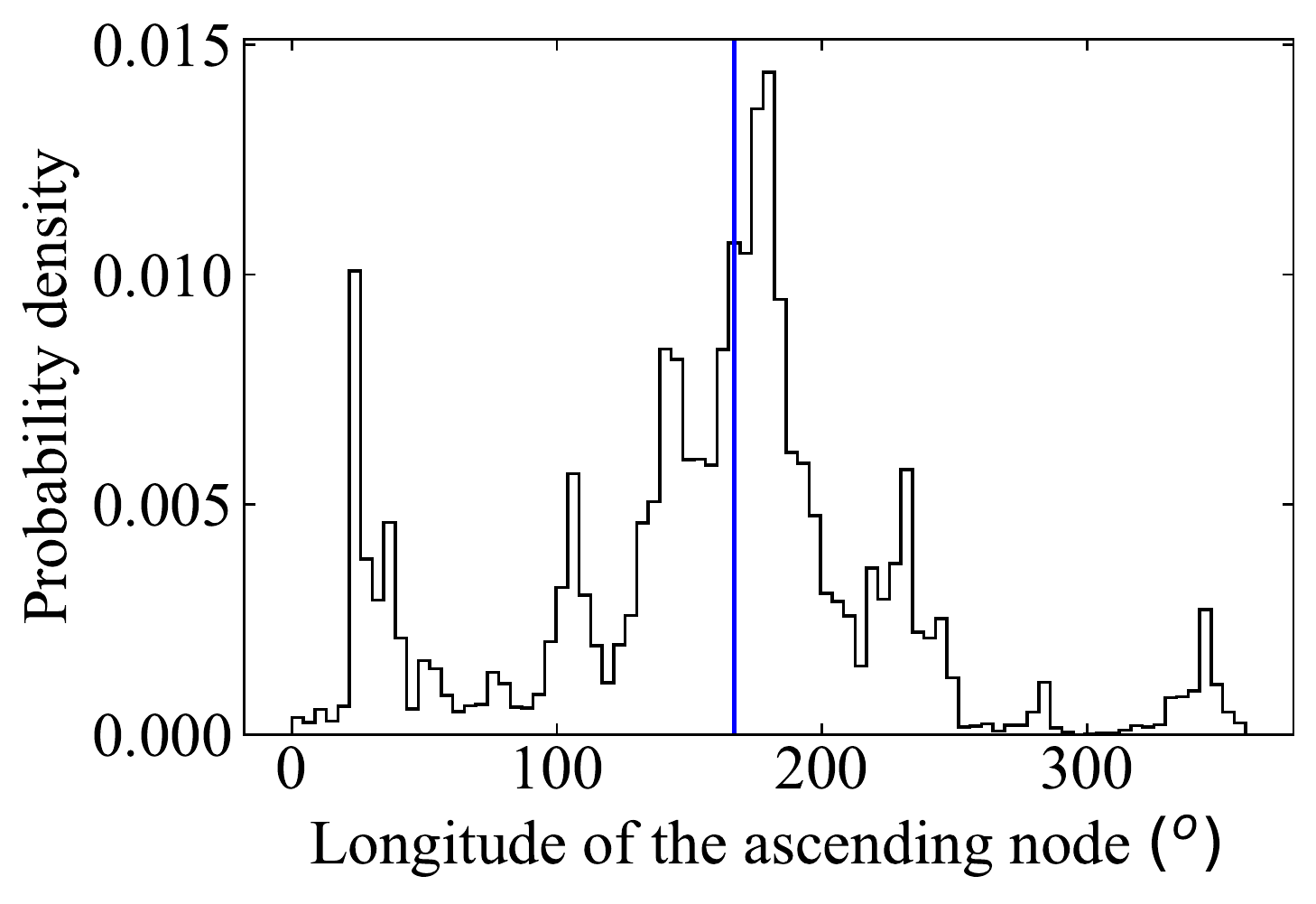}
         \includegraphics[width=0.245\linewidth]{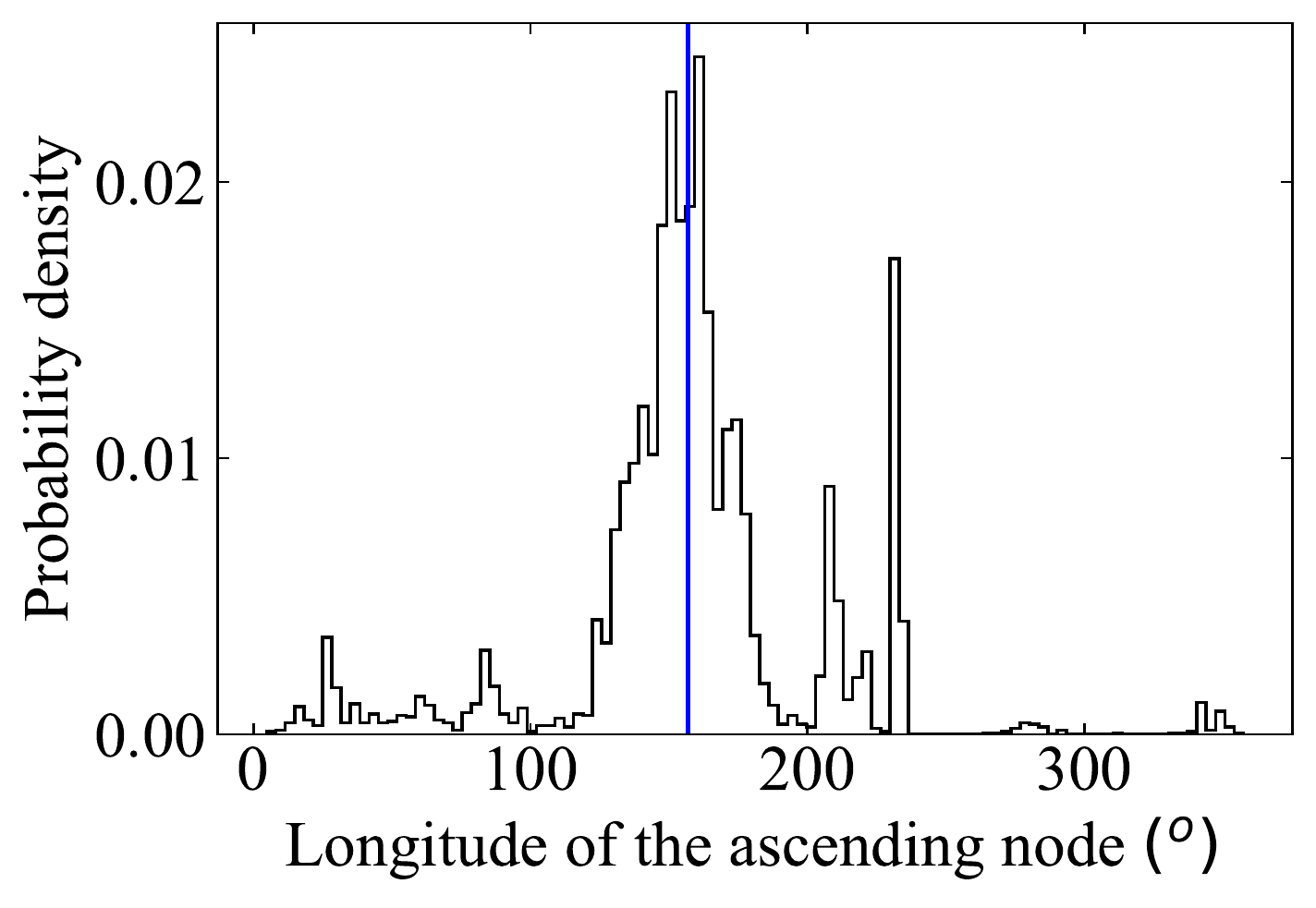}
         \includegraphics[width=0.245\linewidth]{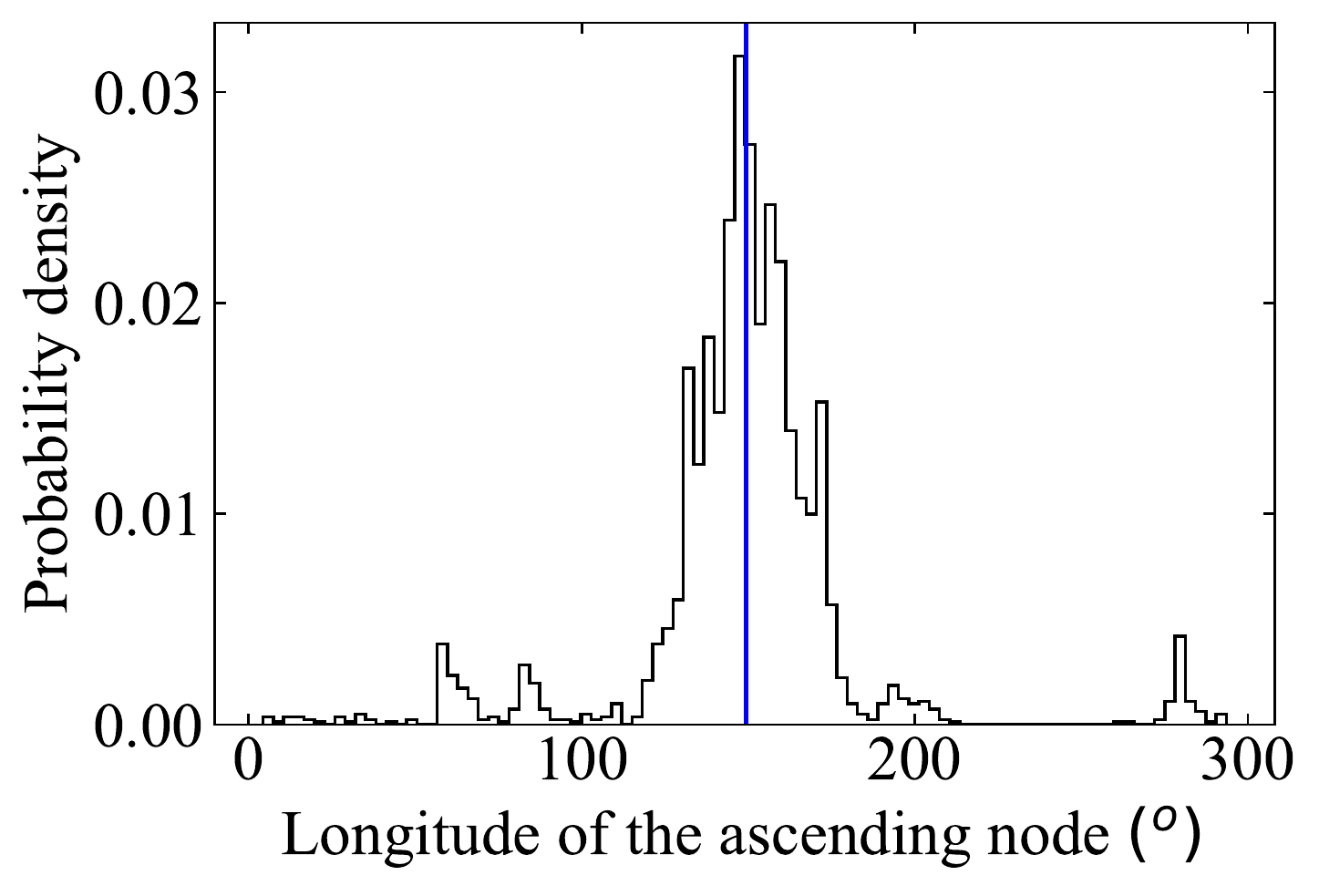}
         \includegraphics[width=0.245\linewidth]{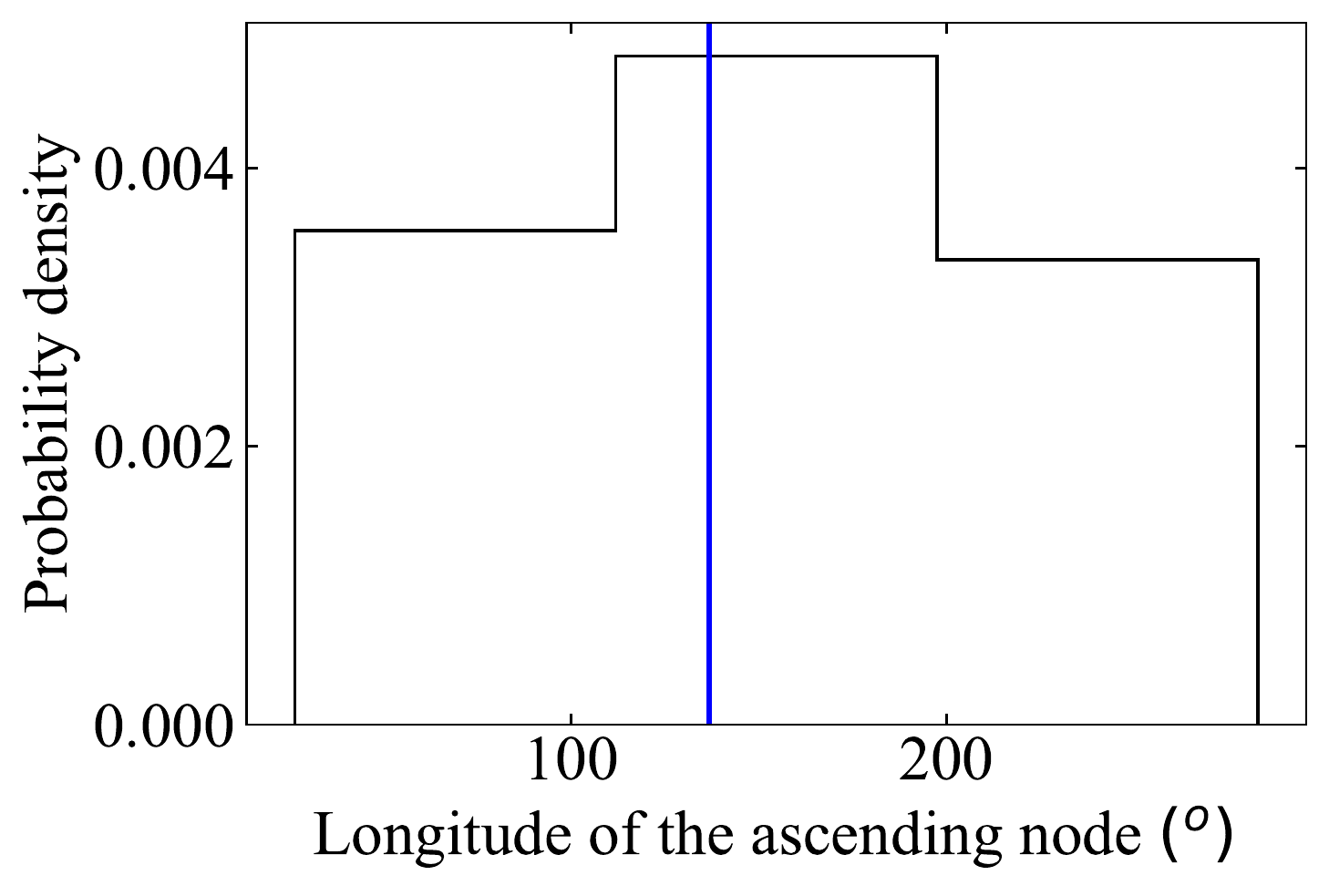}
         \includegraphics[width=0.245\linewidth]{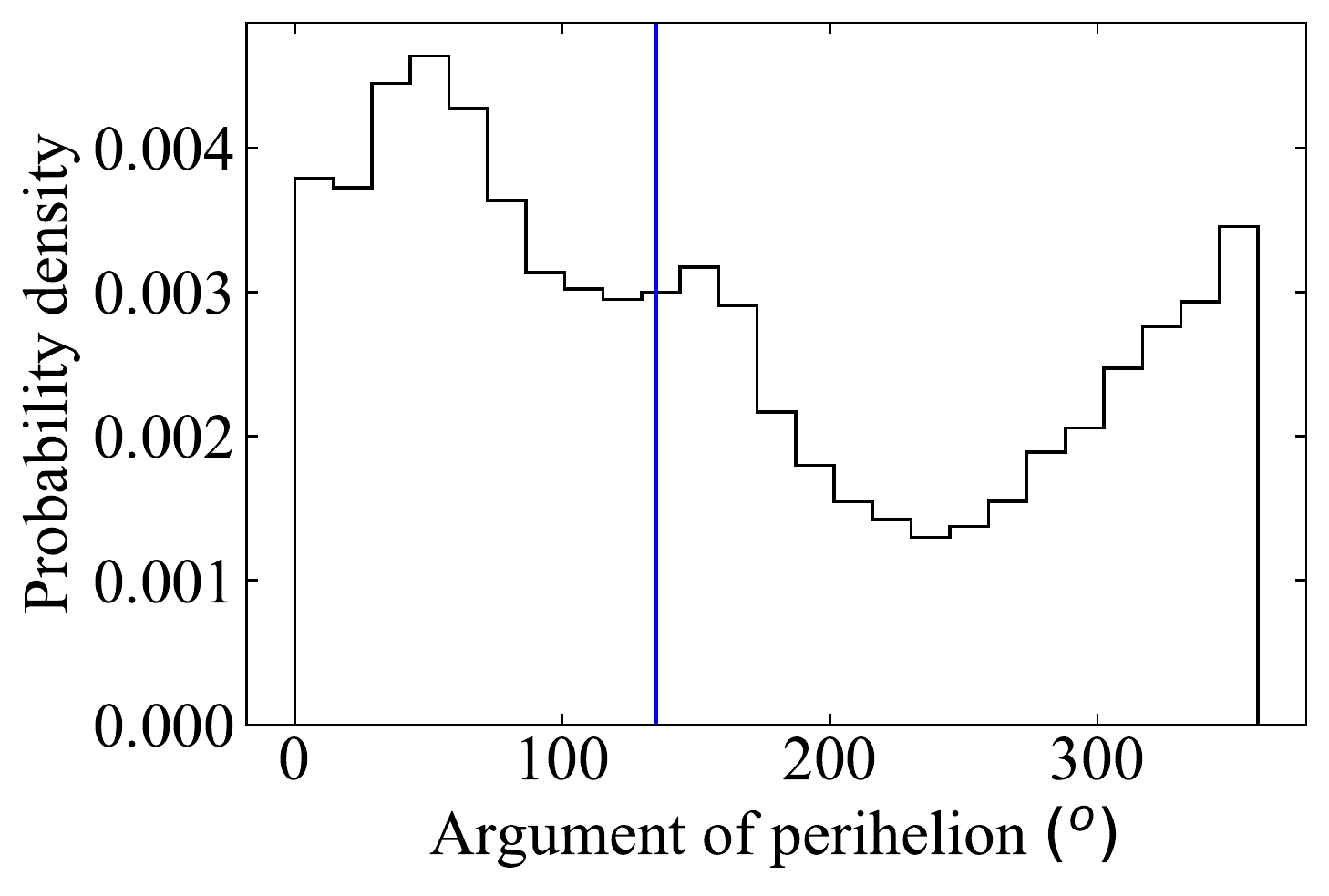}
         \includegraphics[width=0.245\linewidth]{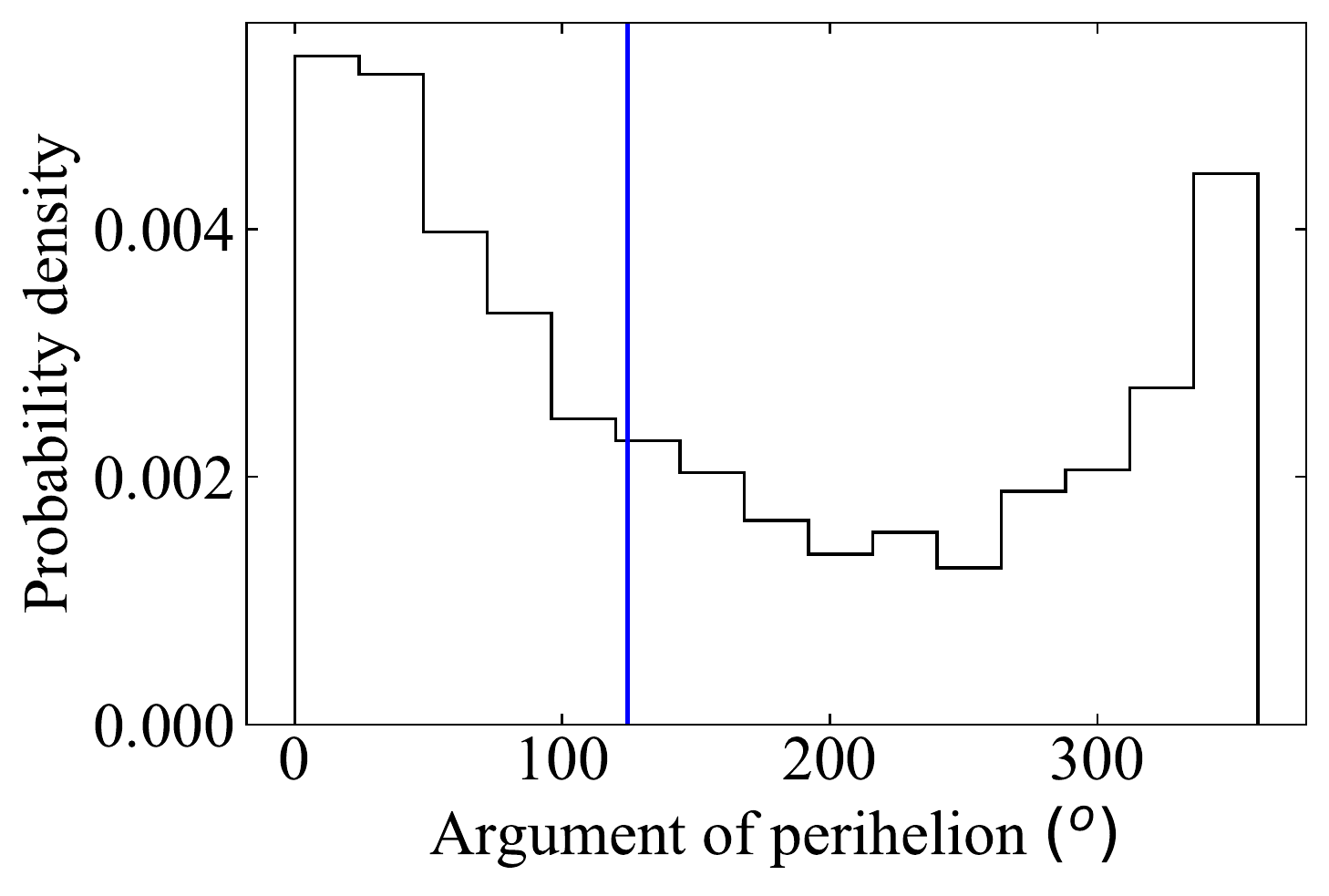}
         \includegraphics[width=0.245\linewidth]{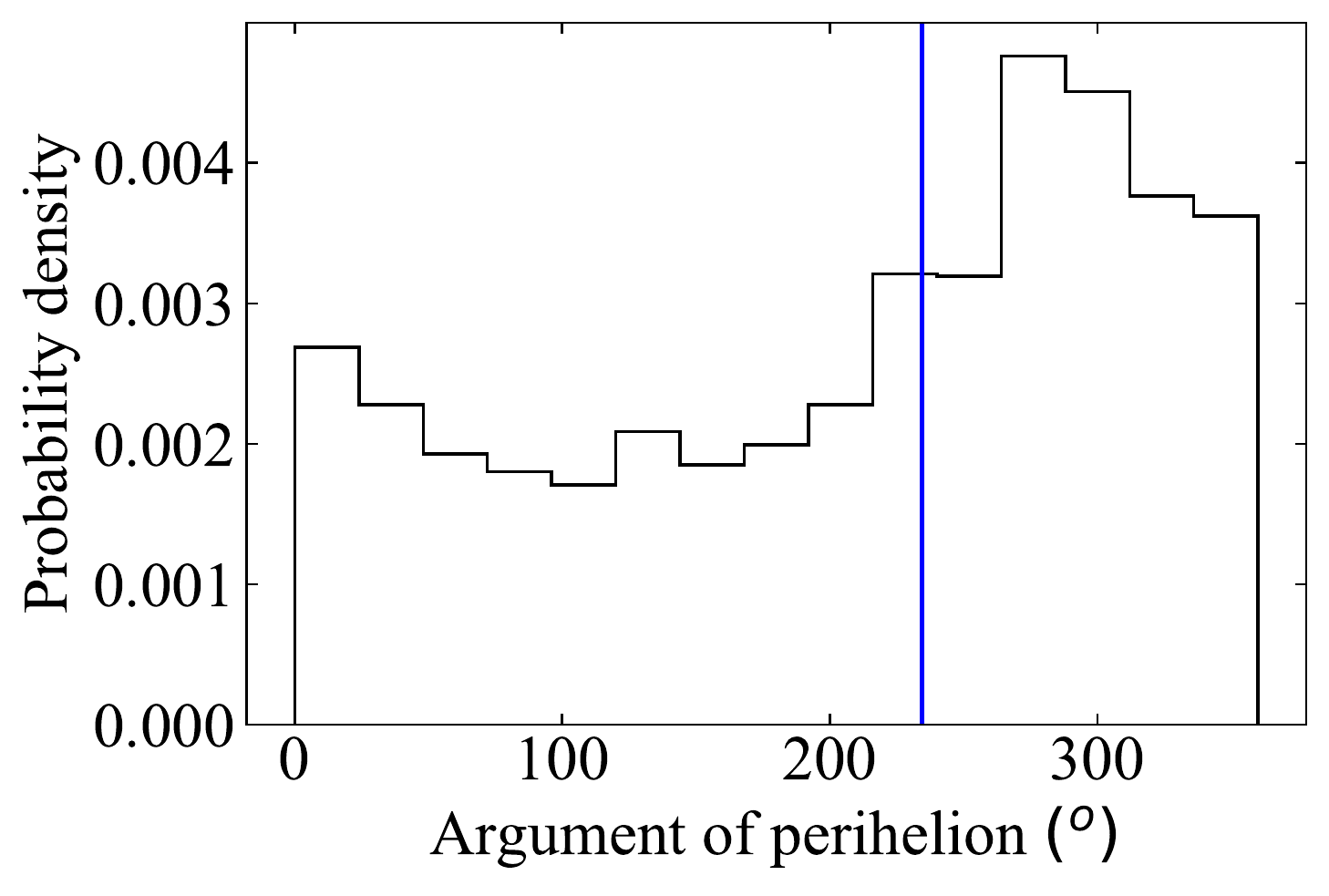}
         \includegraphics[width=0.245\linewidth]{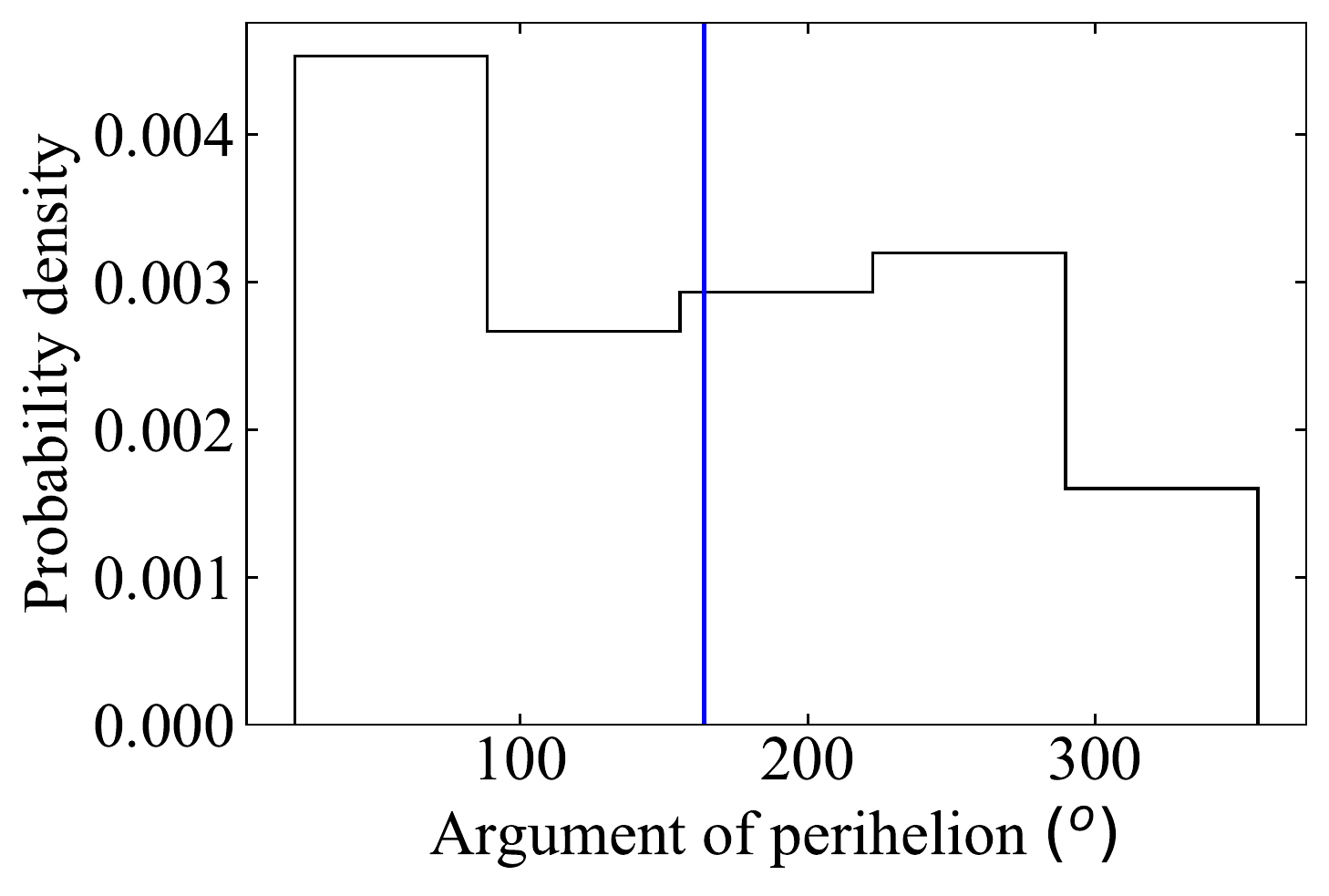}
         \caption{Barycentric orbital elements of putative perturber planets. Distributions of barycentric orbital elements of planetary 
                  orbits that may result in close encounters (under 5~AU for $q_{\rm p}>300$~AU, under 7.5~AU for $q_{\rm p}>400$~AU, and 
                  under 10~AU for $q_{\rm p}>500$~AU and $q_{\rm p}>600$~AU) with five or more present-day extreme trans-Neptunian objects 
                  (ETNOs). Each column of panels shows the cumulative results of 2$\times$10$^{10}$ experiments. From left to right, the panels 
                  show the results of imposing $q_{\rm p}>300$~AU (left column, 20304 orbits), $q_{\rm p}>400$~AU (5671 orbits), 
                  $q_{\rm p}>500$~AU (2635 orbits), and $q_{\rm p}>600$~AU (right column, 56 orbits). Median values are shown as vertical 
                  blue lines. Based on the available data on known ETNOs, the presence of massive perturbers well beyond 600~AU is strongly 
                  excluded within the context of the hypotheses centered in this study.
                 }
         \label{histogramsB}
      \end{figure*}

      As pointed out above, the motivation behind this study is the belief that a fossil record of planetary encounters might be preserved 
      in the distribution of the orbital elements of ETNOs. With this hypothesis in mind, we monitored which planetary orbits had 
      the maximum number of potential close encounters with synthetic ETNOs and analyzed the resulting distributions that are shown in 
      Fig.~\ref{histogramsB}. The first (leftmost) column of panels shows the results based on the assumption that $q_{\rm p}>300$~AU,  including 
      20304 orbits with a number of potential close approaches in the range between 5--7; the next one, shows results for $q_{\rm p}>400$~AU also 
      with a range of potential close approaches of 5--7 for 5671 orbits; the following column of panels displays results for 
      $q_{\rm p}>500$~AU with a range of 5--6 for 2635 orbits; the right column shows panels with results for $q_{\rm p}>600$~AU and the 
      number of potential close approaches between planet and set of ETNOs is just 5 for 56 orbits. From these values and the plots, it is
      increasingly difficult to find consistent perturbers as we move farther away from the Sun.
%
%
%
%
%
%
     \begin{table*}
        \fontsize{8}{12pt}\selectfont
        \tabcolsep 0.15truecm
        \caption{\label{resultsB}Summary of central values and dispersions of optimal barycentric orbits. 
                }
        \centering
        \begin{tabular}{lcccc}
           \hline\hline
            Orbital parameter                                         & $q_{\rm p}>300$~AU            & $q_{\rm p}>400$~AU 
                                                                      & $q_{\rm p}>500$~AU            & $q_{\rm p}>600$~AU            \\
           \hline
            Semi-major axis, $a_{\rm p}$ (AU)                         & 375$_{-44}^{+111}$ (376)      & 489$_{-64}^{+130}$ (420)
                                                                      & 569$_{-31}^{+55}$ (581)       & 762$_{-120}^{+279}$ (680)     \\
            Eccentricity, $e_{\rm p}$                                 & 0.10$_{-0.07}^{+0.21}$ (0.01) & 0.10$_{-0.08}^{+0.19}$ (0.01)  
                                                                      & 0.05$_{-0.03}^{+0.09}$ (0.02) & 0.16$_{-0.13}^{+0.19}$ (0.07) \\
            Inclination, $i_{\rm p}$ (\degr)                          & 20$_{-8}^{+34}$ (14)          & 24$_{-10}^{+22}$ (14)     
                                                                      & 14$_{-3}^{+9}$ (12)           & 21$_{-6}^{+52}$ (24)          \\
            Longitude of the ascending node, $\Omega_{\rm p}$ (\degr) & 167$_{-70}^{+44}$ (180)       & 157$_{-23}^{+48}$ (161) 
                                                                      & 149$_{-17}^{+18}$ (148)       & 137$_{-55}^{+141}$ (156)      \\
            Argument of perihelion, $\omega_{\rm p}$ (\degr)          & 135$_{-94}^{+170}$ (51)       & 124$_{-94}^{+193}$ (12) 
                                                                      & 234$_{-165}^{+81}$ (275)      & 164$_{-106}^{+109}$ (56)      \\
           \hline
        \end{tabular}
        \tablefoot{Median values and 16th and 84th percentiles (absolute maximum in parentheses) from the Monte Carlo random searches whose 
                   distributions are shown in Fig.~\ref{histogramsB}.
                  }
     \end{table*}
%
%

      Table~\ref{resultsB} shows a summary of the central values and dispersions of the orbital parameters of the sample of orbits that may
      experience close encounters with multiple known ETNOs. We consider that results for $q_{\rm p}>300$~AU and $q_{\rm p}>400$~AU are
      statistically consistent and are compatible with $a_{\rm p}\in(331, 489)$~AU, $e_{\rm p}<0.1$, 
      $i_{\rm p}\in(10\degr,20\degr)$, $\Omega_{\rm p}$$\sim$180{\degr}, and $\omega_{\rm p}\in(-60{\degr}, 60{\degr})$. The value of 
      $a_{\rm p}$ is probably $\sim$400~AU but the uncertainty is significant, the value of the eccentricity is well constrained and it has
      to be low, the inclination is perhaps the best constrained value and it is $\sim$14{\degr}, the value of $\Omega_{\rm p}$ is very 
      likely $\sim$180{\degr}, but the value of $\omega_{\rm p}$ is poorly constrained, perhaps $\sim$50{\degr}. If scattering is the main 
      source of orbital modification for the group of ETNOs that mainly move at (300, 400)~AU from the Sun, the orbit of the putative
      perturber is relatively well-constrained and according to \citet{2020A&A...640A...6F}, it must have a mass $<5~M_{\oplus}$.

      Table~\ref{resultsB} shows that our approach is far less conclusive in the cases of $q_{\rm p}>500$~AU and $q_{\rm p}>600$~AU
as these values produce distributions of the orbital parameters that may not be statistically compatible. In any case, we must emphasize that $>50$\%
      of the known ETNOs cannot interact directly with a perturber with $q_{\rm p}>500$~AU because their aphelion distances, $Q=a\,(1+e)$,
      are below 500~AU: 9 ETNOs have $Q<300$~AU, 18 have $Q<400$~AU, and 23 have $Q<500$~AU. Therefore, the distribution in $Q$ and our 
      calculations suggest that more than one perturber is required if scattering is the main source of orbital modification for the known 
      ETNOs. Perturbers might not be located farther than 600~AU and they have to follow moderately eccentric and inclined orbits to be capable 
      of experiencing present-day close encounters with multiple known ETNOs.

      At this point, it can be argued that the results in Fig.~\ref{histogramsB} may be affected by a statistical artifact. In order to test the
      statistical significance of our results, we  repeated the Monte Carlo random search on an input sample of scrambled data as 
      explained in Appendix~\ref{significance}. By randomly reassigning the values of the orbital elements of the ETNOs, we preserve the 
      original distributions of the parameters, but destroy any possible correlations that may have been induced by close encounters with 
      massive perturbers (or the effects of hypothetical mean-motion or secular resonances). The results of this significance test are shown 
      in Fig.~\ref{histogramsCHKB}: the distribution of $i_{\rm p}$ becomes flat and almost the same happens to the distribution of 
      $\Omega_{\rm p}$. In other words, for the scrambled data, it is not possible to find a statistically significant orbital solution 
      that is compatible with the starting hypothesis and we conclude that the results in Fig.~\ref{histogramsB} are unlikely to be due to 
      statistical artifacts. 

   \section{Summary and conclusions\label{Conclusions}}
      When considering the well-studied case of Neptune and the regular trans-Neptunian objects, we observe that these objects are not part
      of a single population, but they are organized into several dynamical classes. Some objects never experience close encounters with 
      Neptune due to the existence of protection mechanisms such as mean-motion or secular resonances, as in the case of Neptune's 
      Trojans or Pluto (see e.g., \citealt{1989Icar...82..200M,2001AJ....121.1155W}), others undergo close encounters that may send 
      them towards the inner solar system (centaurs) or outwards to become scattered objects (see the recent review by 
      \citealt{2020CeMDA.132...12S}). If trans-Plutonian planets exist, their perturbations may shape the extreme trans-Neptunian space in a 
      similar fashion and the current sample of ETNOs may include the signatures of their presence (see e.g. 
      \citealt{2017CeMDA.129..329S}).

      Our results show that it is possible to find suitable orbits for which the mutual nodal distances of sizeable ($\geq5$, comprising at 
      least 13\% of the sample) groups of ETNOs with their assumed perturber could be small enough for a close encounter to occur, at least 
      in theory (assuming that no protection mechanisms, such as mean-motion or secular resonances, are in place to avoid the flyby). This 
      was our original aim. In addition, the results presented in Sect.~\ref{Results} clearly indicate that the most probable planetary 
      orbit consistent with the starting hypothesis is the one obtained for the experiment with $q_{\rm p}>300$~AU. The number of consistent 
      orbits in this case is 3.6 times higher than the one found for $q_{\rm p}>400$~AU. Our results seem to be incompatible with those attributed to a 
      statistical artifact (see Appendix~\ref{significance}).

      Prior to the announcement of the Planet~9 hypothesis \citep{2016AJ....151...22B}, \citet{2014Natur.507..471T} had already argued for 
      the existence of a planet at 250~AU within the context of the ETNOs --- driving von~Zeipel-Lidov-Kozai librations on 2012~VP$_{113}$ --- 
      and \citet{2014MNRAS.443L..59D} suggested that the limited data available at the time were more compatible with the presence of two 
      massive perturbers, one of them close to 300~AU. These massive perturbers were initially proposed based on data corresponding to a 
      sample of 13 objects, whereas the current sample has tripled this number. If we repeat the experiment discussed in Sect.~\ref{Results} for
      $q_{\rm p}>200$~AU (see Appendix~\ref{200au}), we find that the number of consistent orbits, although larger than the one generated in
      the experiment for $q_{\rm p}>400$~AU, is still lower than that of the most statistically significant case, namely, 8234 versus 20304 for 
      $q_{\rm p}>300$~AU. The most likely orbit is still similar, in terms of shape and orientation in space, to the most probable one in the 
      $q_{\rm p}>300$~AU case.

      Our results are consistent with the conclusions of the study presented in \citet{2017MNRAS.471L..61D}. Although our approach has not been able to single out a statistically significant, unique planetary orbit that may be responsible for the orbital architecture observed 
      in extreme trans-Neptunian space, we provide narrow ranges for the orbital parameters of putative planets that may have 
      experienced orbit-changing encounters with known ETNOs. Some sections of the available orbital parameter space are strongly 
      excluded by the findings of our analysis.

   \begin{acknowledgements}
      We thank the anonymous referee for a constructive and timely report, S.~J. Aarseth, J. de Le\'on, J. Licandro, A. Cabrera-Lavers, 
      J.-M. Petit, M.~T. Bannister, D.~P. Whitmire, G. Carraro, E. Costa, D. Fabrycky, A.~V. Tutukov, S. Mashchenko, S. Deen and J. Higley 
      for comments on ETNOs and A.~I. G\'omez de Castro for providing access to computing facilities. This work was partially supported by 
      the Spanish `Ministerio de Econom\'{\i}a y Competitividad' (MINECO) under grant ESP2017-87813-R. In preparation of this letter, we made 
      use of the NASA Astrophysics Data System and the MPC data server.
   \end{acknowledgements}

   \bibliographystyle{aa}

   \begin{appendix}
      \section{Mutual nodal distance: formulae\label{nodaleqns}}
         Our statistical analyses are based on studying the distribution of nodal distances between two Keplerian trajectories (one ETNO and 
         one hypothetical planet) with a common focus; therefore, the core of our approach is purely geometrical and gravitational 
         interactions are not directly taken into account. The mutual nodal distance between the orbits of a small body (an ETNO in our 
         case) and an arbitrary planet can be written as (see Eqs.~16 and 17 in \citealt{2017CeMDA.129..329S}):
         \begin{equation}
            {\Delta}d_{\rm pn} = \frac{a \ (1 - e^2)}{1 \pm e \ \cos{\varpi}} - \frac{a_{\rm p} \ (1 - e_{\rm p}^2)}{1 \pm e_{\rm p} 
                                  \ \cos{\varpi_{\rm p}}} \,,
            \label{noddis}
         \end{equation}
         where
         \begin{equation}
            \resizebox{0.9\hsize}{!}{$
            \cos{\varpi} = \frac{\cos{\omega} \ (\sin{i}\ \cos{i_{\rm p}} - \cos{i}\ \sin{i_{\rm p}}\ \cos{\Delta\Omega_{\rm p}}) + 
                                  \sin{\omega}\ \sin{i_{\rm p}}\ \sin{\Delta\Omega_{\rm p}}}
                                {\sqrt{1 - (\cos{i}\ \cos{i_{\rm p}} + \sin{i}\ \sin{i_{\rm p}}\ \cos{\Delta\Omega_{\rm p}})^2}} 
                                  \,,$}
            \label{cosw}
         \end{equation}
         \begin{equation}
            \resizebox{0.9\hsize}{!}{$
            \cos{\varpi_{\rm p}} = \frac{-\cos{\omega_{\rm p}} \ (\sin{i_{\rm p}}\ \cos{i} - \cos{i_{\rm p}}\ \sin{i}\ \cos{\Delta\Omega_{\rm p}}) +
                                           \sin{\omega_{\rm p}}\ \sin{i}\ \sin{\Delta\Omega_{\rm p}}}
                                        {\sqrt{1 - (\cos{i}\ \cos{i_{\rm p}} + \sin{i}\ \sin{i_{\rm p}}\ \cos{\Delta\Omega_{\rm p}})^2}} 
                                          \,,$}
            \label{cosvarpi}
         \end{equation} 
         $\Delta\Omega_{\rm p} = \Omega - \Omega_{\rm p}$, and $a$, $e$, $i$, $\Omega$ and $\omega$ are the orbital elements of the small 
         body, and $a_{\rm p}$, $e_{\rm p}$, $i_{\rm p}$, $\Omega_{\rm p}$, and $\omega_{\rm p}$ those of the arbitrary planet. For each
         random search experiment (each analysis consists of 2$\times$10$^{10}$), we compute the orbital elements of the putative perturber
         using the expressions:
         \begin{equation}
            \begin{aligned}
               q_{\rm p} & = x + (1000 - x)\,r_{1} \\
               e_{\rm p} & = 0.9\,r_{2} \\
               a_{\rm p} & = q_{\rm p}/(1 - e_{\rm p}) \\
               i_{\rm p} & = 80\,r_{3} \\
               \Omega_{\rm p} & = 360\,r_{4} \\
               \omega_{\rm p} & = 360\,r_{5} \,,
               \label{planet}
            \end{aligned}
         \end{equation}
         where $x=300, 400, 500,$ and 600~AU (see Sect.~\ref{Data}, or 200~AU for Appendix~\ref{200au}), and $r_j$ with $j=1, 5$, are random 
         numbers in the interval (0, 1) with a uniform distribution.

      \section{Extreme trans-Neptunian objects: Data\label{ETNOelements}}
         The barycentric orbit determinations used as input data in the uniform Monte Carlo random search discussed in Sect.~\ref{Results} 
         are shown in Table~\ref{etnosB}. The data are referred to epoch 2459000.5 Barycentric Dynamical Time (TDB) and they have been 
         retrieved (as of 11-January-2021) from JPL's SBDB and HORIZONS using tools provided by the Python package Astroquery 
         \citep{2019AJ....157...98G}.
%
%
         \begin{landscape}
         \begin{table*}
            \fontsize{8}{12pt}\selectfont
            \tabcolsep 0.10truecm
            \caption{\label{etnosB}Barycentric orbital elements and 1$\sigma$ uncertainties of known ETNOs with robust orbit determinations.
                    }
            \centering
            \begin{tabular}{lcccccccccc}
            \hline\hline
             Object              &  $a_{\rm b}$ & $\sigma_{a}$   & $e_{\rm b}$ & $\sigma_{e}$            & $i_{\rm b}$ & $\sigma_{i}$            & 
                                   $\Omega_{\rm b}$ & $\sigma_{\Omega}$       & $\omega_{\rm b}$ & $\sigma_{\omega}$  \\      
                                 & (AU)         & (AU)           &             &                         & (\degr)     & (\degr)                 & 
                                   (\degr)          & (\degr)                 & (\degr)          & (\degr)            \\
            \hline
         82158 (2001~FP$_{185}$) &  215.548928  &   0.040005     & 0.841097    & 2.7618$\times$10$^{-5}$ & 30.800299   & 3.259$\times$10$^{-5}$  & 
                                   179.358499       & 4.3407$\times$10$^{-5}$ &   6.874579       &  0.00045074        \\
         90377 Sedna             &  506.424770  &   0.18758      & 0.849551    & 6.2014$\times$10$^{-5}$ & 11.928524   & 3.7219$\times$10$^{-6}$ & 
                                   144.401511       & 0.00055761              & 311.285511       &  0.0035418         \\
        148209 (2000~CR$_{105}$) &  221.976439  &   0.6072       & 0.801226    & 0.00056946              & 22.755910   & 0.00058664              &
                                   128.285827       & 0.00029583              & 316.690118       &  0.011843          \\
        445473 (2010~VZ$_{98}$)  &  153.432684  &   0.011897     & 0.776116    & 1.7043$\times$10$^{-5}$ &  4.510568   & 9.5639$\times$10$^{-6}$ & 
                                   117.394668       & 0.00038454              & 313.728058       &  0.00069884        \\
        474640 (2004~VN$_{112}$) &  327.713834  &   1.5617       & 0.855596    & 0.00067208              & 25.547962   & 0.00027987              &
                                    66.022244       & 0.00043186              & 326.987949       &  0.00928245        \\
        496315 (2013~GP$_{136}$) &  150.177337  &   0.18997      & 0.726731    & 0.00037413              & 33.538905   & 0.00064333              &
                                   210.727262       & 0.00010633              &  42.569086       &  0.03727751        \\
        505478 (2013~UT$_{15}$)  &  200.156721  &   0.80063      & 0.780530    & 0.0010167               & 10.652044   & 0.0010304               &
                                   191.954169       & 0.00038493              & 252.123877       &  0.032992          \\
        506479 (2003~HB$_{57}$)  &  159.594669  &   0.36652      & 0.761283    & 0.00050502              & 15.500384   & 0.00027768              &
                                   197.871006       & 0.00037845              &  10.837414       &  0.008816          \\
        508338 (2015~SO$_{20}$)  &  164.785707  &   0.020652     & 0.798712    & 2.404$\times$10$^{-5}$  & 23.410688   & 3.432$\times$10$^{-5}$  & 
                                    33.634139       & 1.895$\times$10$^{-5}$  & 354.827026       &  0.00055221        \\
        523622 (2007~TG$_{422}$) &  502.465990  &   0.2381       & 0.929226    & 3.4842$\times$10$^{-5}$ & 18.595364   & 3.1757$\times$10$^{-5}$ & 
                                   112.910531       & 0.00012321              & 285.664088       &  0.00090846        \\
        527603 (2007~VJ$_{305}$) &  192.002016  &   0.047887     & 0.816750    & 4.288$\times$10$^{-5}$  & 11.983654   & 6.5092$\times$10$^{-5}$ & 
                                    24.382506       & 3.1038$\times$10$^{-5}$ & 338.356121       &  0.00089984        \\
        541132 Leleakuhonua      & 1077.120640  & 111.53         & 0.939677    & 0.0063597               & 11.671280   & 0.00063416              &
                                   300.993671       & 0.0071997               & 117.944434       &  0.3158            \\
               2002~GB$_{32}$    &  206.586429  &   0.51121      & 0.828932    & 0.00038669              & 14.192110   & 0.00015226              &
                                   177.043472       & 0.00033413              &  37.047917       &  0.0046496         \\
               2003~SS$_{422}$   &  203.255204  & 148.31         & 0.806561    & 0.16161                 & 16.781709   & 0.14714                 &
                                   151.041519      & 0.17403                 & 211.597889        & 43.173             \\
               2005~RH$_{52}$    &  153.565757  &   0.21296      & 0.746072    & 0.00032216              & 20.445652   & 0.00036748              & 
                                   306.109904      & 0.00088931              &  32.512598        &  0.00767           \\
               2010~GB$_{174}$   &  351.380200  &  22.505        & 0.861811    & 0.010215                & 21.562661   & 0.0051779               &
                                   130.715273      & 0.019788                & 347.236662        &  0.36426           \\    
               2012~VP$_{113}$   &  262.065255  &   1.4232       & 0.692739    & 0.0019759               & 24.052063   & 0.0023167               &
                                    90.802707      & 0.0056                  & 293.924997        &  0.37222           \\
               2013~FS$_{28}$    &  191.761394  &  98.598        & 0.821344    & 0.0969                  & 13.068232   & 0.02486                 &
                                   204.638126      & 0.016117                & 102.176514        &  2.4905            \\
               2013~FT$_{28}$    &  294.523622  &  10.063        & 0.852396    & 0.005059                & 17.375255   & 0.0034261               &
                                   217.722701      & 0.0048316               &  40.696900        &  0.16454           \\
               2013~RF$_{98}$    &  363.869961  &  13.352        & 0.900799    & 0.0036143               & 29.538492   & 0.0033747               &
                                    67.635472      & 0.0053263               & 311.757065        &  0.66841           \\
               2013~RA$_{109}$   &  462.902530  &   2.2525       & 0.900602    & 0.00046151              & 12.399719   & 0.00011973              &
                                   104.798695      & 0.005524                & 262.917813        &  0.15134           \\
               2013~SY$_{99}$    &  729.233540  &  24.996        & 0.931474    & 0.0024376               &  4.225431   & 0.0011969               &
                                    29.509257      & 0.0052081               &  32.141187        &  0.11464           \\
               2013~SL$_{102}$   &  314.359140  &   0.70133      & 0.878709    & 0.00025008              &  6.504915   & 7.5898$\times$10$^{-5}$ & 
                                    94.730847      & 0.0056732               & 265.496106        &  0.055127          \\
               2013~UH$_{15}$    &  173.746767  &   8.3586       & 0.798455    & 0.011307                & 26.080589   & 0.0057534               &
                                   176.542268      & 0.0071993               & 282.865545        &  0.27819           \\    
               2014~FE$_{72}$    & 1548.667877  & 440.14         & 0.976632    & 0.0087352               & 20.632449   & 0.0028763               &
                                   336.829059      & 0.0051919               & 133.959137        &  0.064528          \\
               2014~SR$_{349}$   &  298.050876  &  19.924        & 0.840461    & 0.010668                & 17.967979   & 0.0017476               &
                                    34.886583      & 0.014948                & 341.238334        &  0.6312            \\
               2014~WB$_{556}$   &  279.905880  &   2.5421       & 0.847464    & 0.0013182               & 24.157500   & 0.00019211              &
                                   114.891154      & 0.0033985               & 235.333796        &  0.056364          \\
               2015~BP$_{519}$   &  448.721566  &   8.3963       & 0.921446    & 0.0016417               & 54.110682   & 9.9809$\times$10$^{-5}$ & 
                                   135.213150      & 0.0018968               & 348.058562        &  0.027567          \\
               2015~GT$_{50}$    &  311.140075  &   2.6471       & 0.876545    & 0.0010976               &  8.794999   & 0.0012014               &
                                    46.064057      & 0.0028595               & 128.986870        &  0.11107           \\
               2015~KG$_{163}$   &  680.351485  &   5.8808       & 0.940483    & 0.00037323              & 13.994347   & 0.0011581               &
                                   219.103229      & 0.0017119               &  32.110680        &  0.098552          \\
               2015~KH$_{163}$   &  152.922061  &   0.59238      & 0.738806    & 0.0011085               & 27.137445   & 0.0013782               &
                                    67.572913      & 0.00060792              & 230.813811        &  0.048894          \\
               2015~RX$_{245}$   &  423.304136  &   4.7739       & 0.892386    & 0.0013068               & 12.138092   & 0.0019317               &
                                     8.605207      & 0.0001895               &  65.120605        &  0.045144          \\ 
               2015~RY$_{245}$   &  225.301756  &   4.5912       & 0.861023    & 0.0028285               &  6.030530   & 0.0010408               &
                                   341.532143      & 0.005986                & 354.533793        &  0.24642           \\
               2016~GA$_{277}$   &  154.842684  &   6.8842       & 0.767982    & 0.011305                & 19.421077   & 0.00034091              &
                                   112.841139      & 0.019248                & 178.521618        &  0.11573           \\
               2016~QU$_{89}$    &  171.465345  &   0.33356      & 0.794425    & 0.00040081              & 16.975710   & 0.00046385              &
                                   102.898107      & 0.0035642               & 303.349937        &  0.082868          \\
               2016~QV$_{89}$    &  171.616289  &   0.22913      & 0.767183    & 0.00036545              & 21.387605   & 0.00041616              &
                                   173.215227      & 0.0011962               & 281.086675        &  0.019706          \\
               2016~SG$_{58}$    &  232.948888  &   0.40228      & 0.849292    & 0.00027193              & 13.220892   & 4.9315$\times$10$^{-5}$ &
                                   118.978173      & 0.0023883               & 296.291522        &  0.035779          \\ 
               2016~TP$_{120}$   &  174.578585  &  24.948        & 0.771693    & 0.038036                & 32.639476   & 0.0005895               &
                                   126.728289      & 0.027144                & 350.990565        &  0.39834           \\
               2018~VM$_{35}$    &  261.464338  &  64.008        & 0.827901    & 0.050434                &  8.479536   & 0.0033767               &
                                   192.409778      & 0.058367                & 302.700112        &  2.6758            \\ 
        \hline
        \end{tabular}
        \tablefoot{The orbit determinations have been computed at epoch JD 2459000.5 that corresponds to 00:00:00.000 TDB on 2020 May 31 
                   (J2000.0 ecliptic and equinox). Input data source: JPL's SBDB.
                  }
     \end{table*}
     \end{landscape}
%
%

         The orbits of the synthetic ETNOs are obtained using the expressions:
         \begin{equation}
            \begin{aligned}
               a_{\rm s} & = a_{\rm b} + \sigma_{a}\,R_{1} \\
               e_{\rm s} & = e_{\rm b} + \sigma_{e}\,R_{2} \\
               i_{\rm s} & = i_{\rm b} +  \sigma_{i}\,R_{3} \\
               \Omega_{\rm s} & = \Omega_{\rm b} +  \sigma_{\Omega}\,R_{4} \\
               \omega_{\rm s} & = \omega_{\rm b} +  \sigma_{\omega}\,R_{5} \,,
               \label{etnosyn}
            \end{aligned}
         \end{equation}
         where $R_j$ with $j=1, 5$, are univariate Gaussian random numbers.

      \section{Heliocentric orbit determinations: results\label{heliocentric}}
\begin{figure*}
        \centering
         \includegraphics[width=0.245\linewidth]{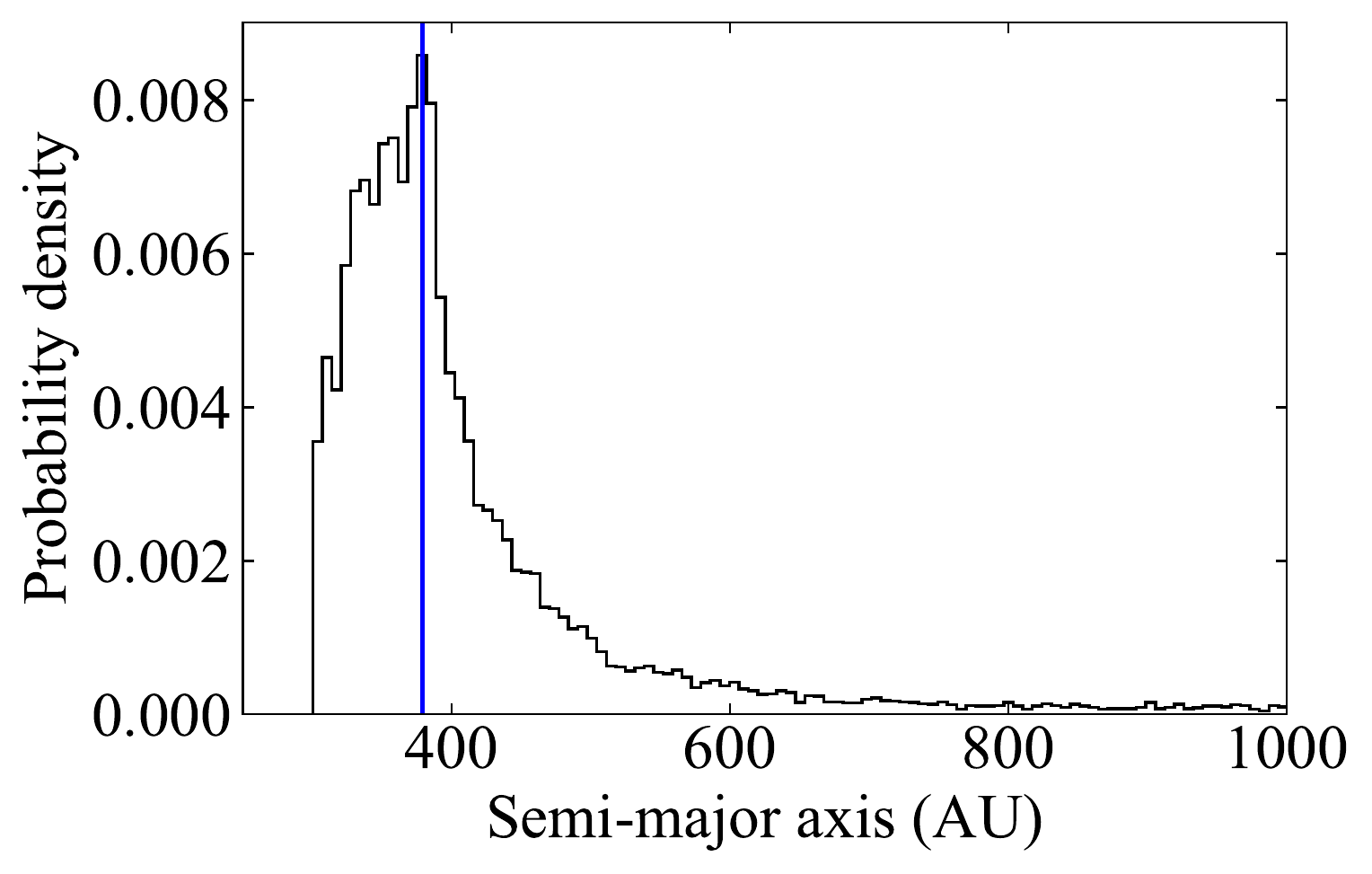}
         \includegraphics[width=0.245\linewidth]{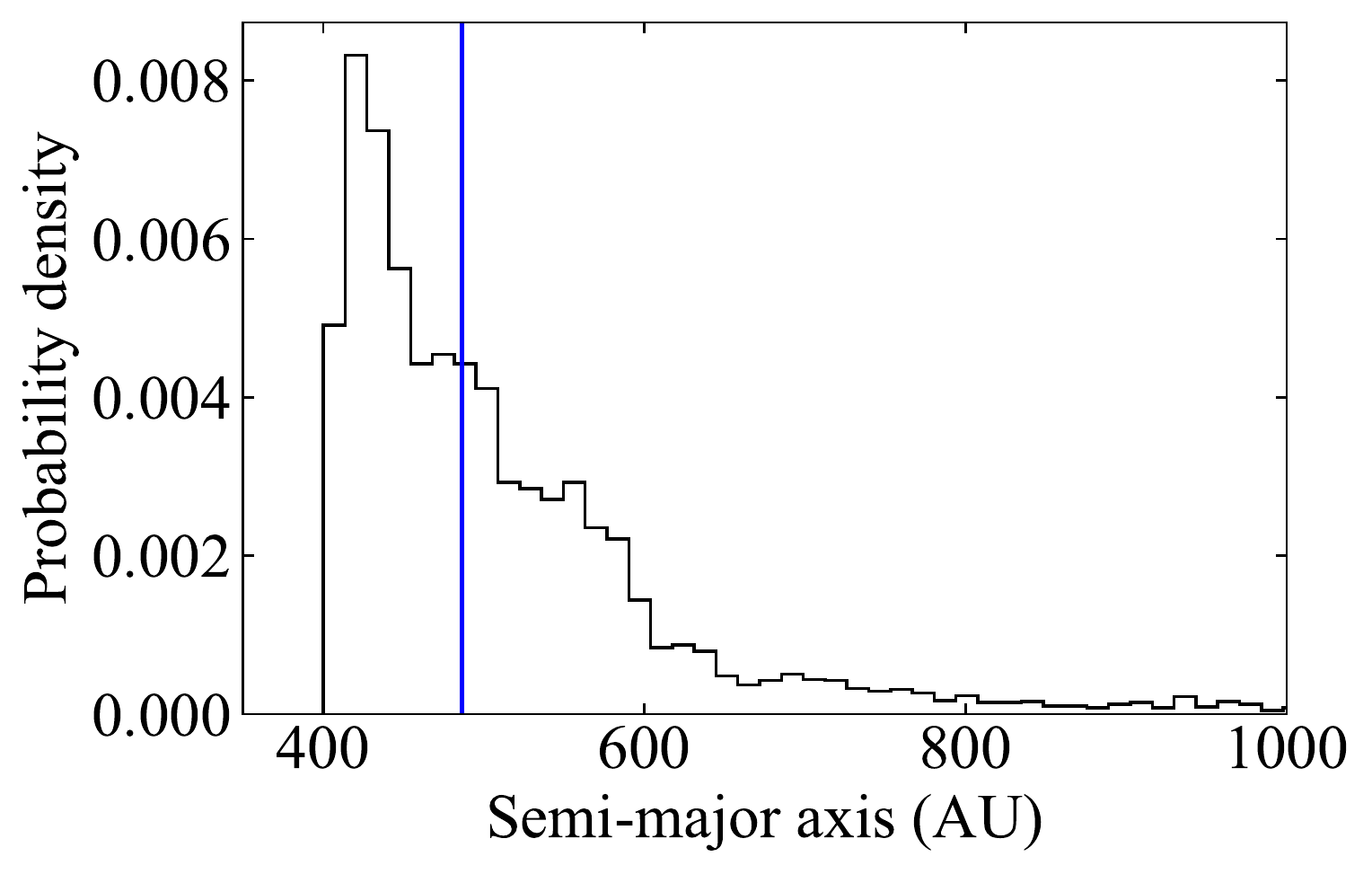}
         \includegraphics[width=0.245\linewidth]{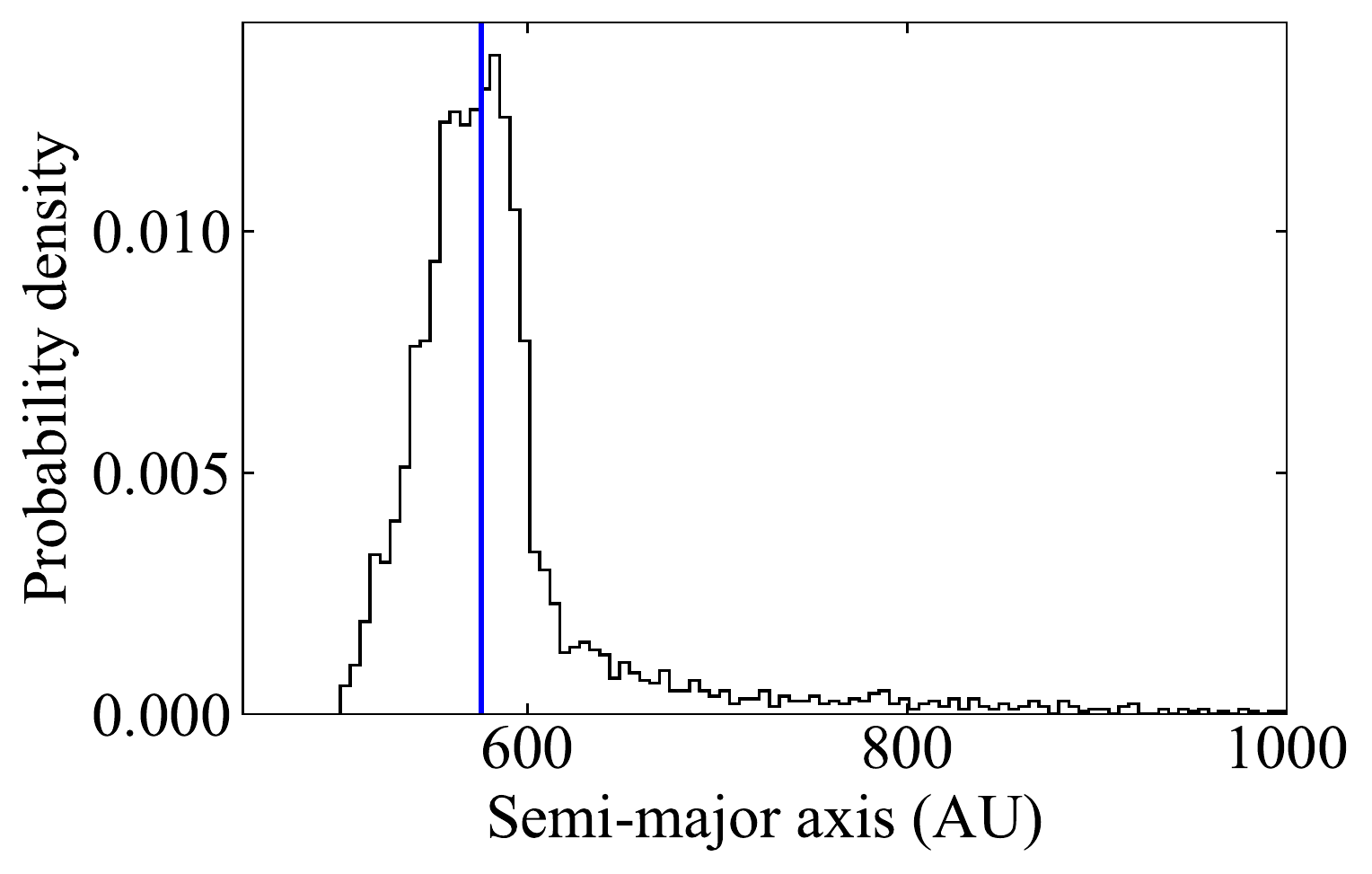}
         \includegraphics[width=0.245\linewidth]{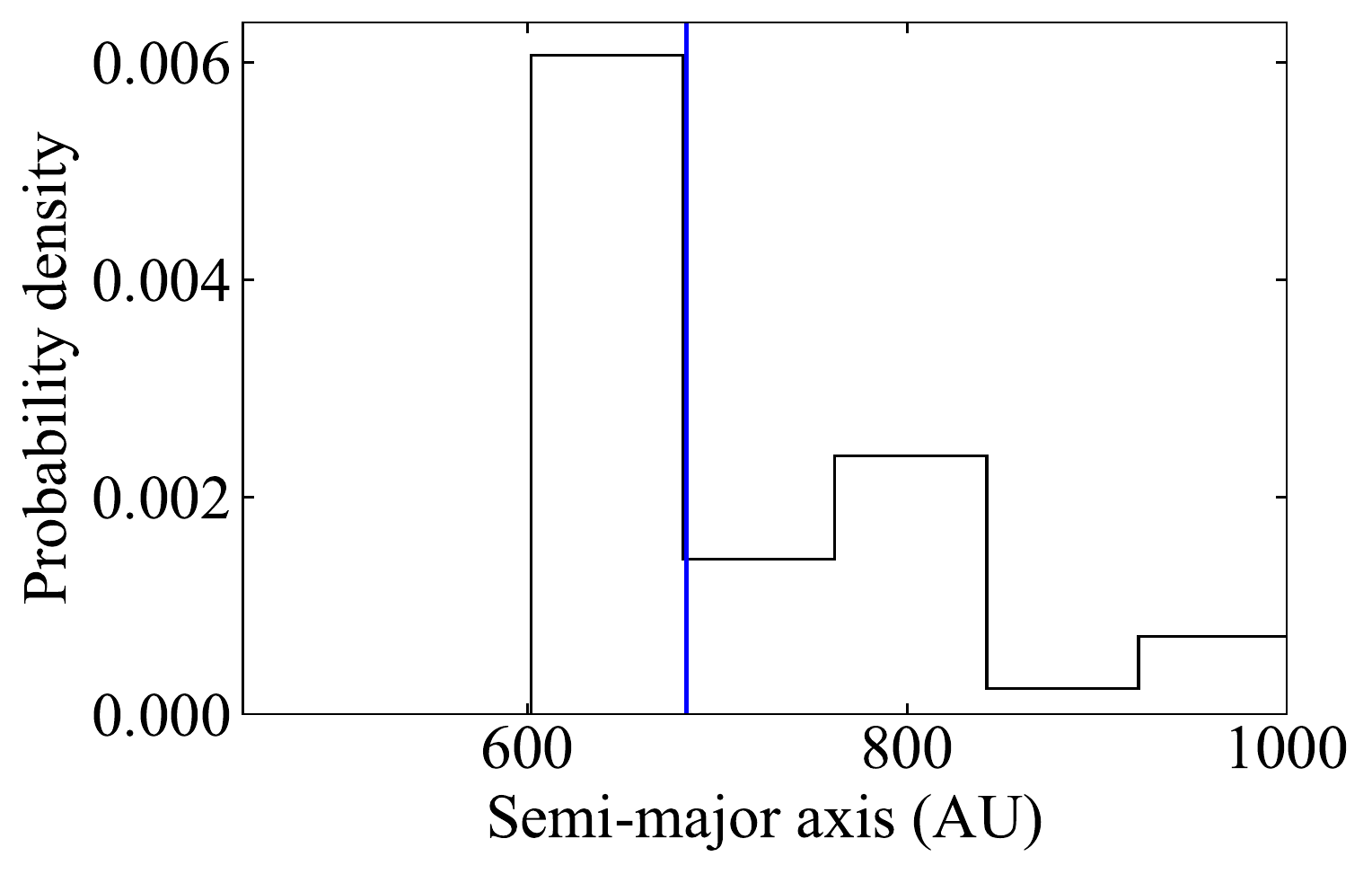}
         \includegraphics[width=0.245\linewidth]{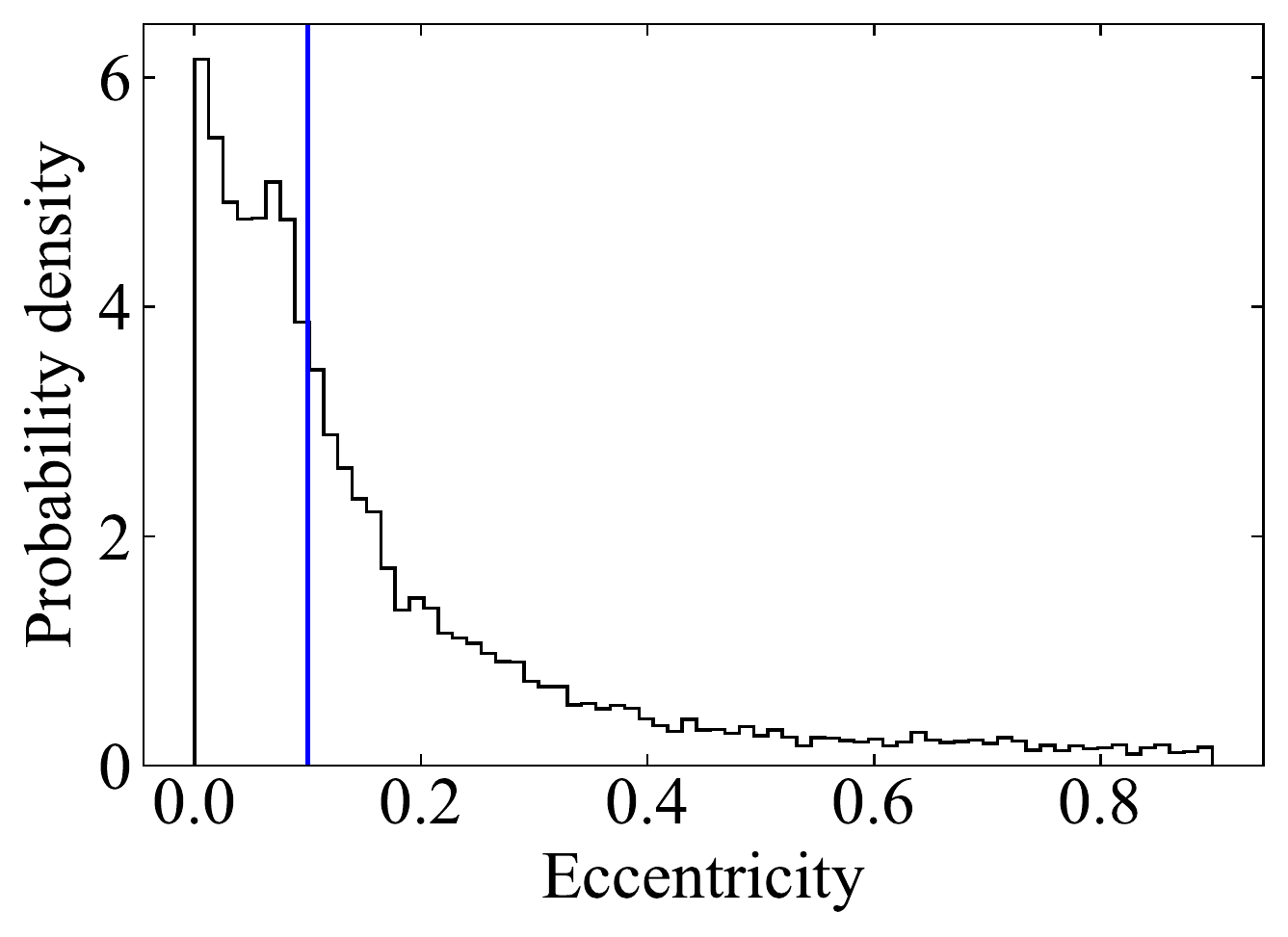}
         \includegraphics[width=0.245\linewidth]{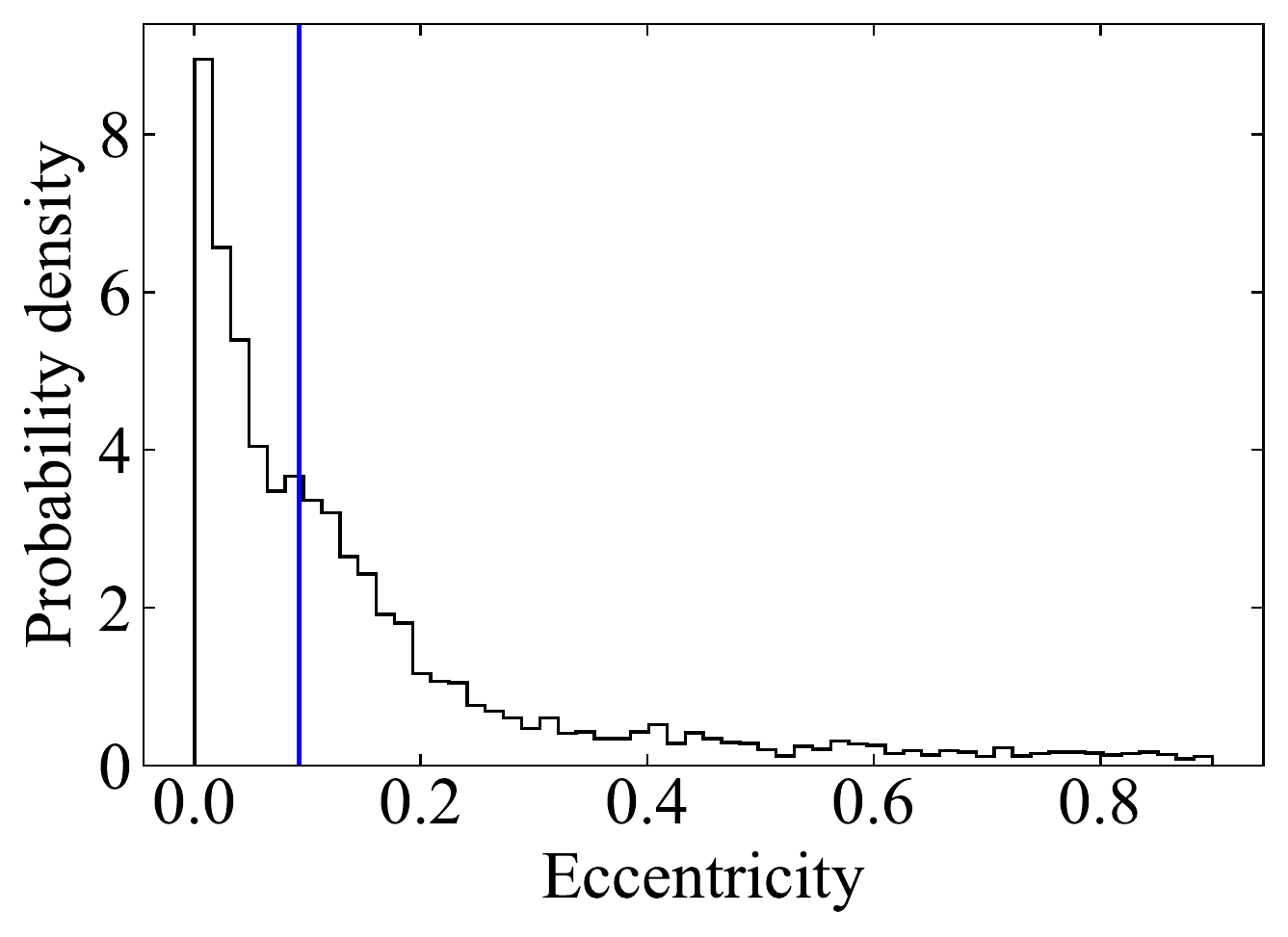}
         \includegraphics[width=0.245\linewidth]{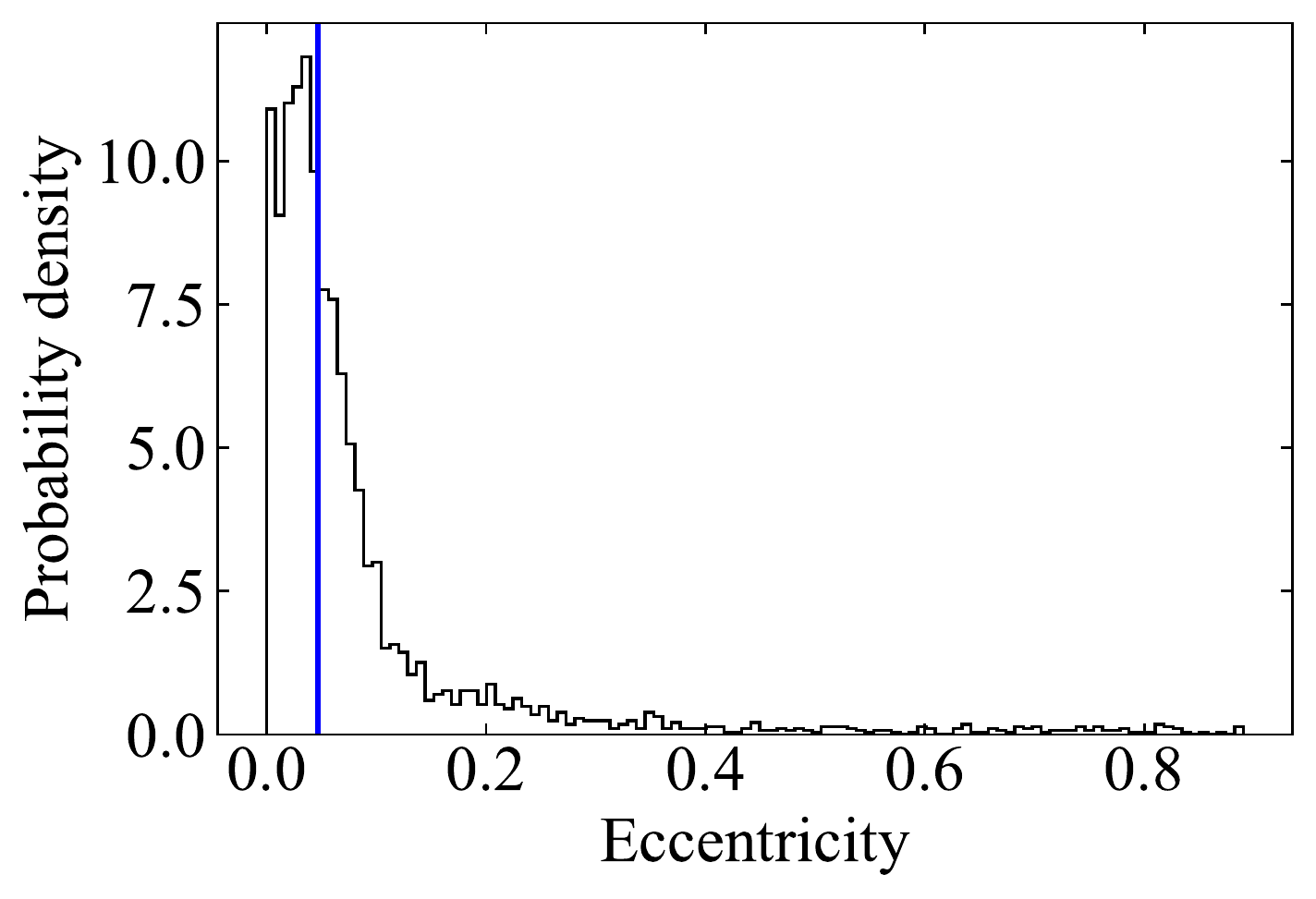}
         \includegraphics[width=0.245\linewidth]{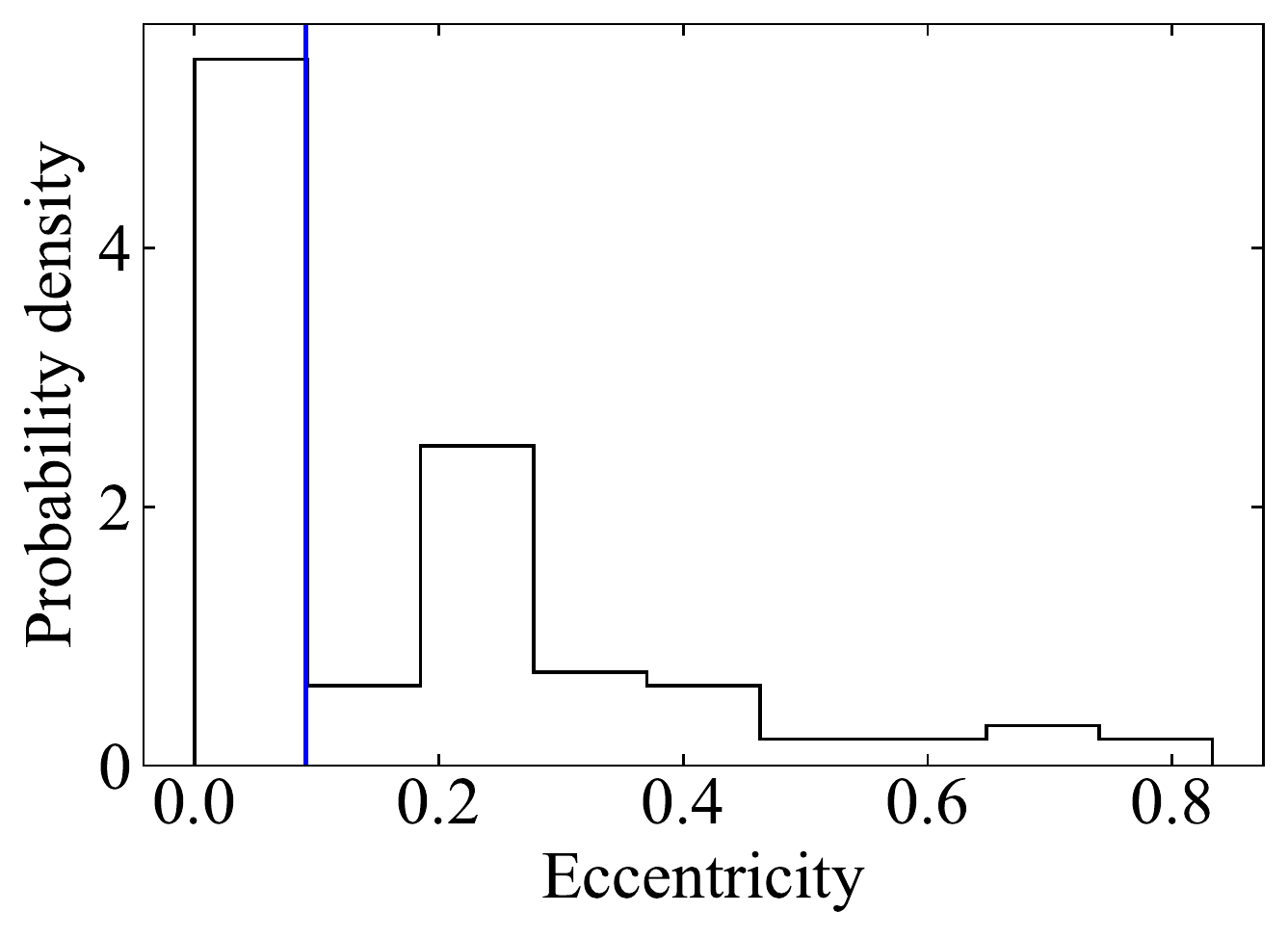}
         \includegraphics[width=0.245\linewidth]{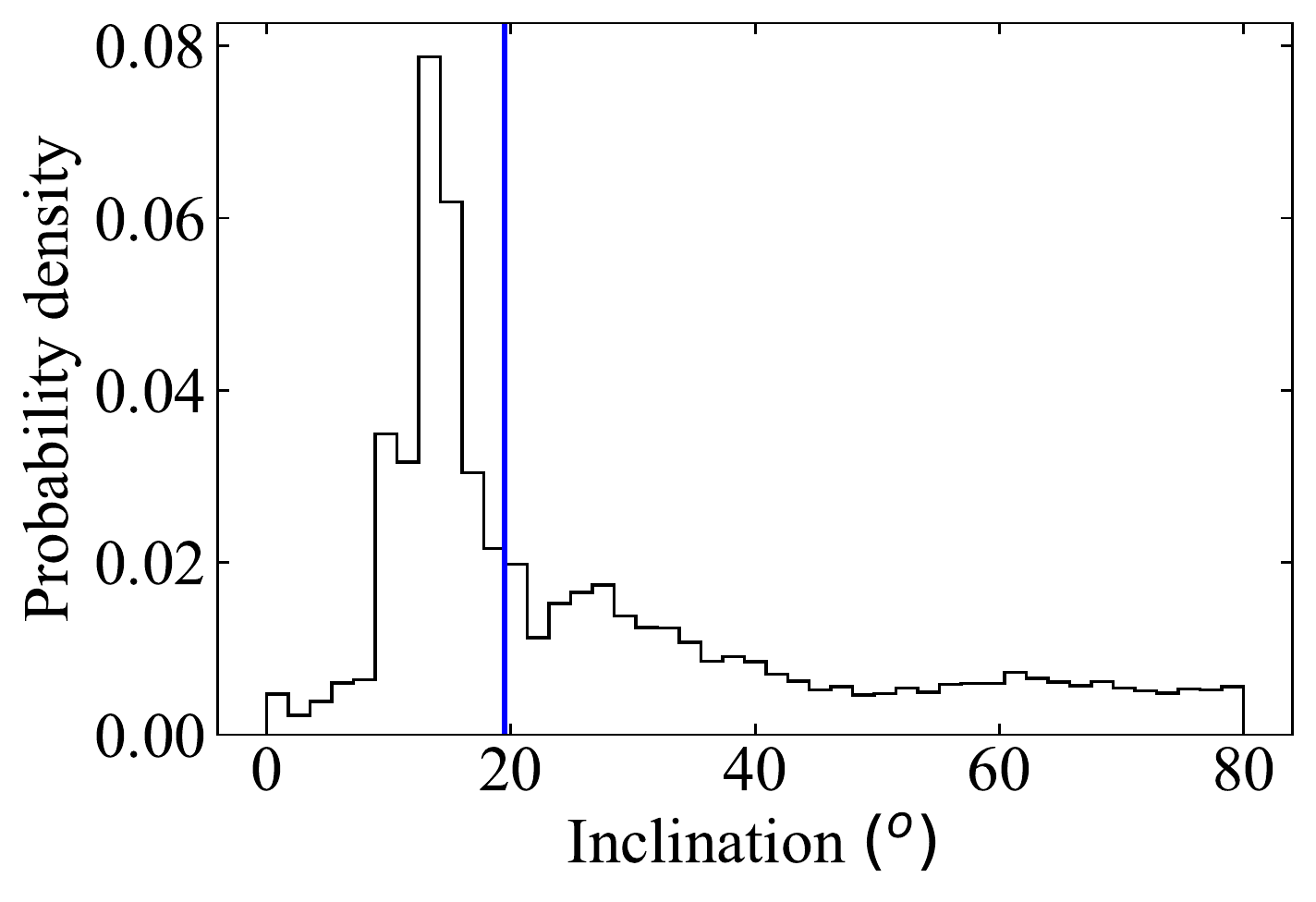}
         \includegraphics[width=0.245\linewidth]{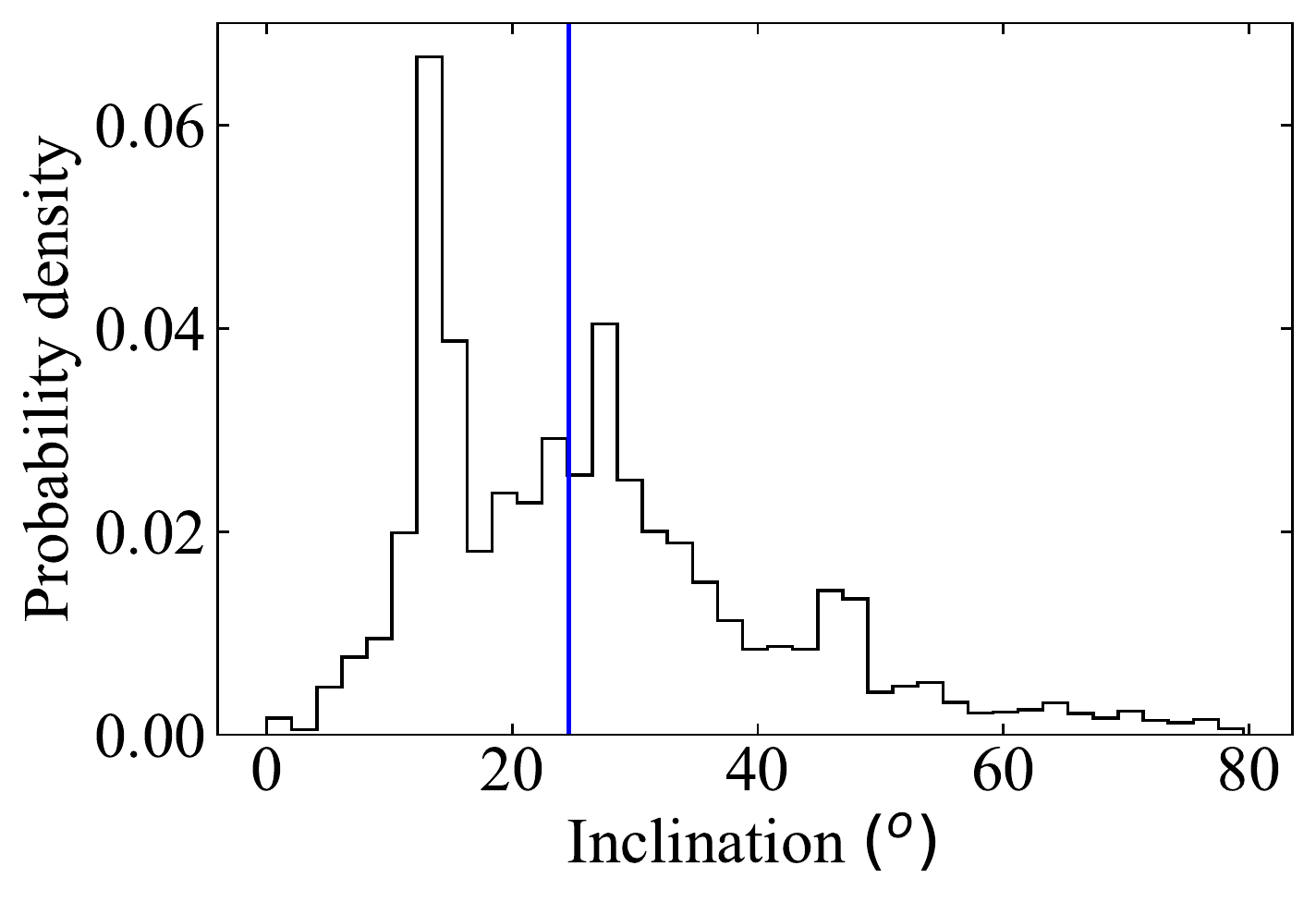}
         \includegraphics[width=0.245\linewidth]{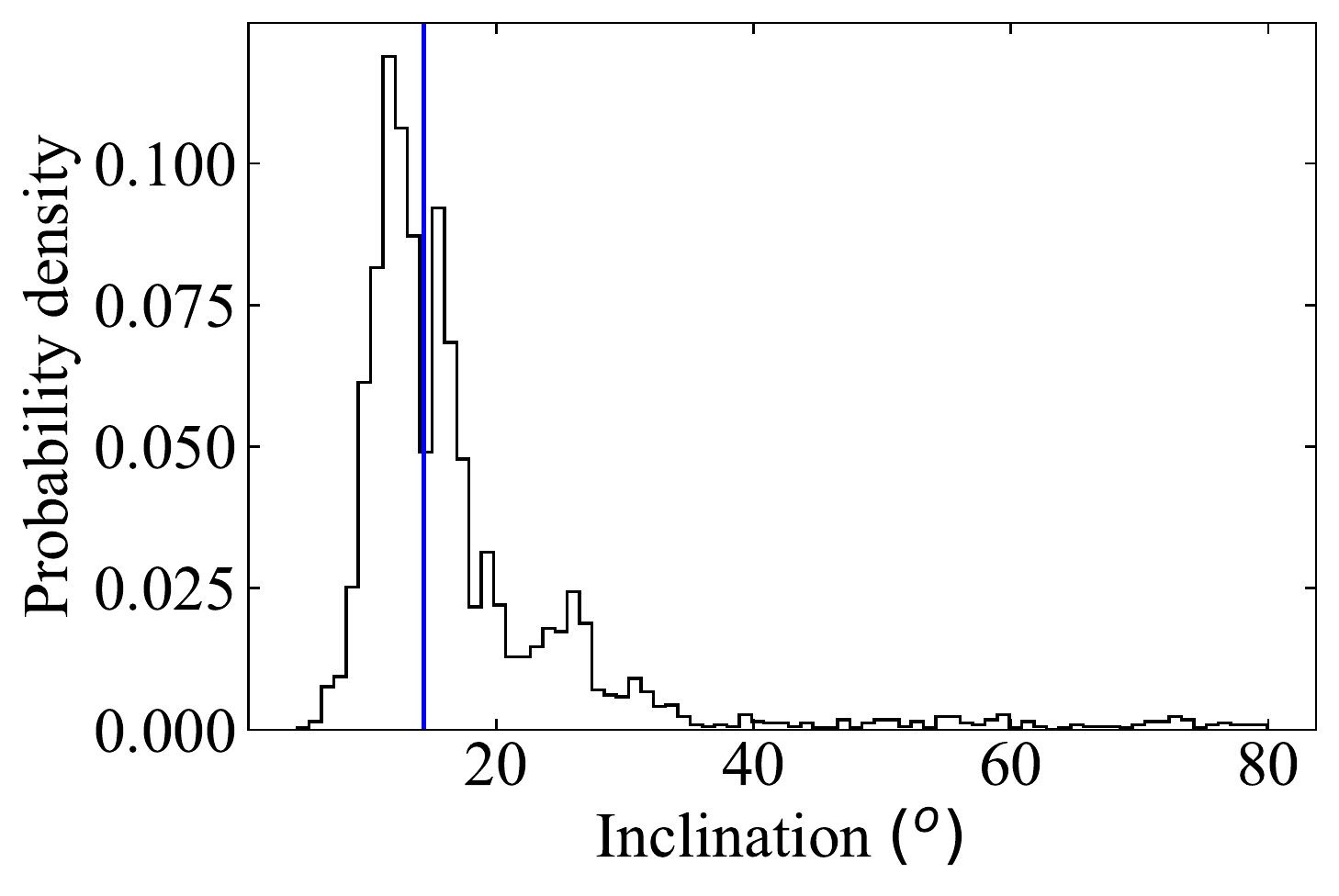}
         \includegraphics[width=0.245\linewidth]{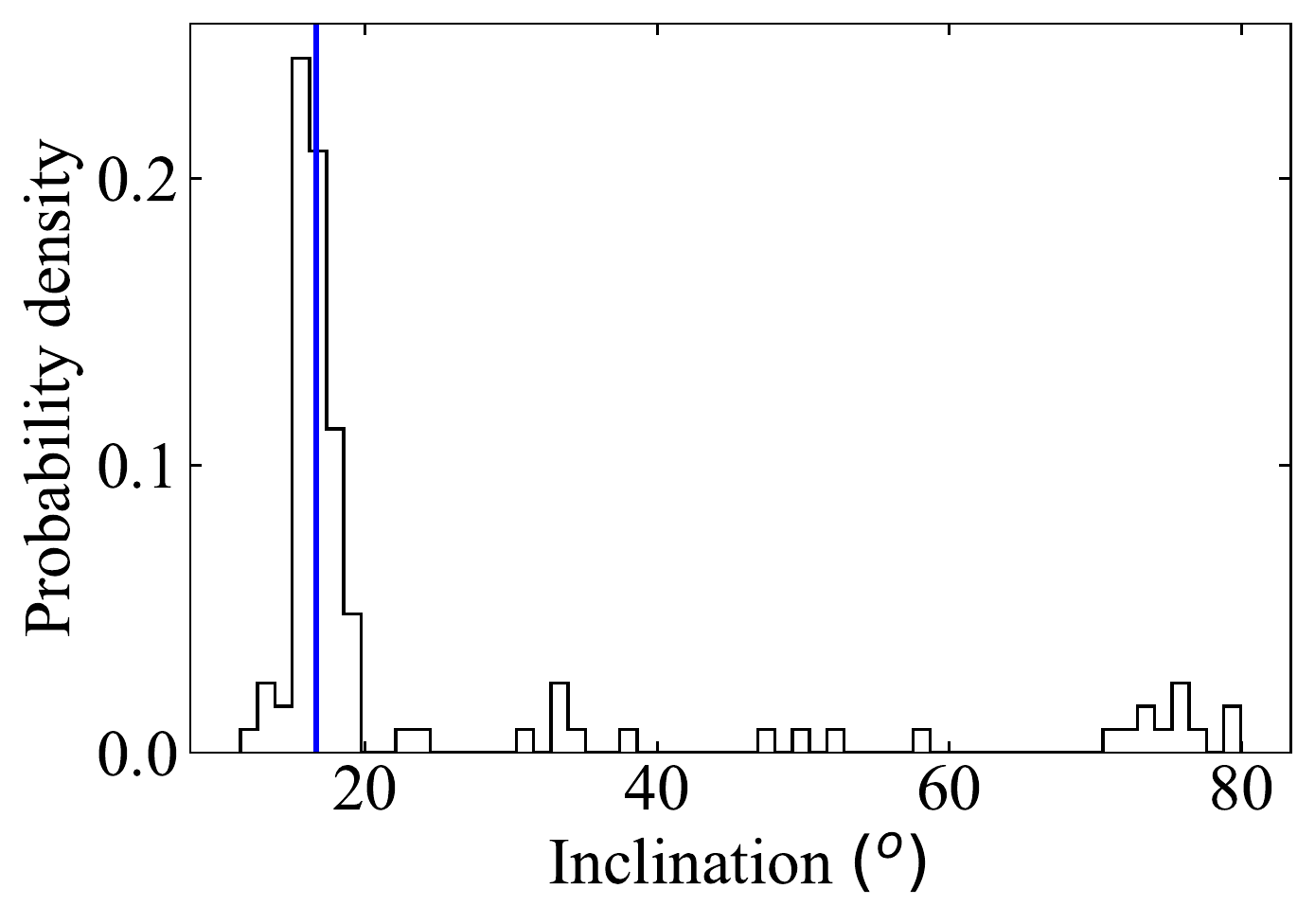}
         \includegraphics[width=0.245\linewidth]{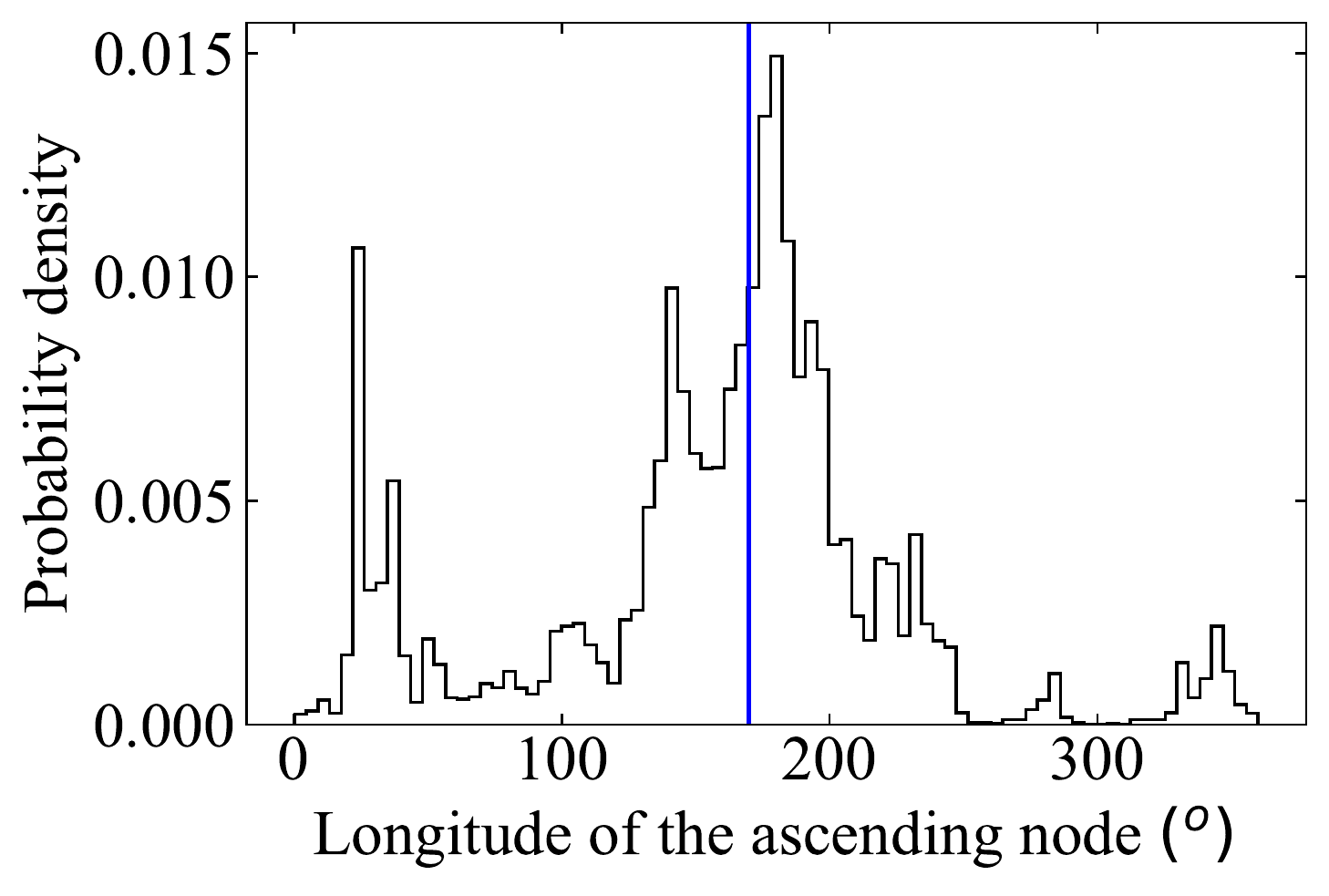}
         \includegraphics[width=0.245\linewidth]{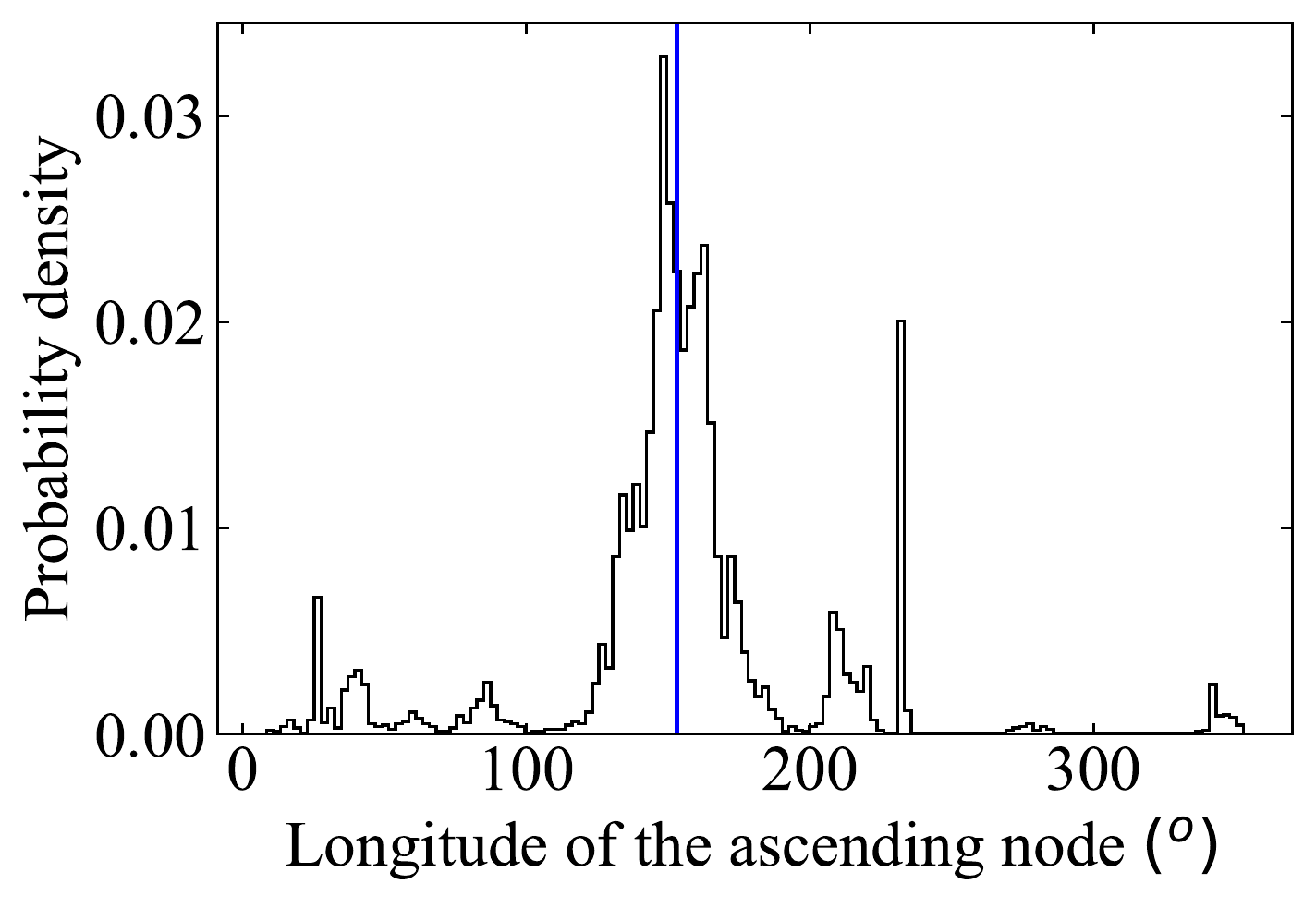}
         \includegraphics[width=0.245\linewidth]{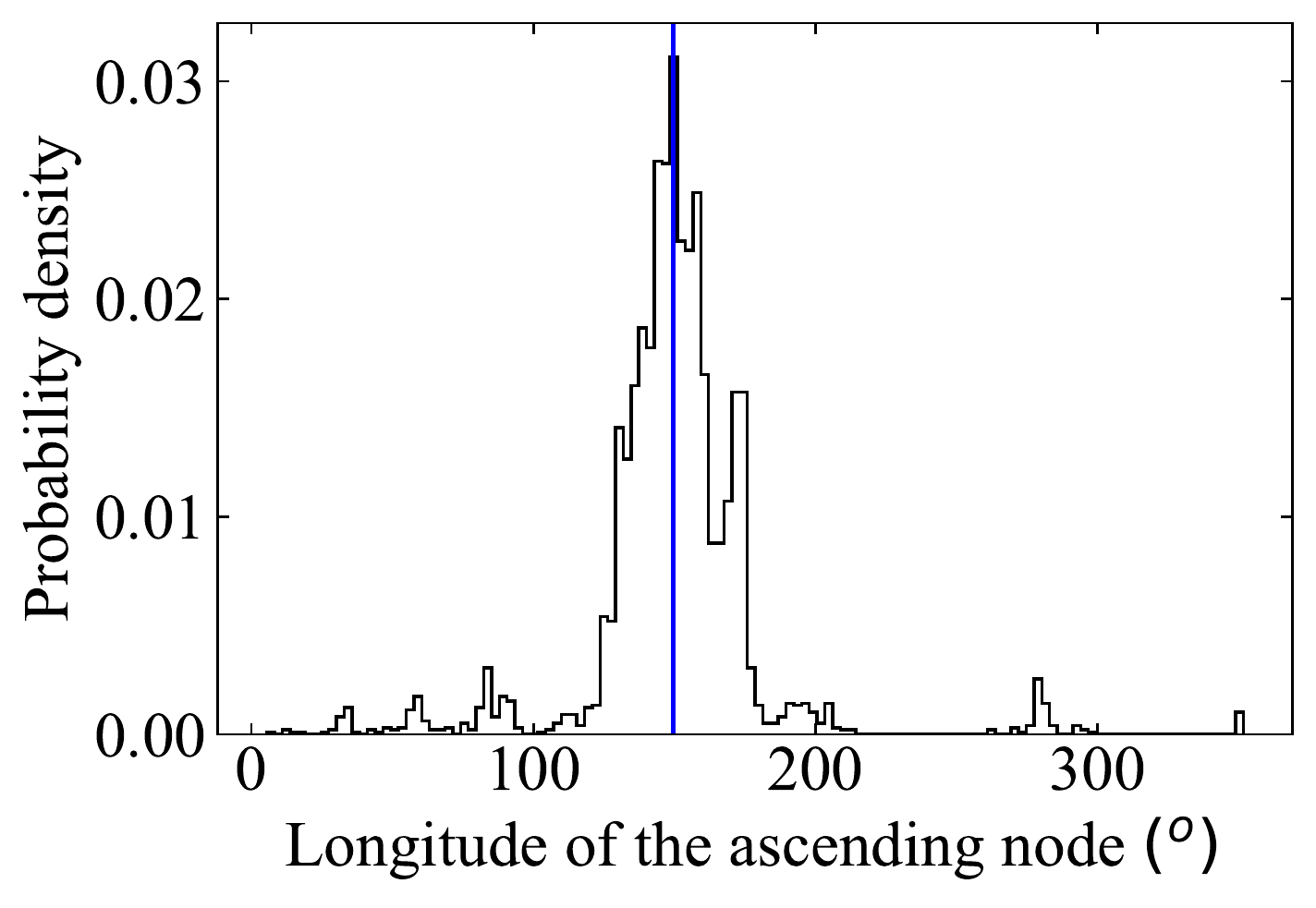}
         \includegraphics[width=0.245\linewidth]{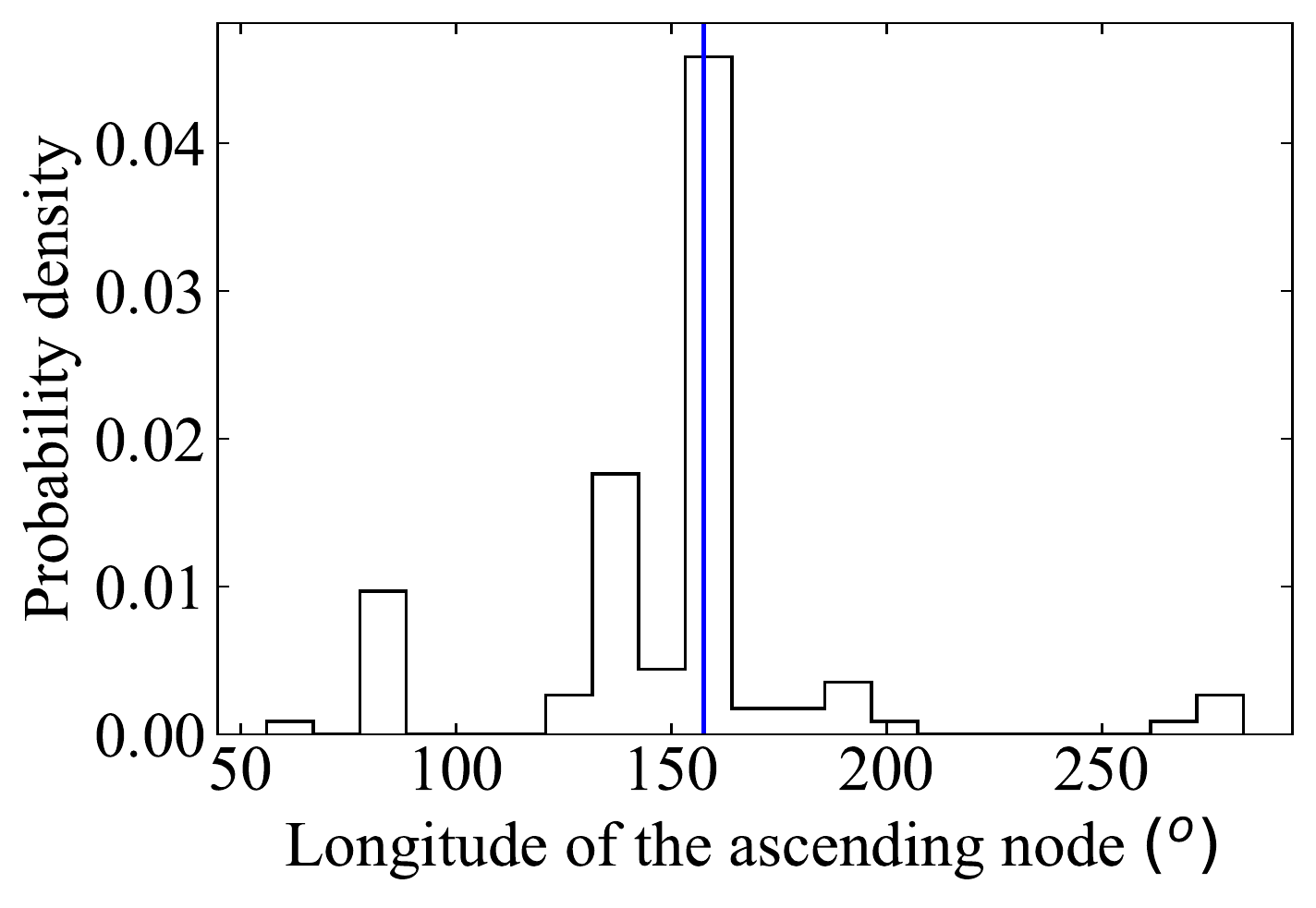}
         \includegraphics[width=0.245\linewidth]{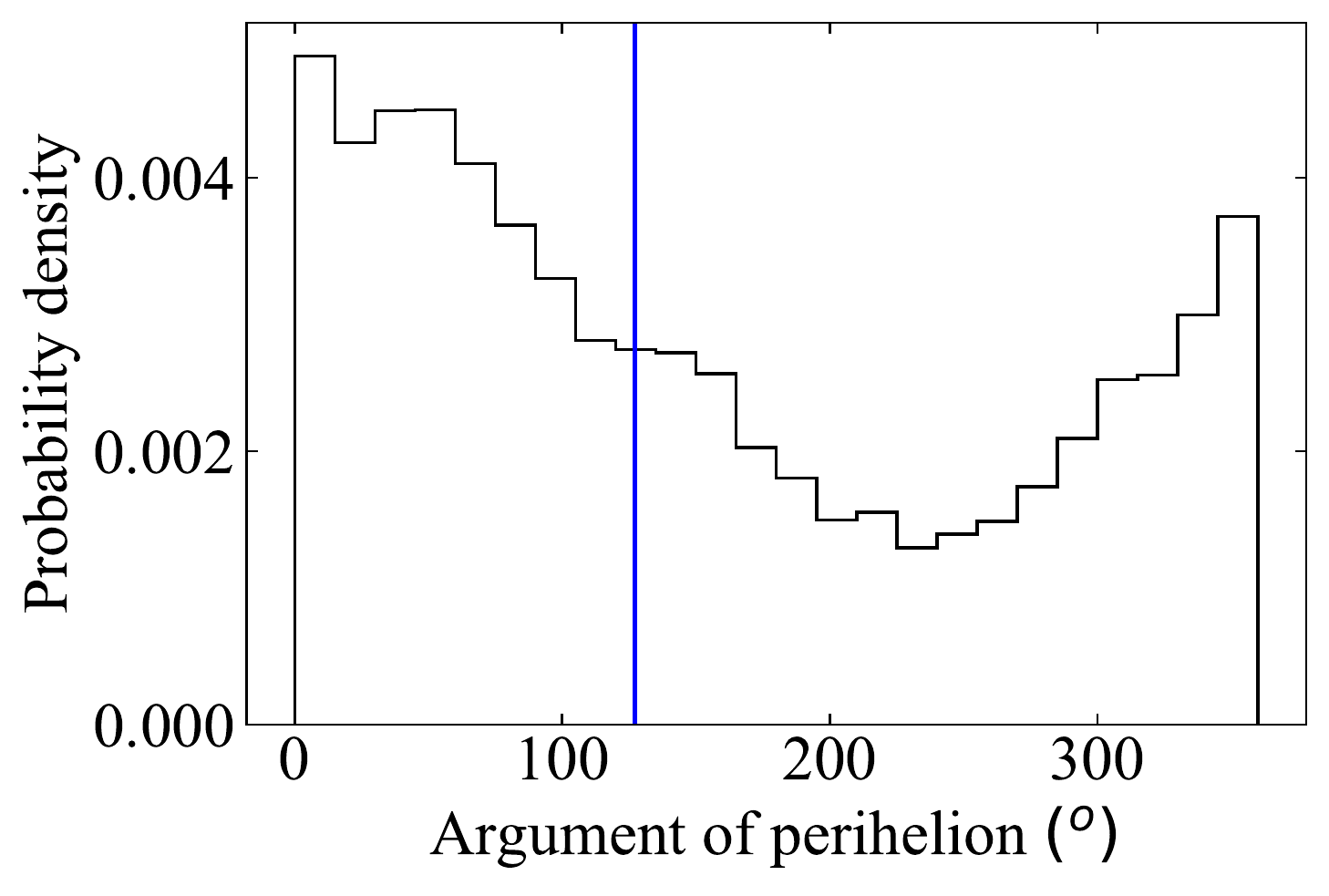}
         \includegraphics[width=0.245\linewidth]{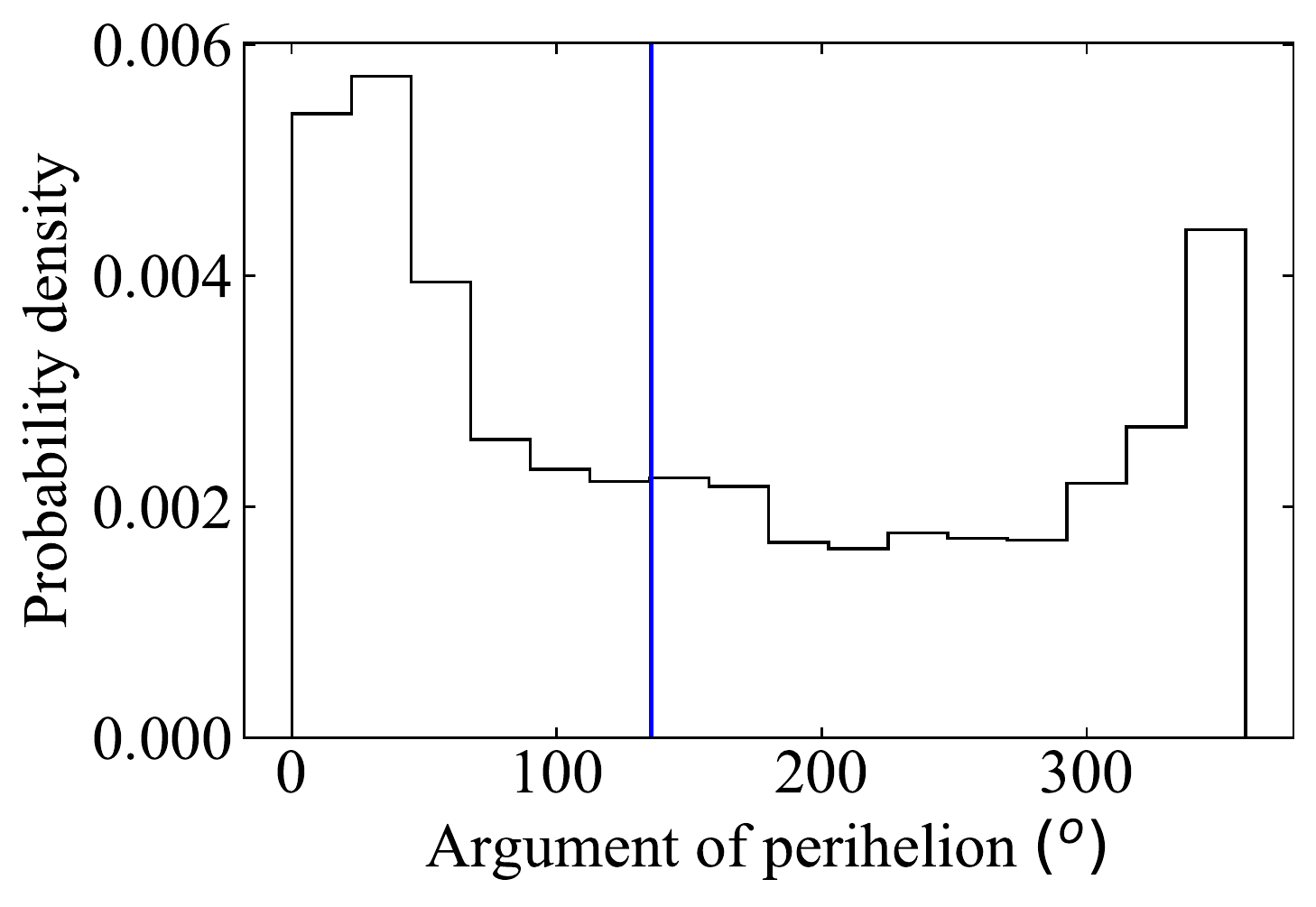}
         \includegraphics[width=0.245\linewidth]{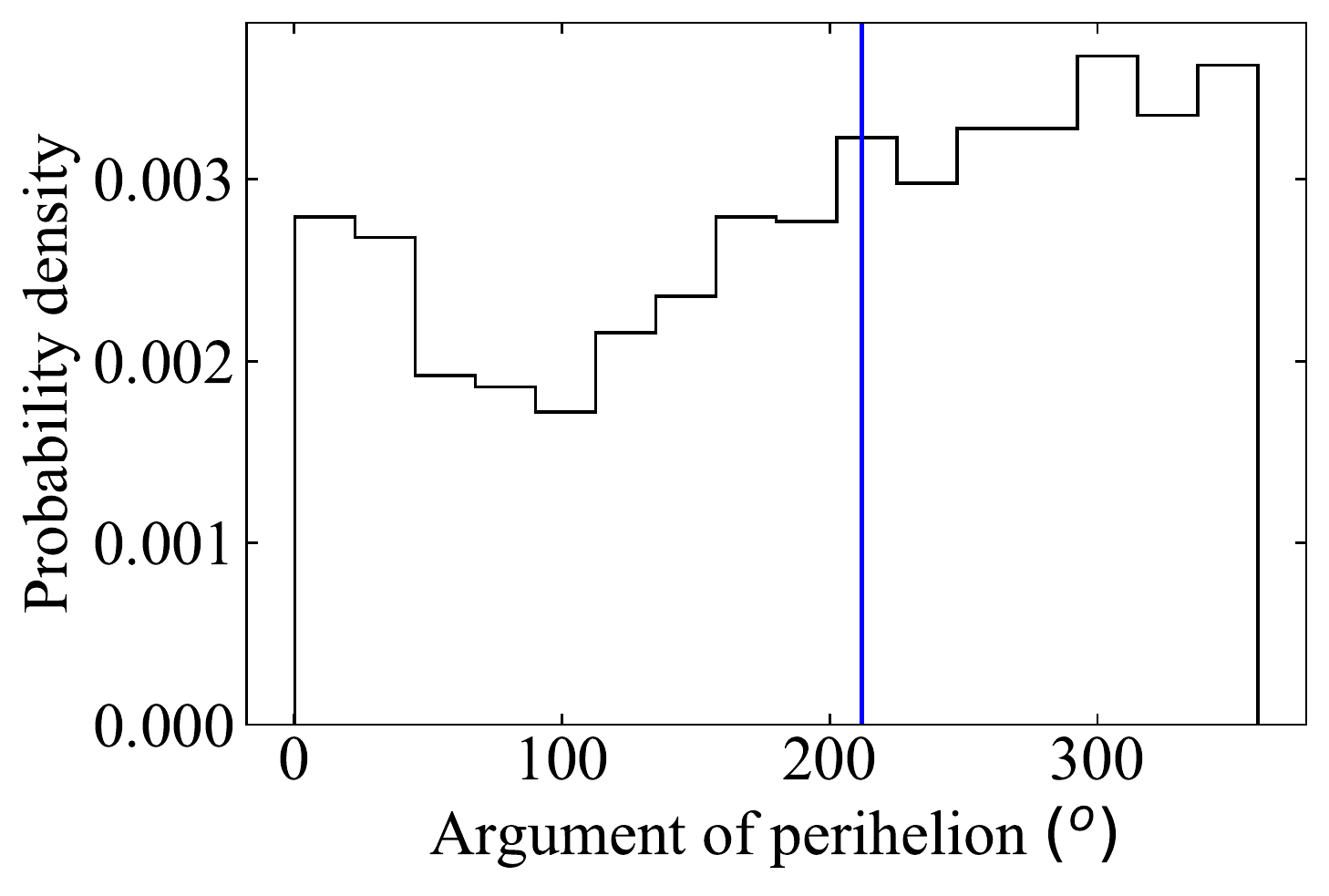}
         \includegraphics[width=0.245\linewidth]{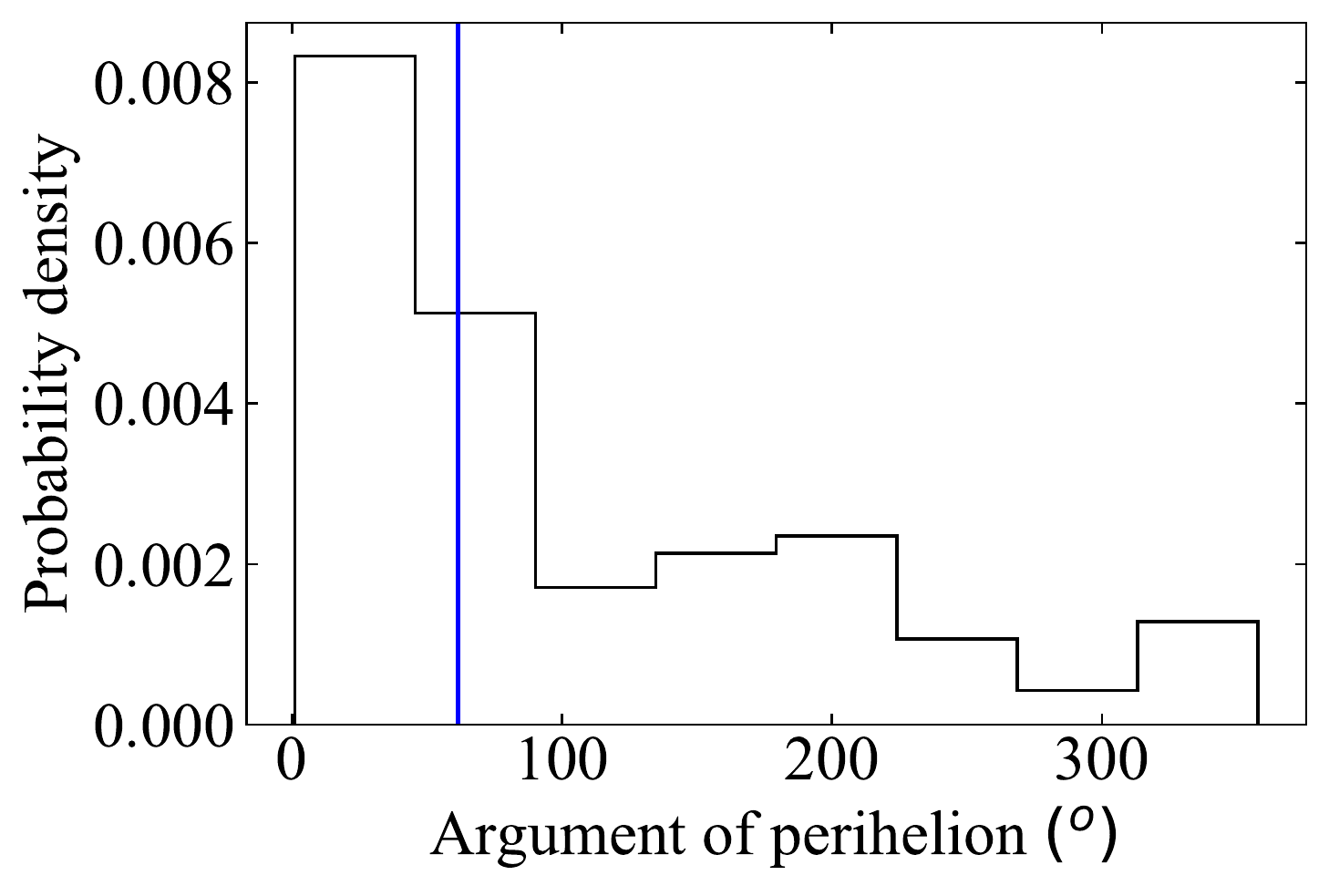}
         \caption{Heliocentric orbital elements of putative perturber planets. Distributions of heliocentric orbital elements of planetary
                  orbits that may result in close encounters (under 5~AU for $q_{\rm p}>300$~AU, under 7.5~AU for $q_{\rm p}>400$~AU, and
                  under 10~AU for $q_{\rm p}>500$~AU and $q_{\rm p}>600$~AU) with five or more present-day extreme trans-Neptunian objects
                  (ETNOs). Each column of panels shows the cumulative results of 2$\times$10$^{10}$ experiments. From left to right, the
                  panels show the results of imposing $q_{\rm p}>300$~AU (left column, 18369 orbits), $q_{\rm p}>400$~AU (6594 orbits),
                  $q_{\rm p}>500$~AU (3567 orbits), and $q_{\rm p}>600$~AU (right column, 105 orbits). Median values are shown as vertical
                  blue lines. Based on the available data on known ETNOs, the presence of massive perturbers well beyond 600~AU is strongly
                  excluded within the context of the hypotheses centered in this study.
                 }
         \label{histograms}
      \end{figure*}

         If we repeat the calculations discussed in Sect.~\ref{Results} using heliocentric orbit determinations instead of barycentric ones
         as input data, we obtain the distributions in Fig.~\ref{histograms} with central values and dispersions summarized in 
         Table~\ref{resultsH}.
%
%
%
%
%
%
     \begin{table*}
        \fontsize{8}{12pt}\selectfont
        \tabcolsep 0.15truecm
        \caption{\label{resultsH}Summary of central values and dispersions of optimal heliocentric orbits.
                }
        \centering
        \begin{tabular}{lcccc}
           \hline\hline
            Orbital parameter                                         & $q_{\rm p}>300$~AU            & $q_{\rm p}>400$~AU
                                                                      & $q_{\rm p}>500$~AU            & $q_{\rm p}>600$~AU            \\
           \hline
            Semi-major axis, $a_{\rm p}$ (AU)                         & 379$_{-47}^{+112}$ (378)      & 486$_{-62}^{+129}$ (420)
                                                                      & 576$_{-29}^{+45}$ (584)       & 684$_{-65}^{+299}$ (645)      \\
            Eccentricity, $e_{\rm p}$                                 & 0.10$_{-0.07}^{+0.21}$ (0.01) & 0.09$_{-0.07}^{+0.18}$ (0.01)
                                                                      & 0.05$_{-0.03}^{+0.08}$ (0.04) & 0.09$_{-0.08}^{+0.25}$ (0.05) \\
            Inclination, $i_{\rm p}$ (\degr)                          & 19$_{-7}^{+32}$ (13)          & 25$_{-11}^{+17}$ (13)
                                                                      & 14$_{-3}^{+10}$ (12)          & 17$_{-1}^{+17}$ (16)          \\
            Longitude of the ascending node, $\Omega_{\rm p}$ (\degr) & 170$_{-77}^{+37}$ (181)       & 153$_{-19}^{+24}$ (148)
                                                                      & 150$_{-16}^{+19}$ (151)       & 157$_{-26}^{+5}$ (160)        \\
            Argument of perihelion, $\omega_{\rm p}$ (\degr)          & 127$_{-92}^{+180}$ (7)        & 136$_{-107}^{+179}$ (35)
                                                                      & 212$_{-148}^{+102}$ (305)     & 62$_{-33}^{+144}$ (22)        \\
           \hline
        \end{tabular}
        \tablefoot{Median values and 16th and 84th percentiles (absolute maximum in parentheses) from the Monte Carlo random searches whose
                   distributions are shown in Fig.~\ref{histograms}.
                  }
     \end{table*}
%
%

      \section{Statistical significance\label{significance}}
\begin{figure*}
        \centering
         \includegraphics[width=0.245\linewidth]{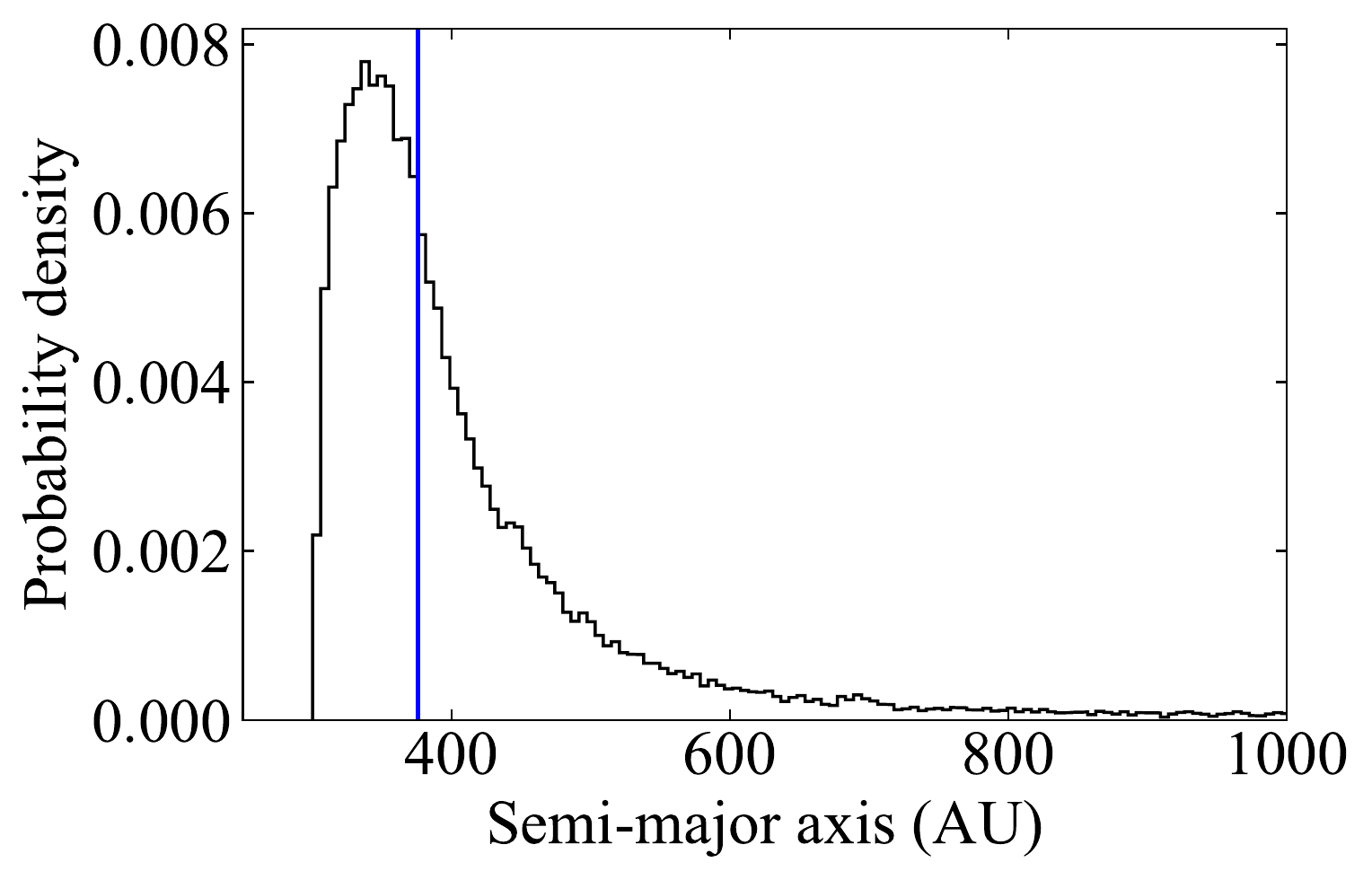}
         \includegraphics[width=0.245\linewidth]{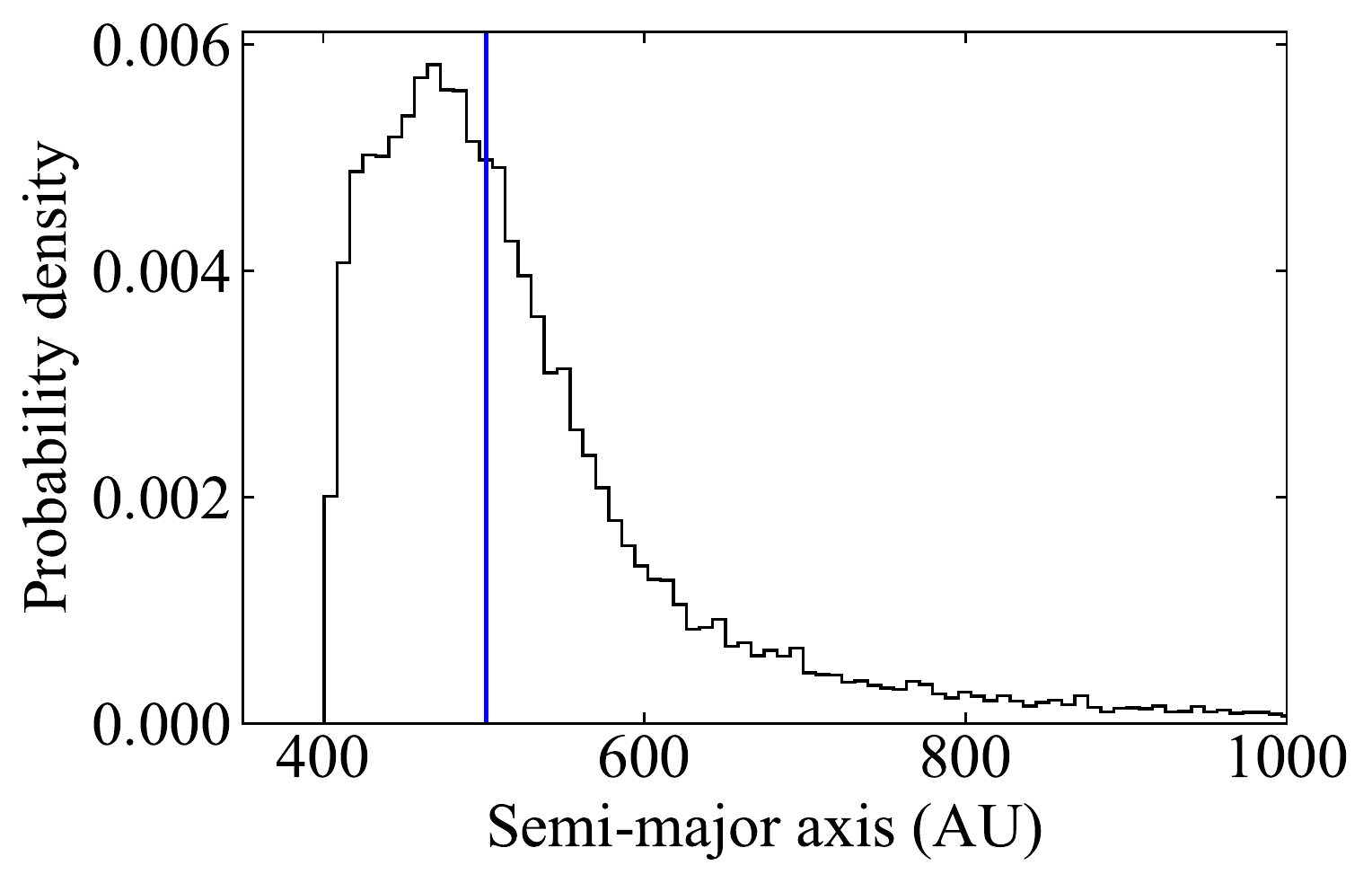}
         \includegraphics[width=0.245\linewidth]{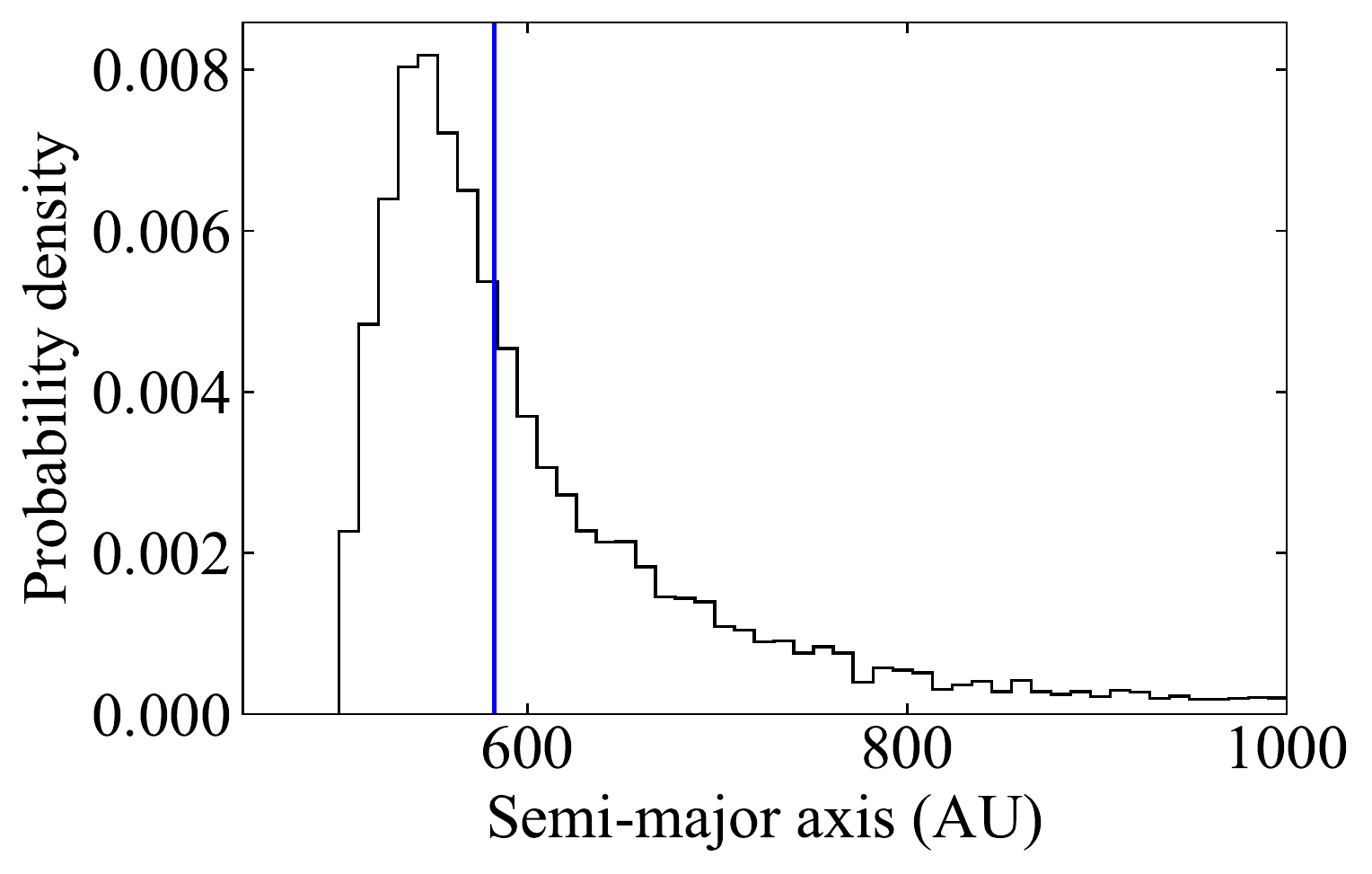}
         \includegraphics[width=0.245\linewidth]{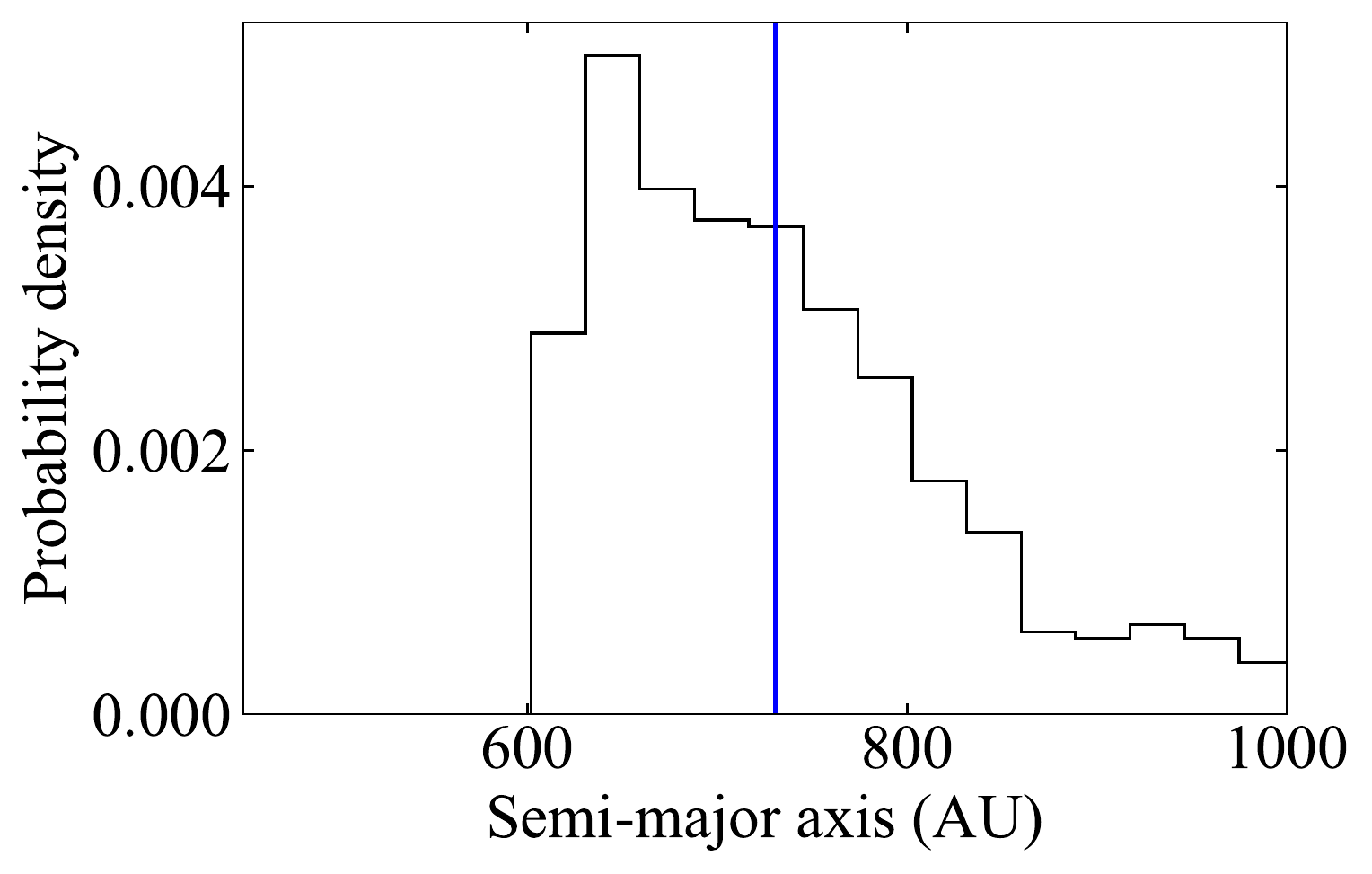}
         \includegraphics[width=0.245\linewidth]{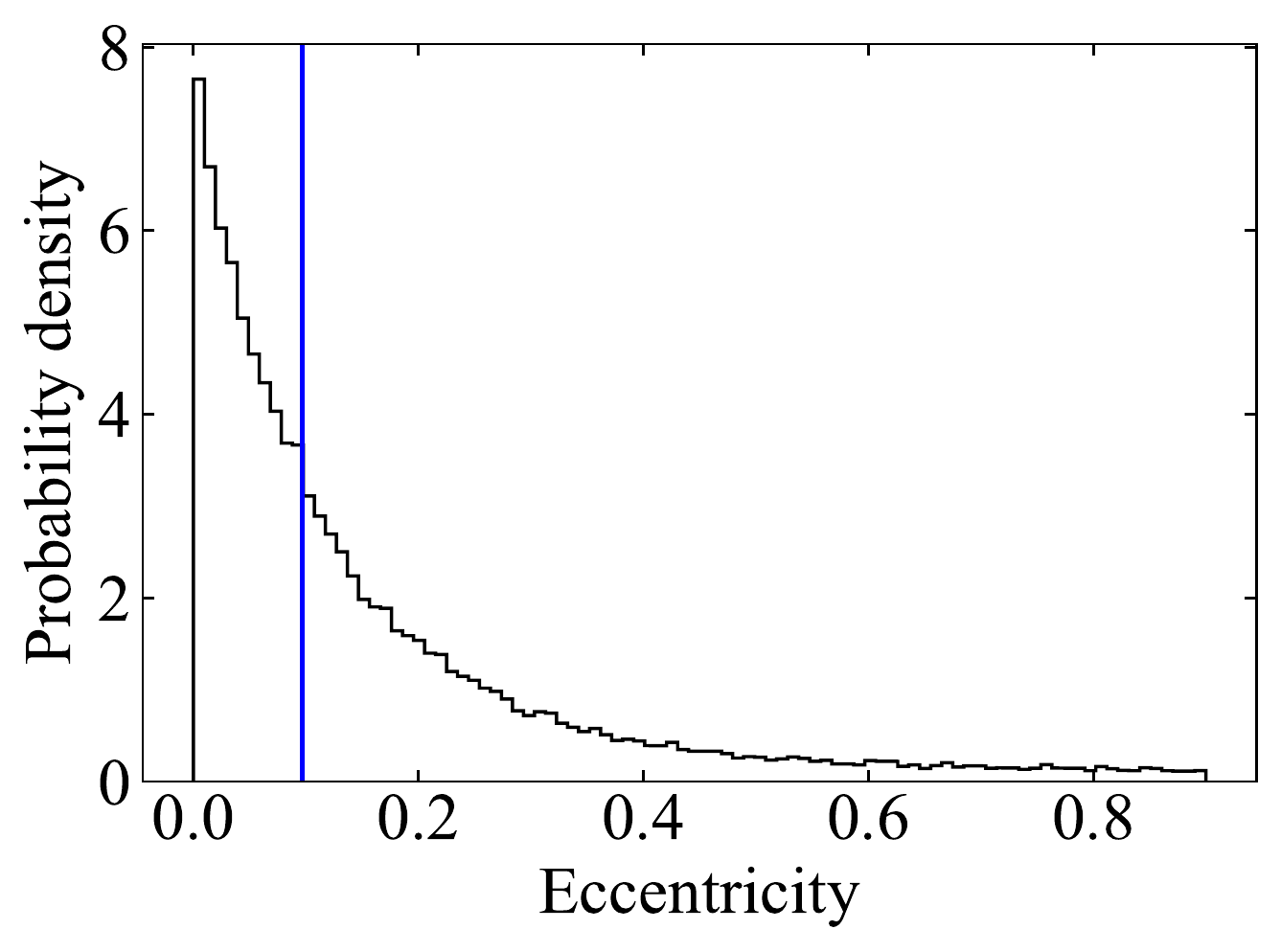}
         \includegraphics[width=0.245\linewidth]{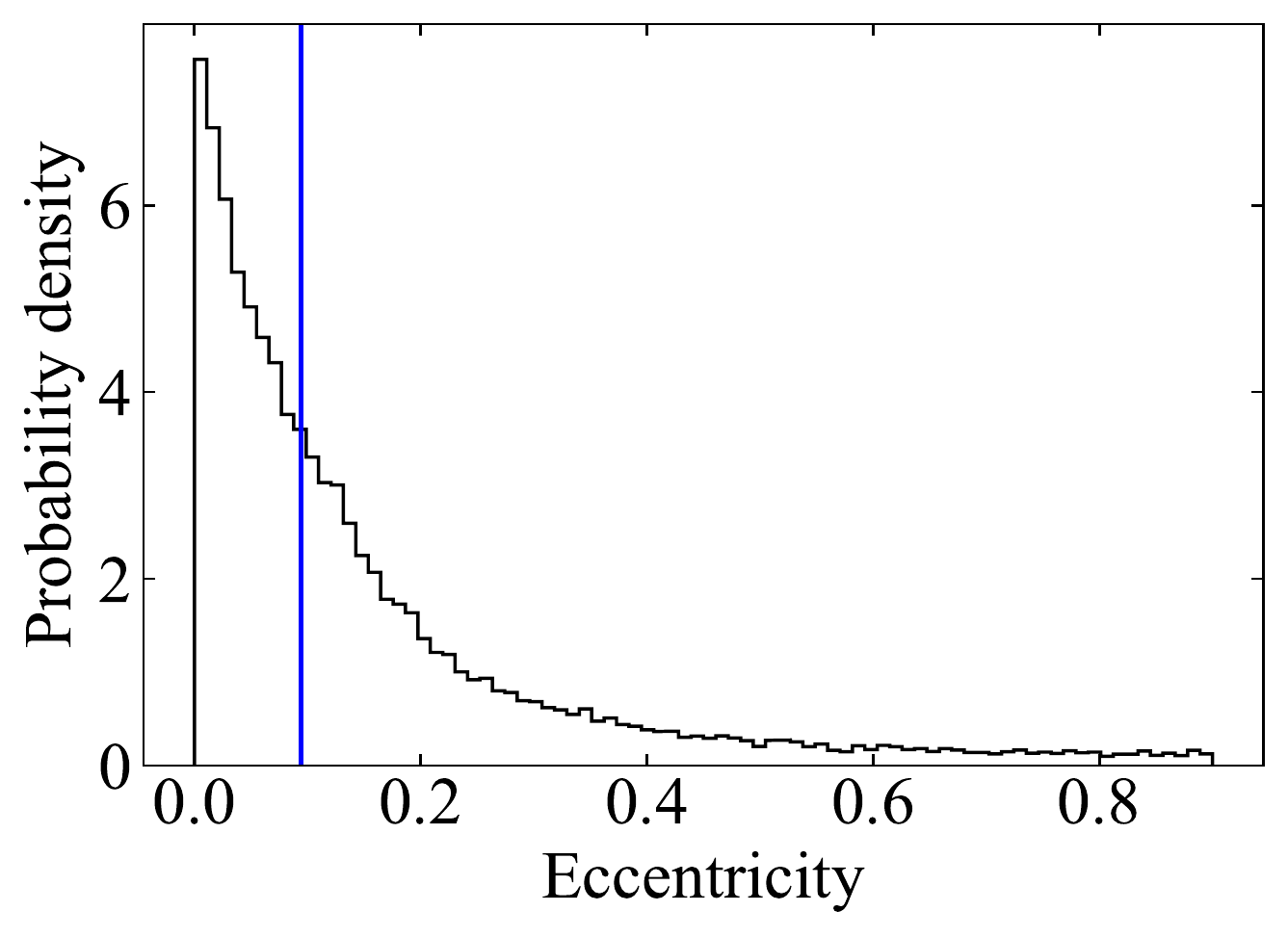}
         \includegraphics[width=0.245\linewidth]{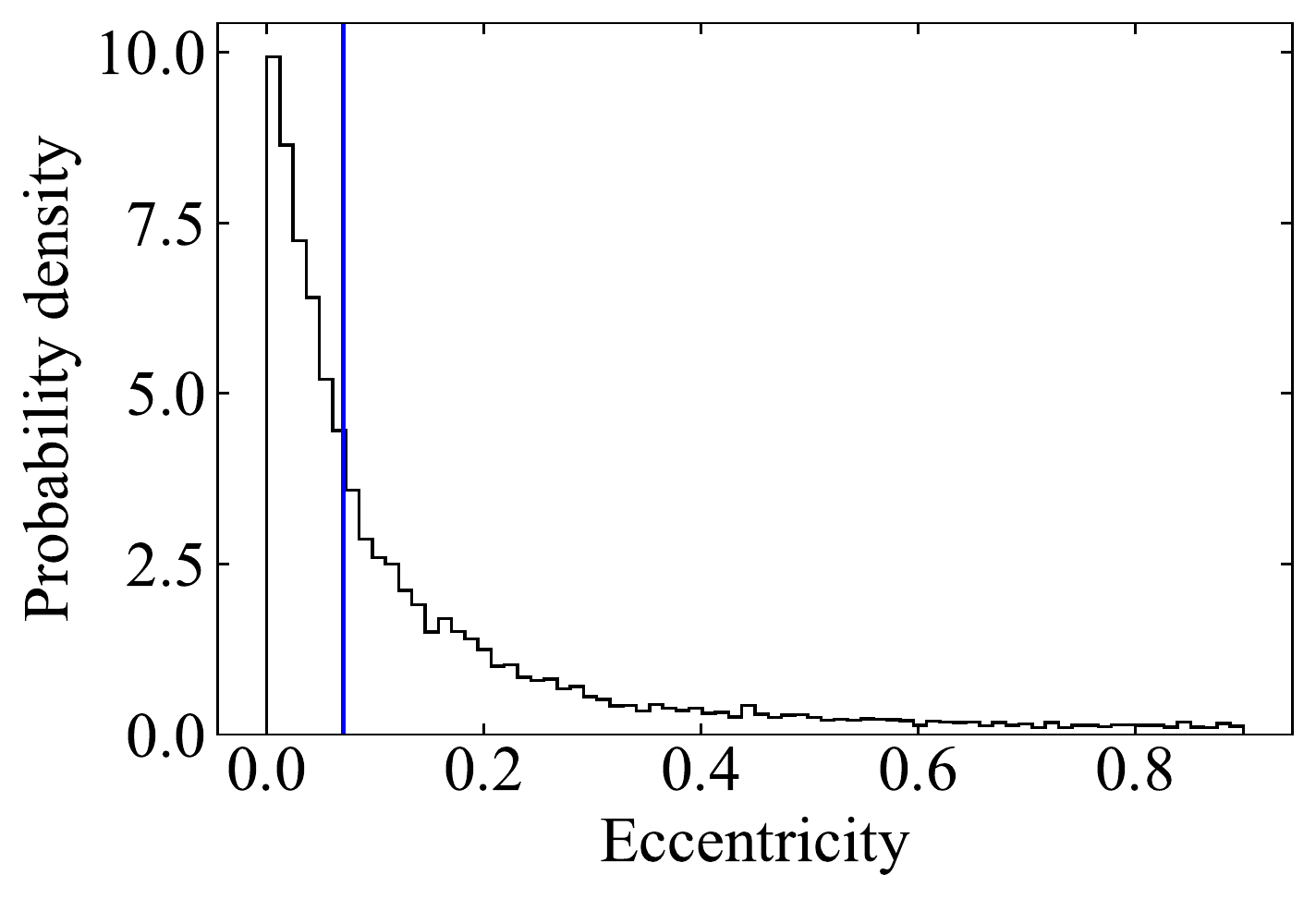}
         \includegraphics[width=0.245\linewidth]{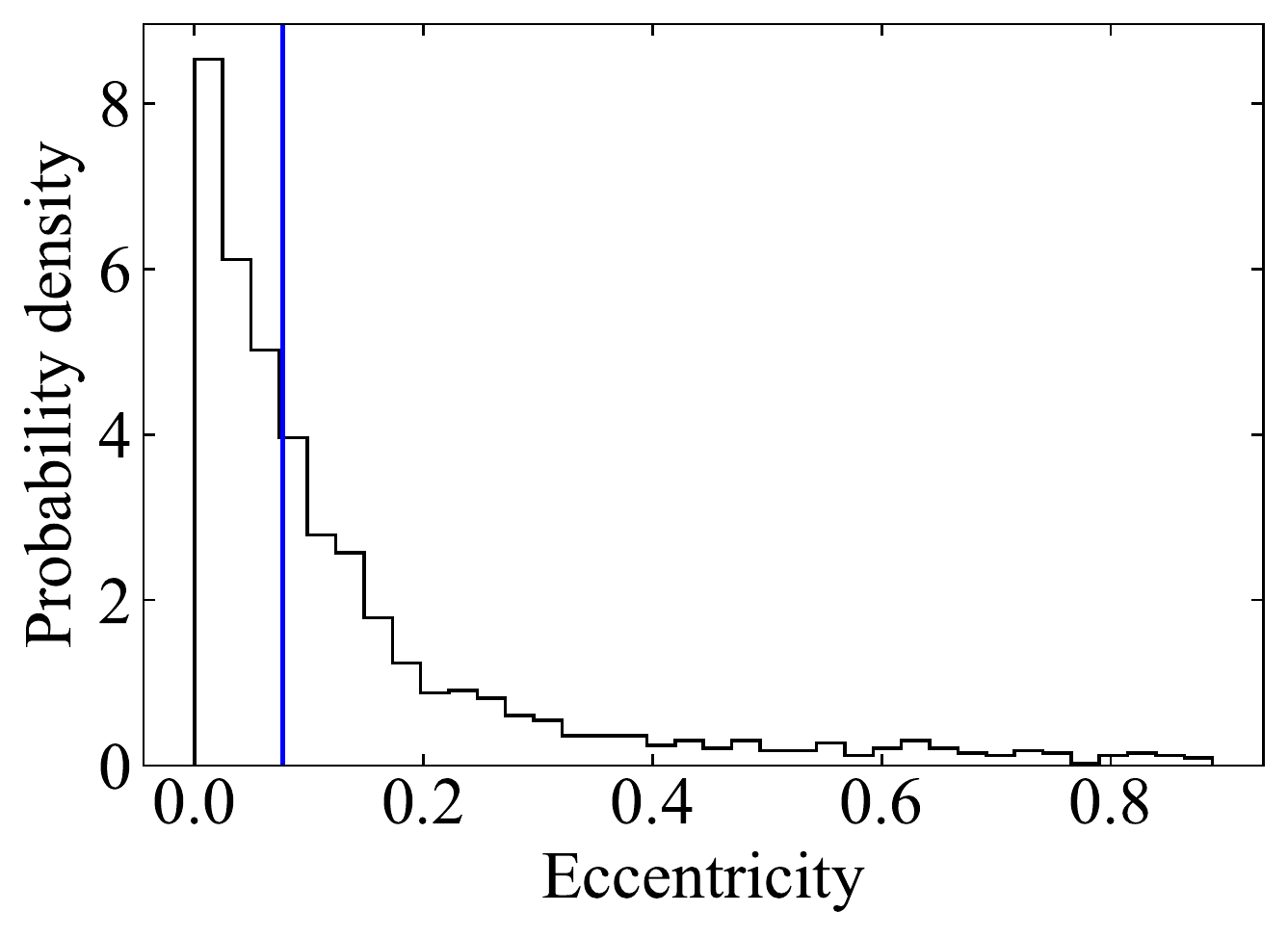}
         \includegraphics[width=0.245\linewidth]{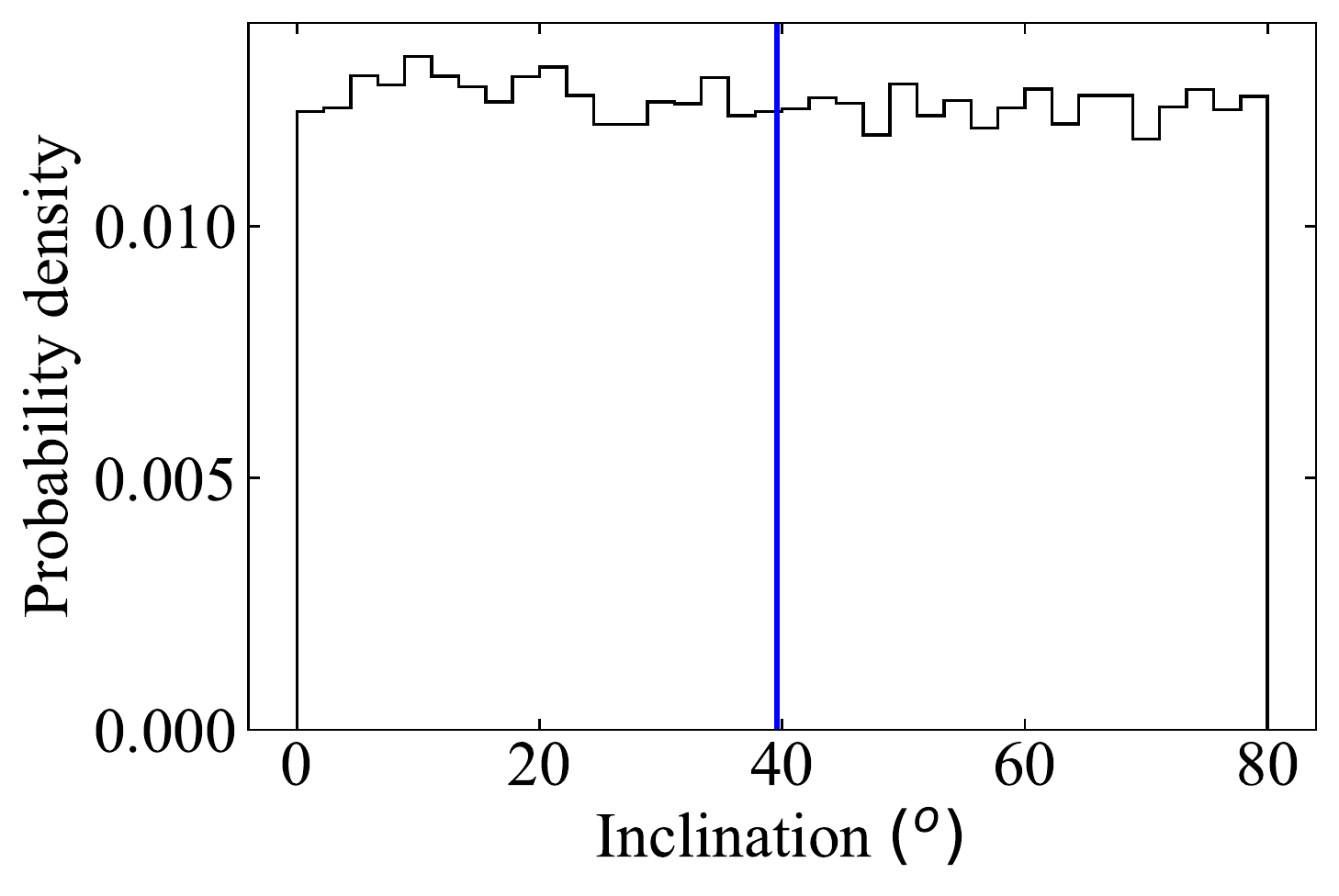}
         \includegraphics[width=0.245\linewidth]{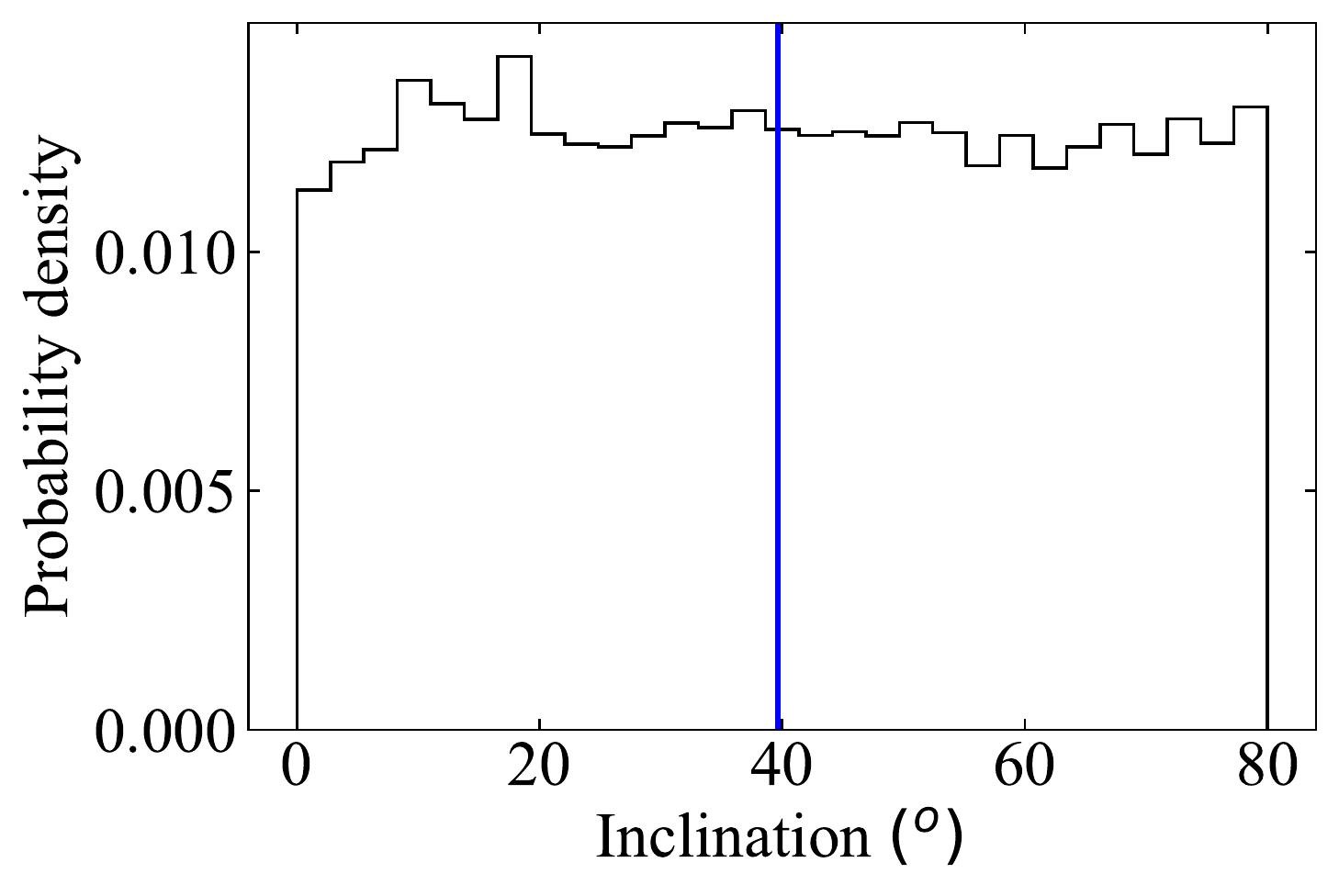}
         \includegraphics[width=0.245\linewidth]{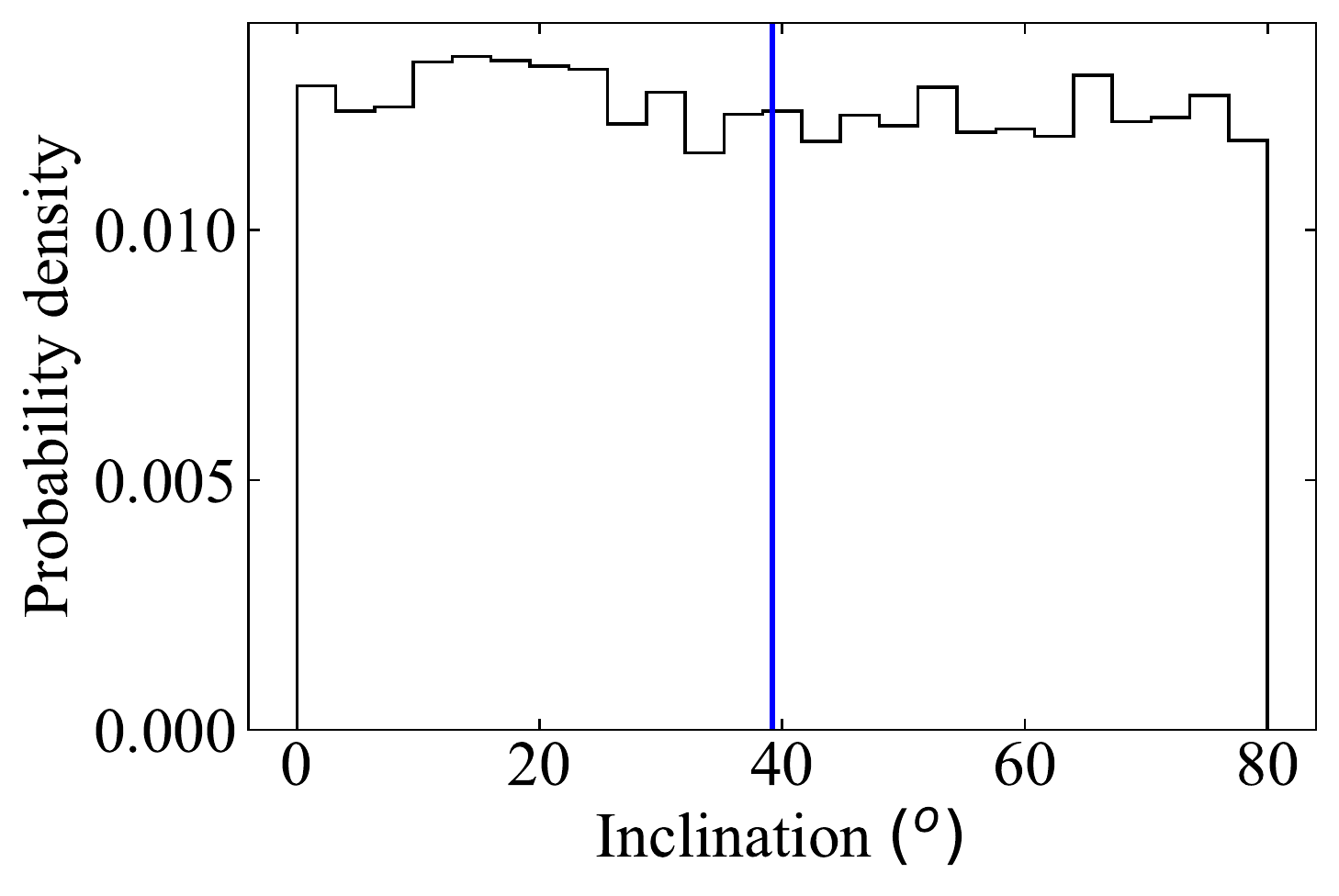}
         \includegraphics[width=0.245\linewidth]{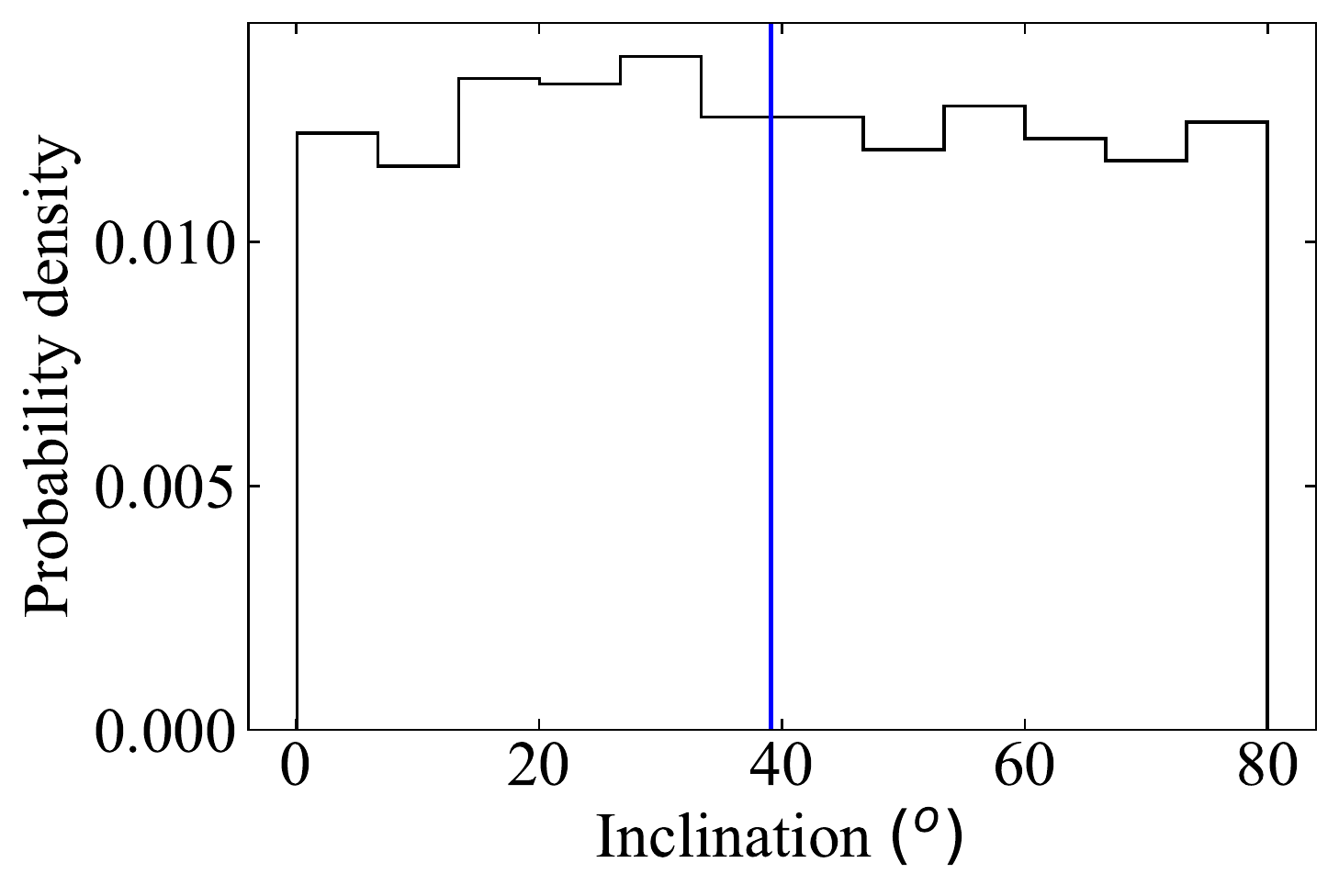}
         \includegraphics[width=0.245\linewidth]{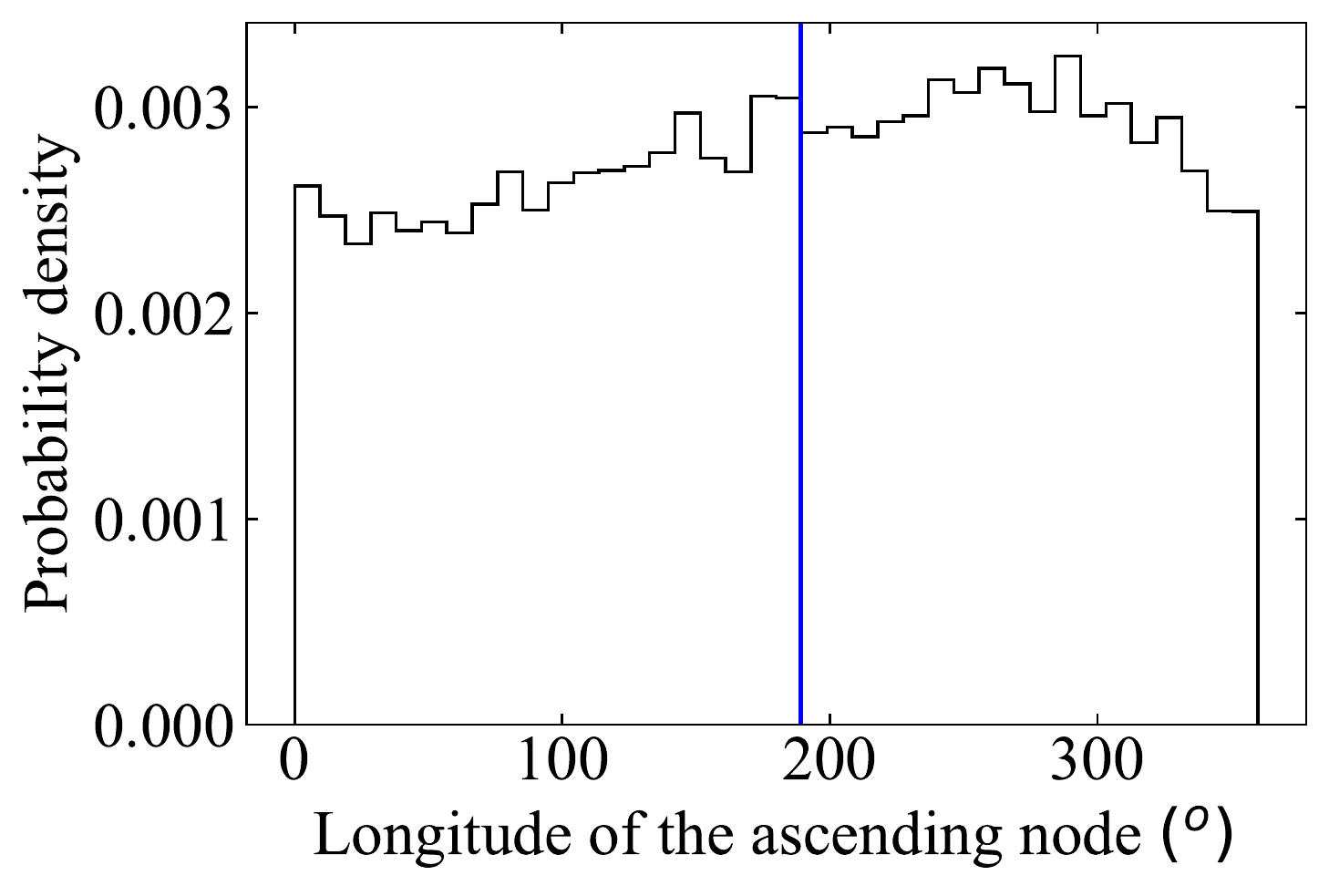}
         \includegraphics[width=0.245\linewidth]{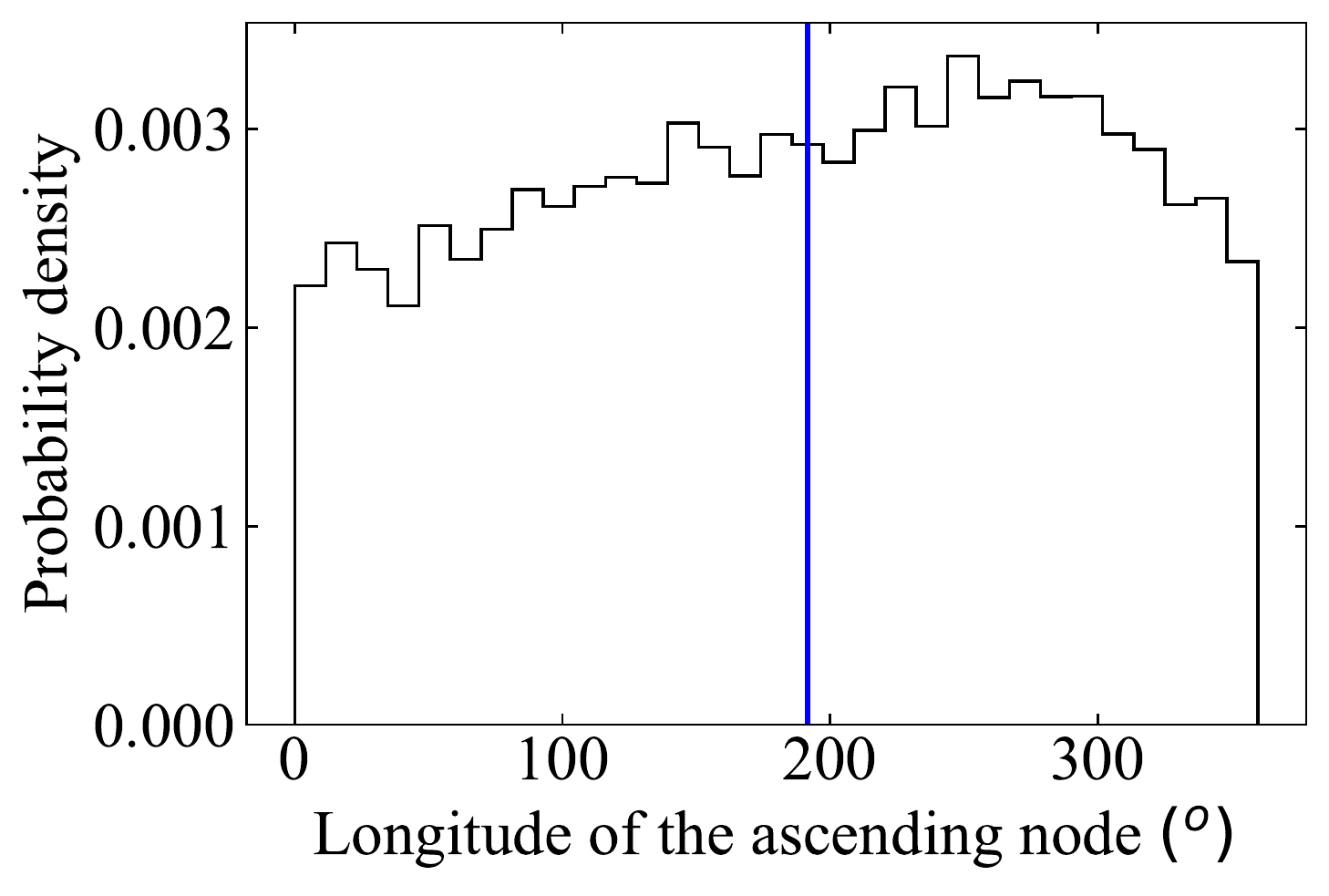}
         \includegraphics[width=0.245\linewidth]{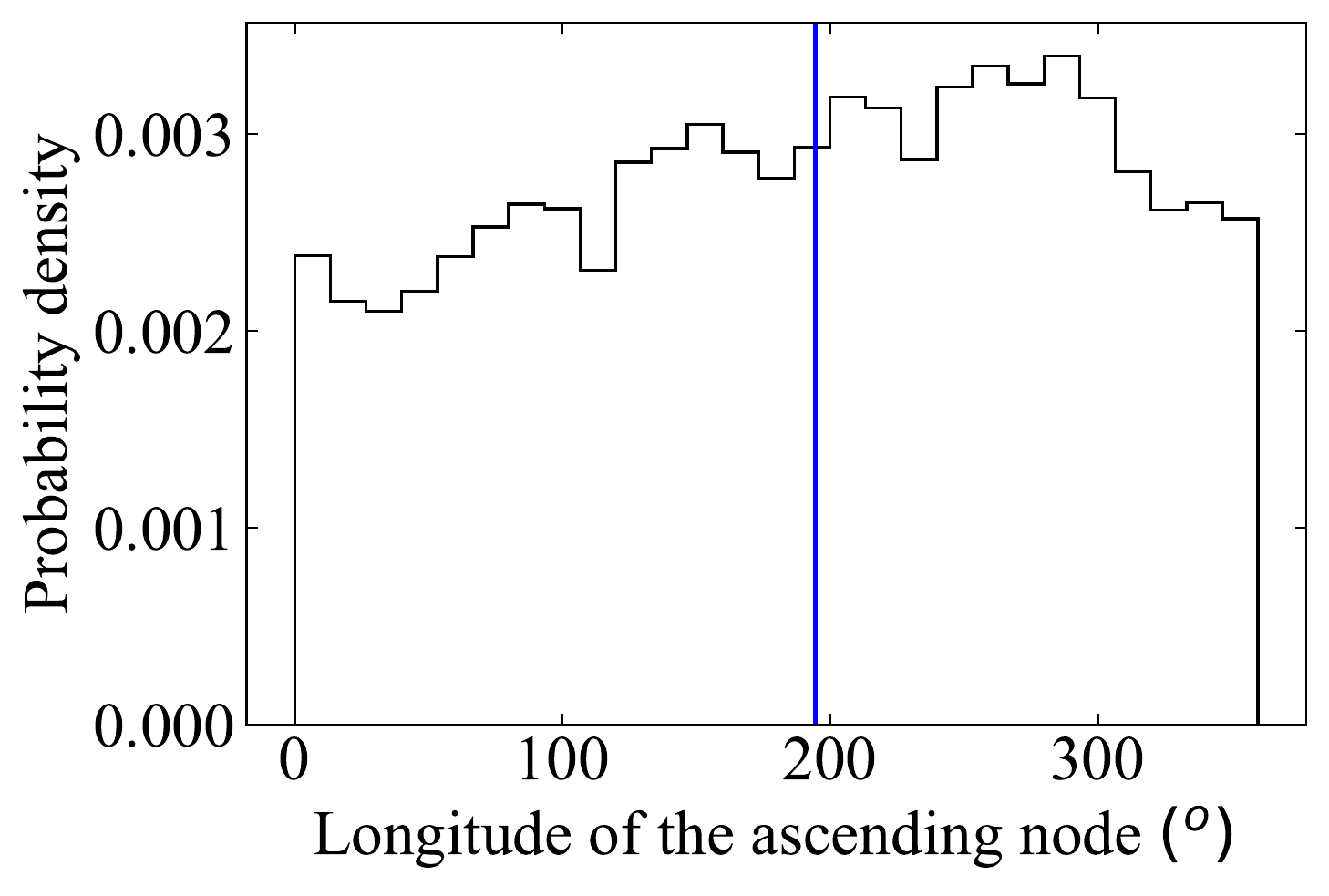}
         \includegraphics[width=0.245\linewidth]{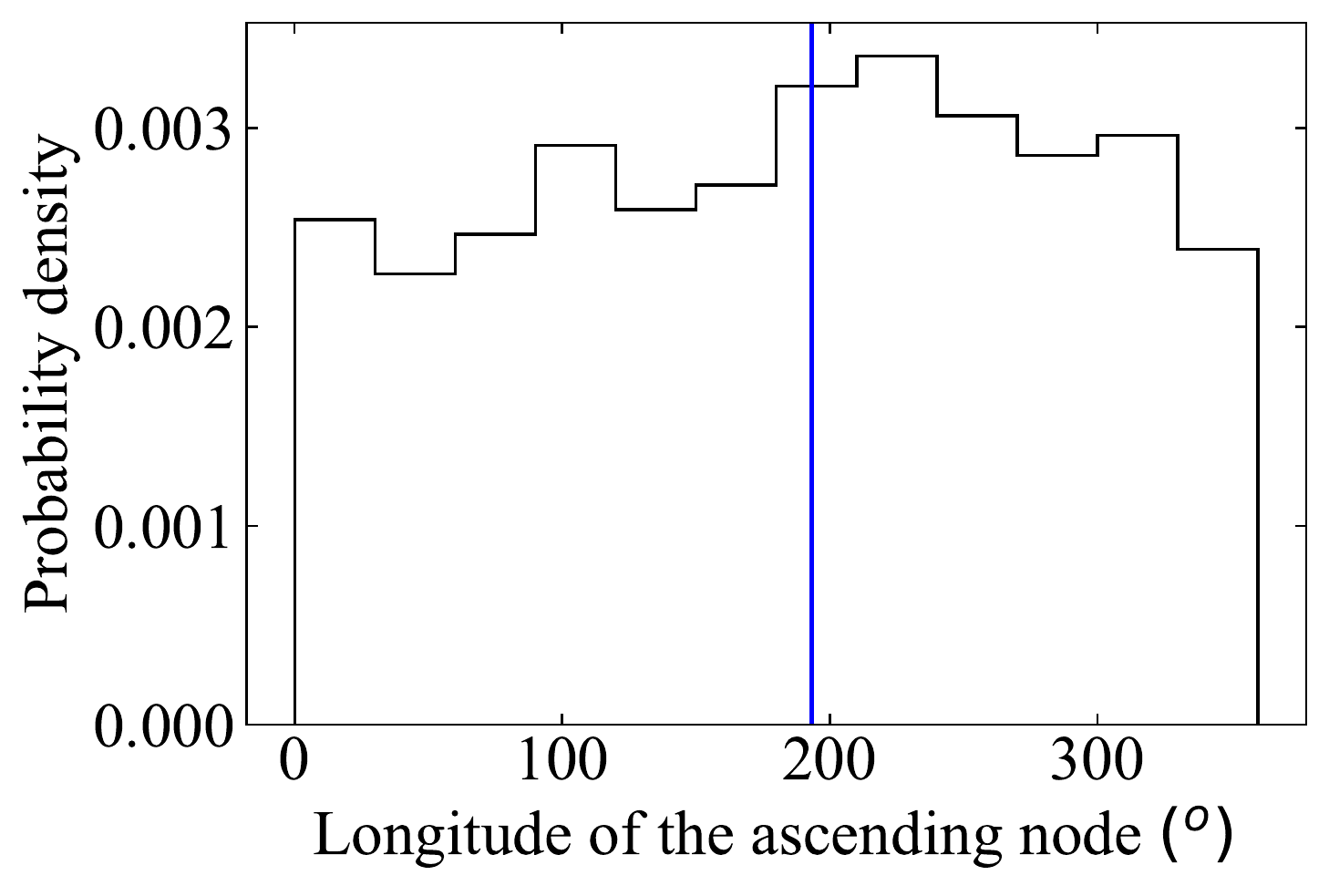}
         \includegraphics[width=0.245\linewidth]{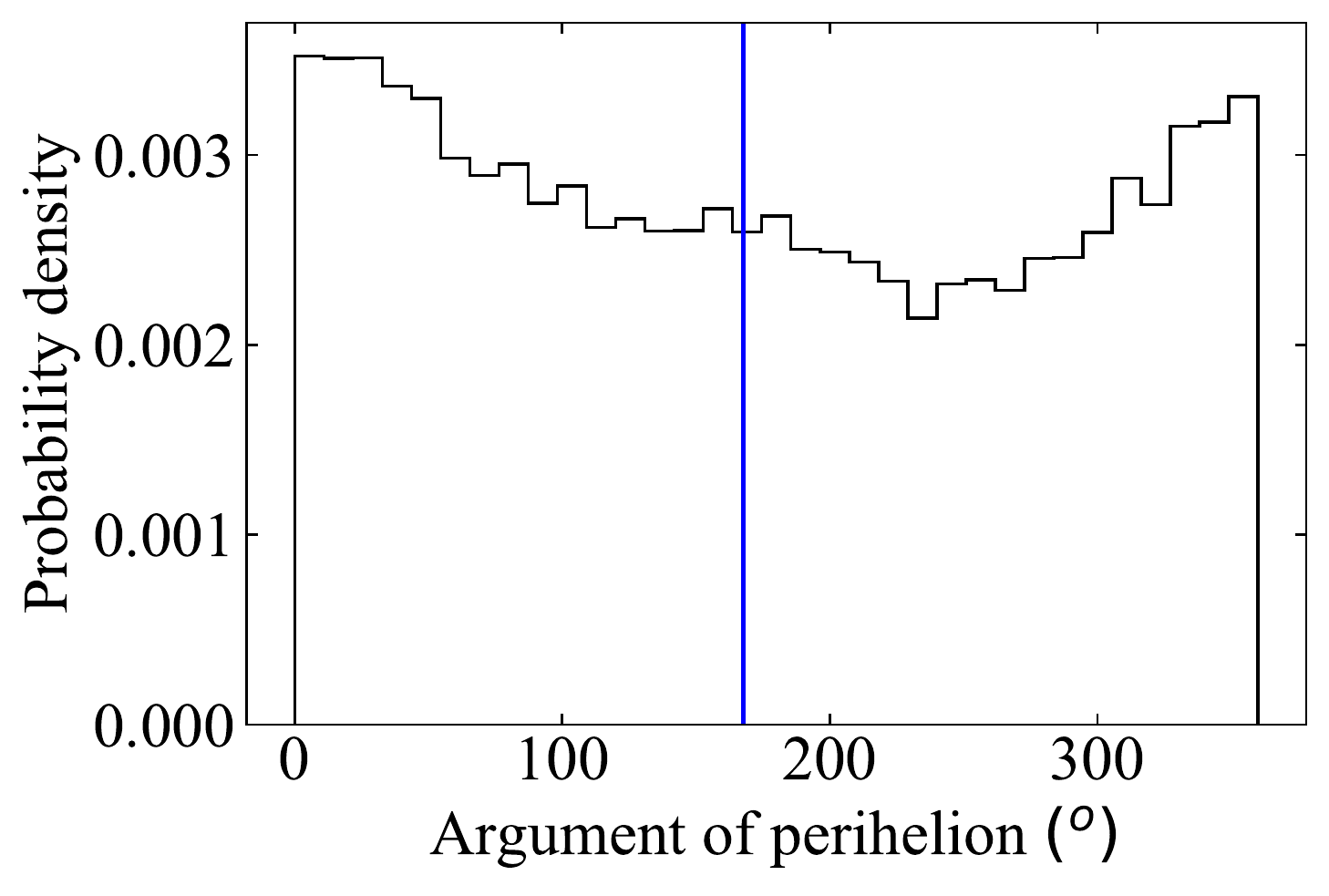}
         \includegraphics[width=0.245\linewidth]{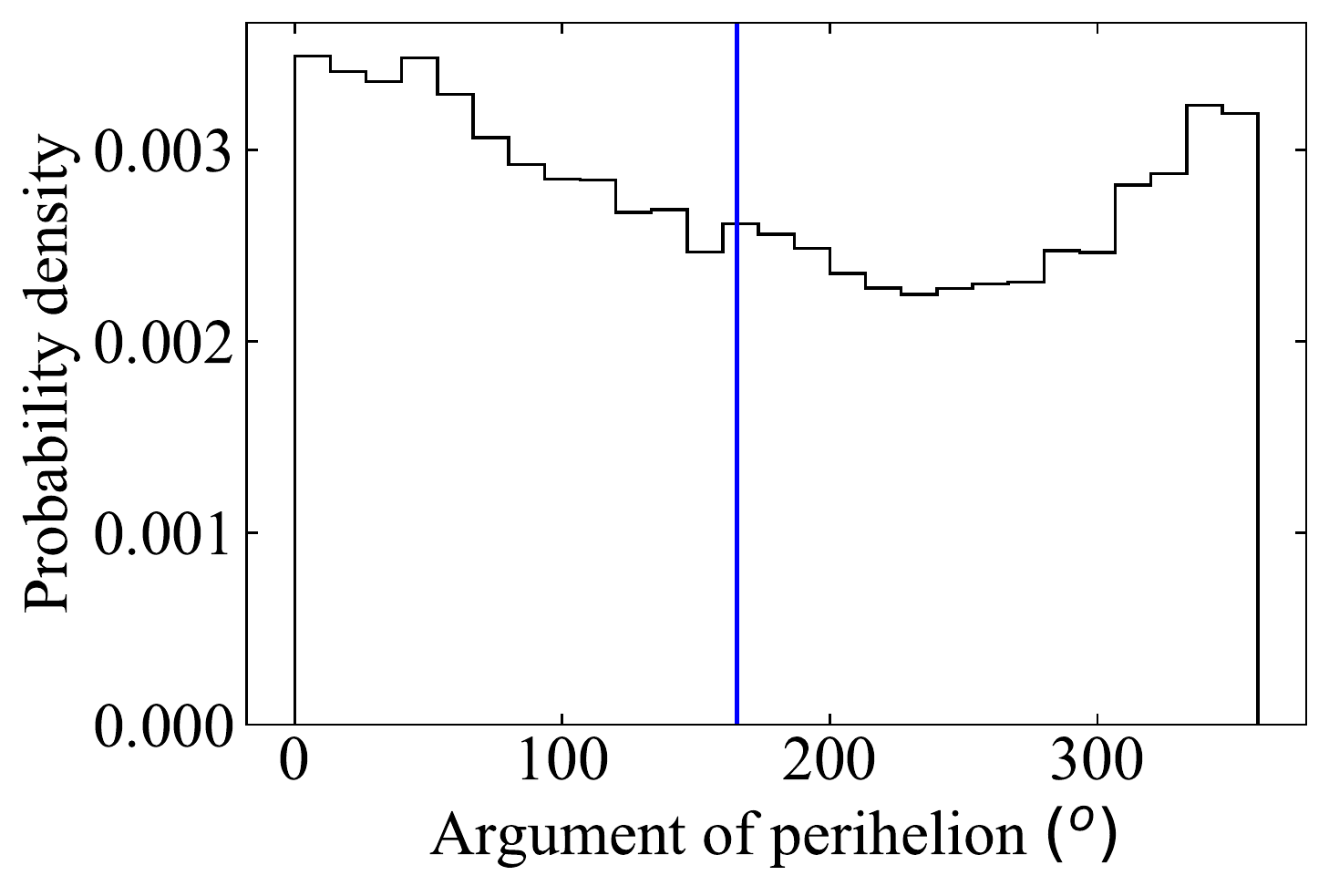}
         \includegraphics[width=0.245\linewidth]{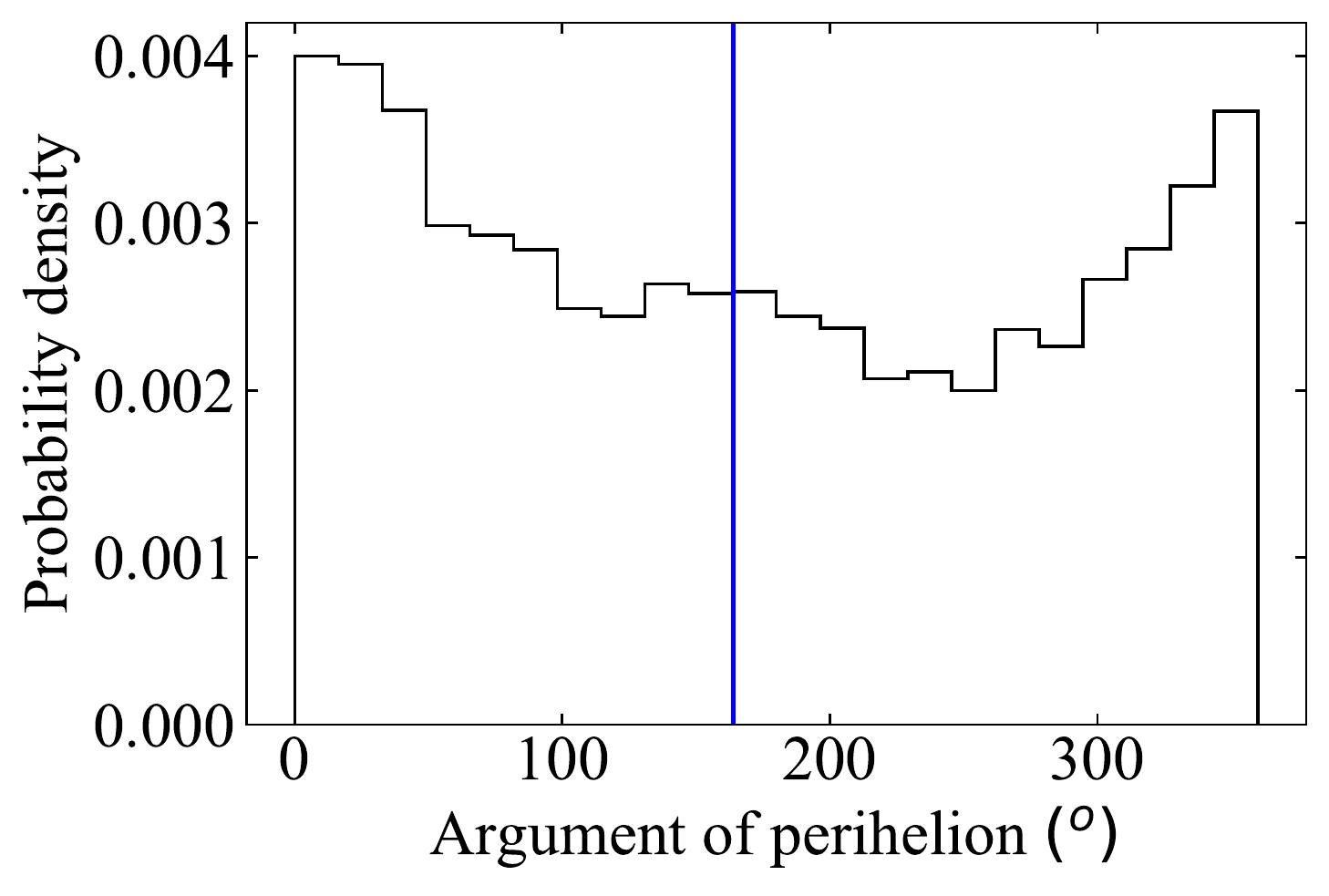}
         \includegraphics[width=0.245\linewidth]{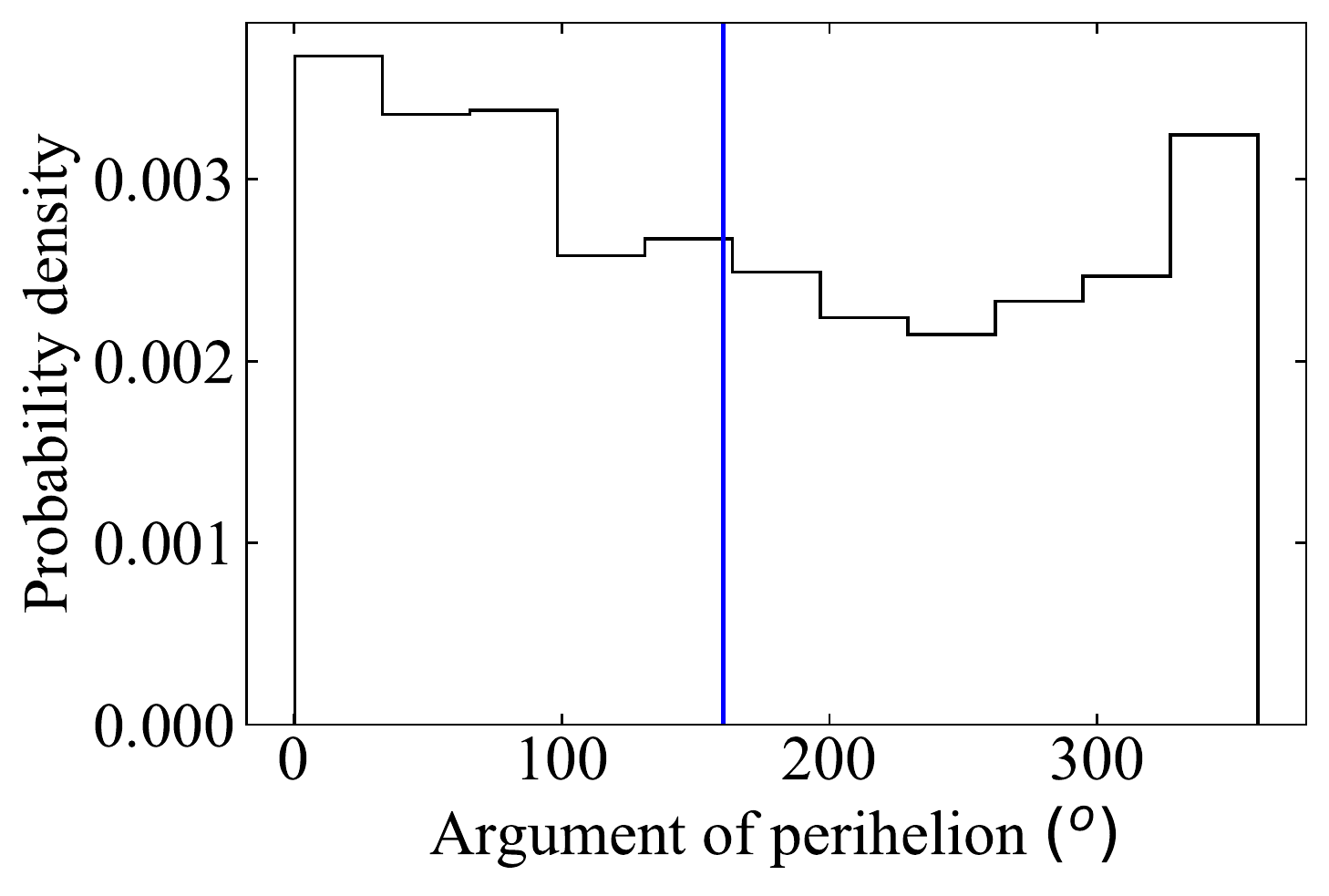}
         \caption{Barycentric orbital elements of putative perturber planets. As Fig.~\ref{histogramsB} but using scrambled data as input. 
                 }
         \label{histogramsCHKB}
      \end{figure*}

\begin{figure*}
        \centering
         \includegraphics[width=0.245\linewidth]{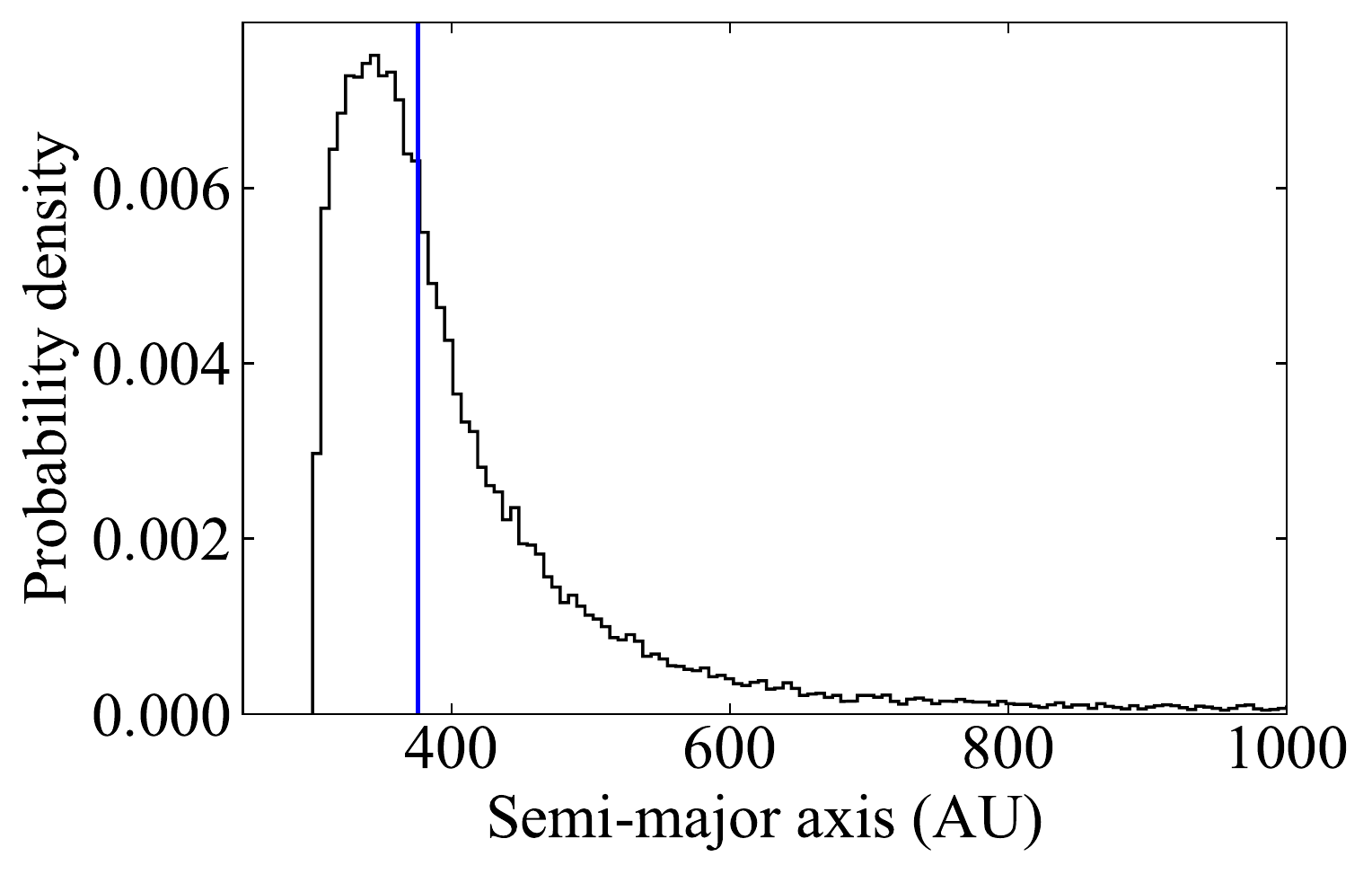}
         \includegraphics[width=0.245\linewidth]{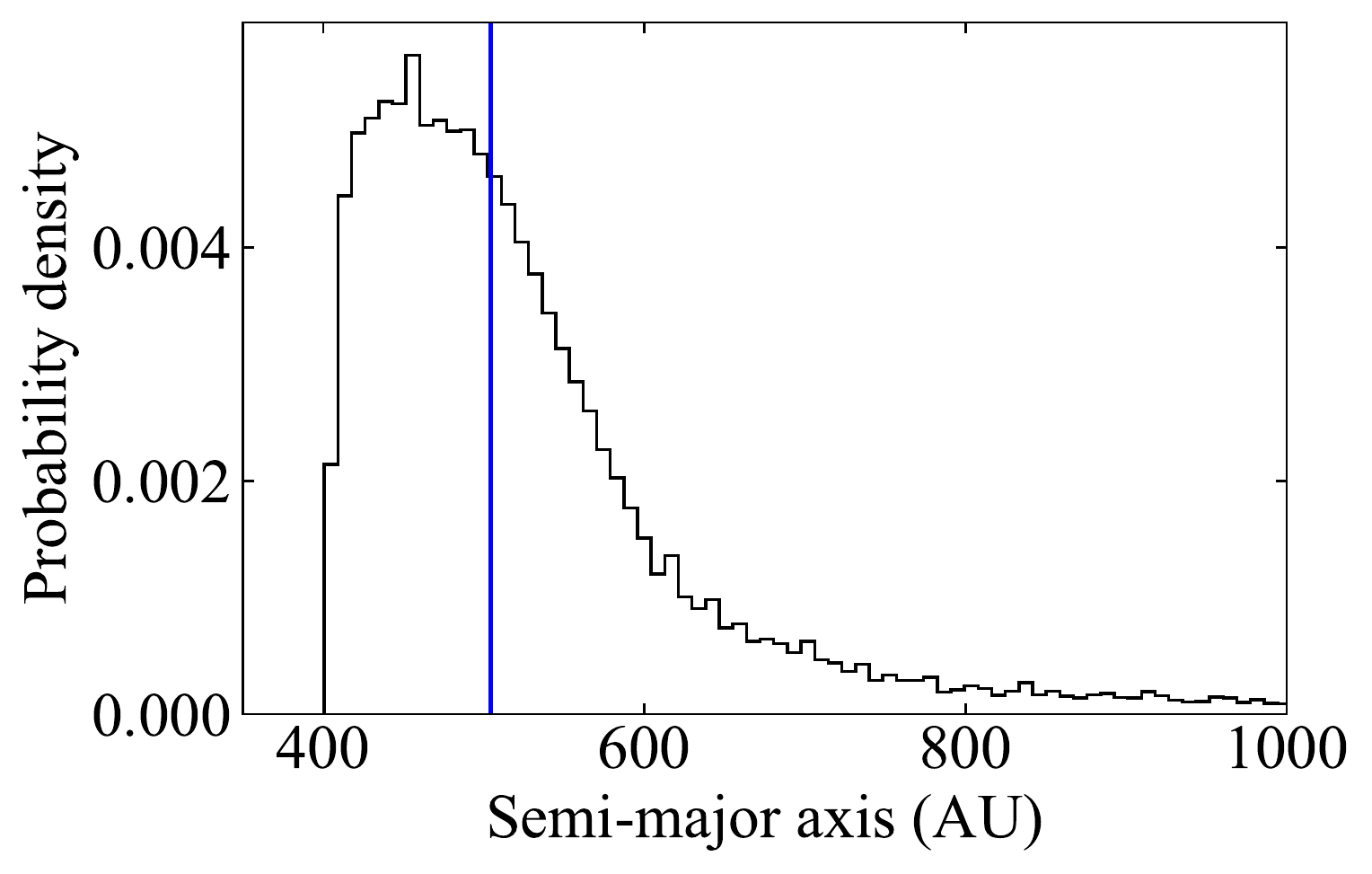}
         \includegraphics[width=0.245\linewidth]{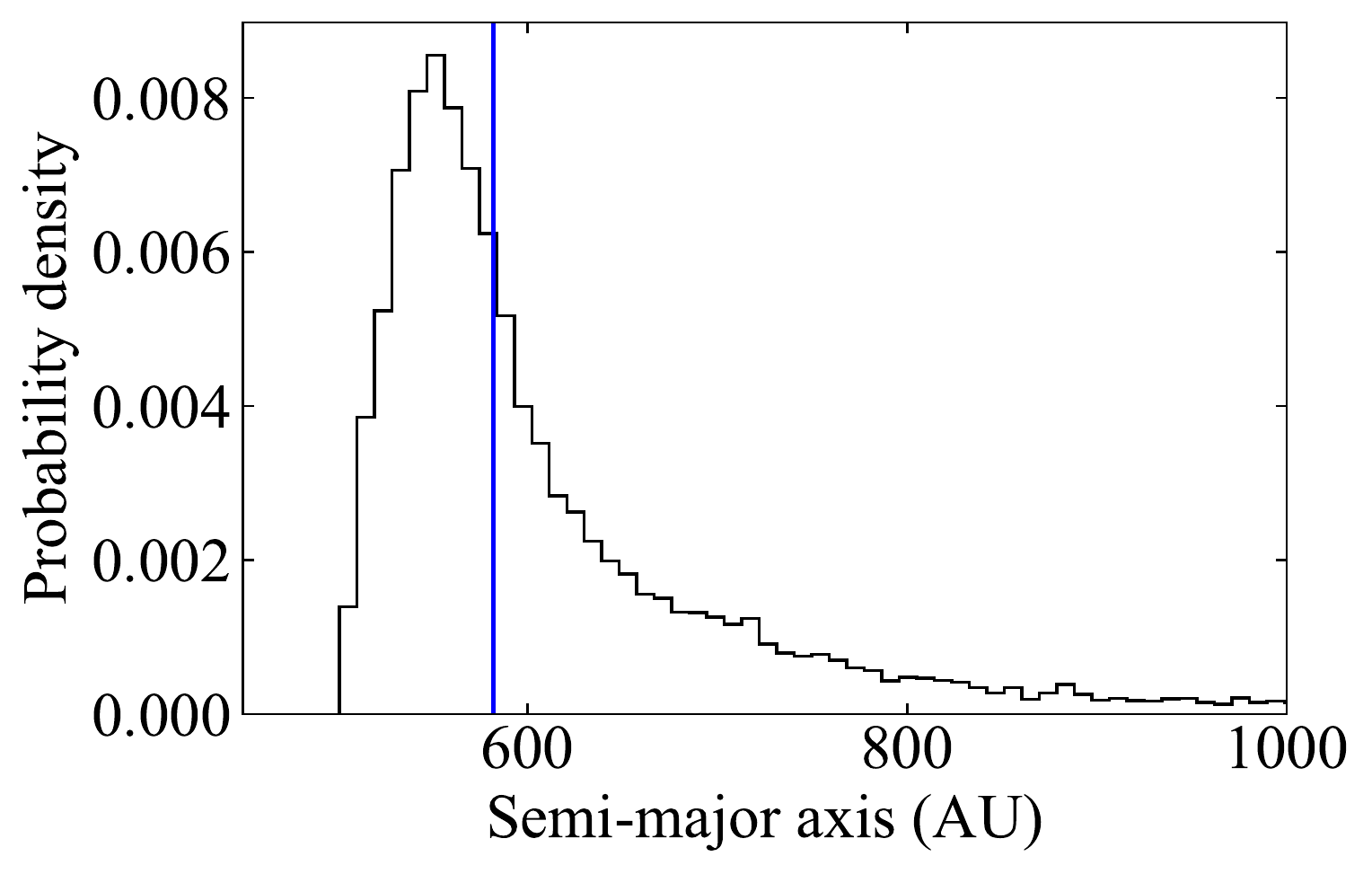}
         \includegraphics[width=0.245\linewidth]{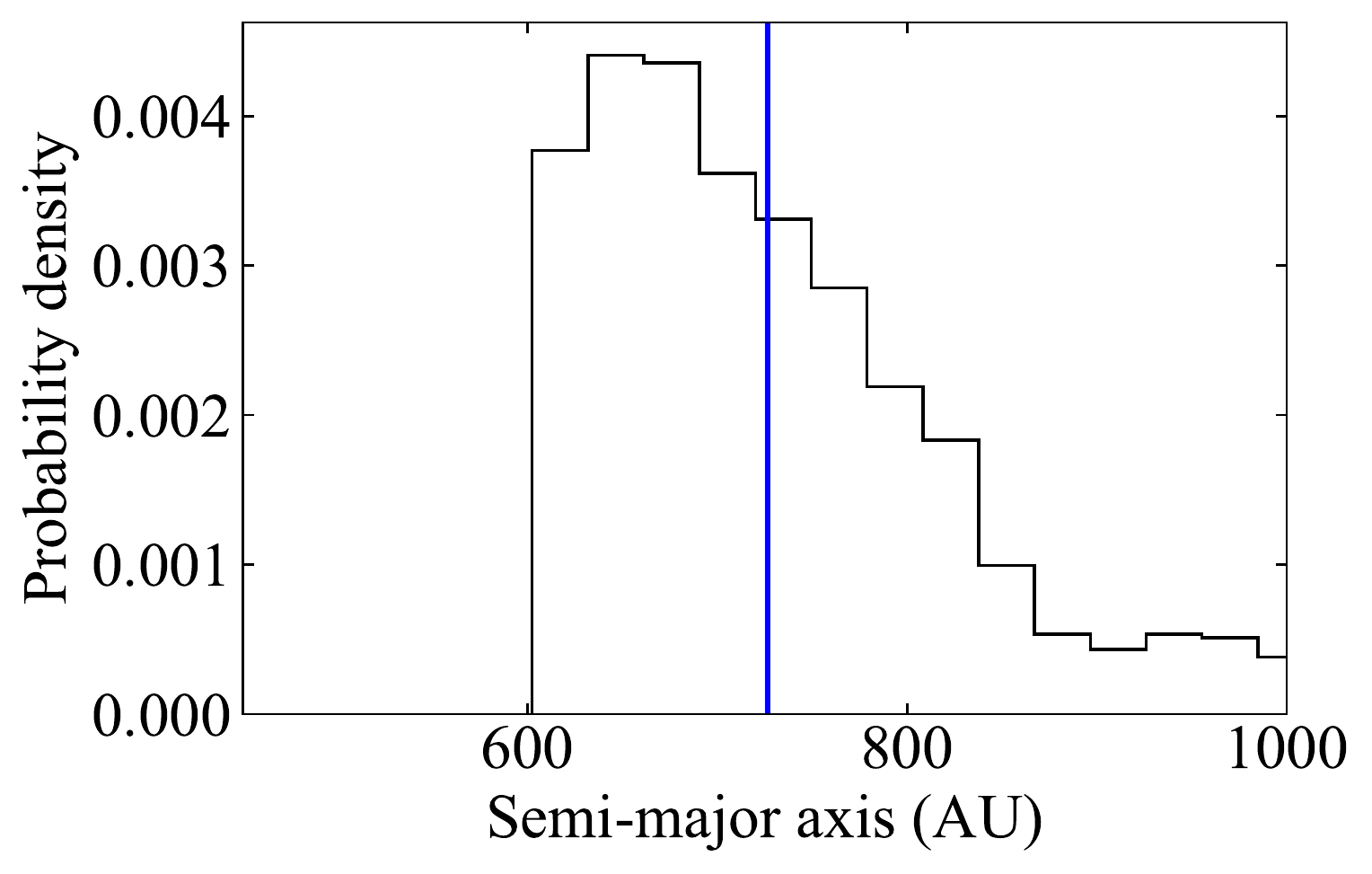}
         \includegraphics[width=0.245\linewidth]{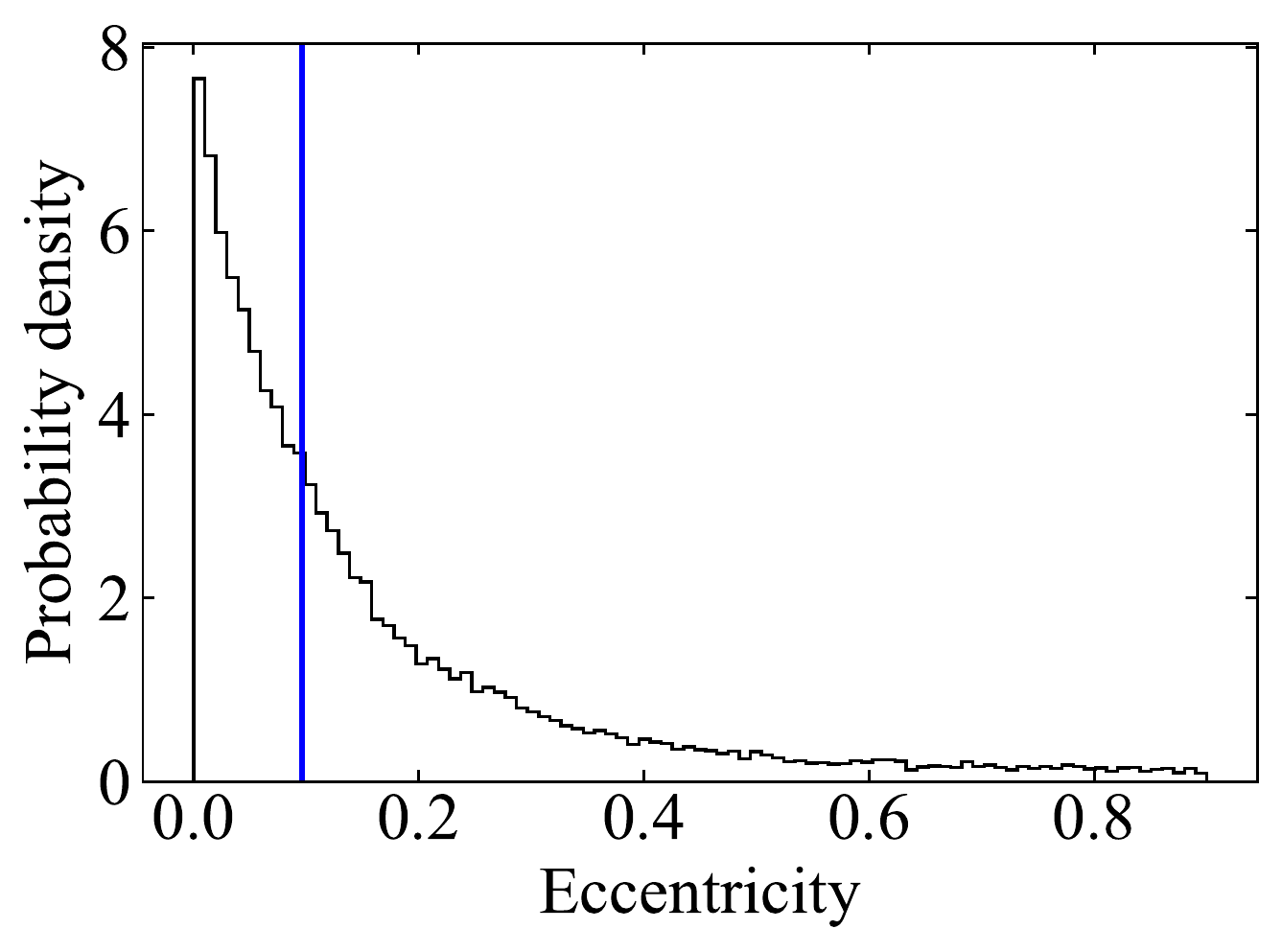}
         \includegraphics[width=0.245\linewidth]{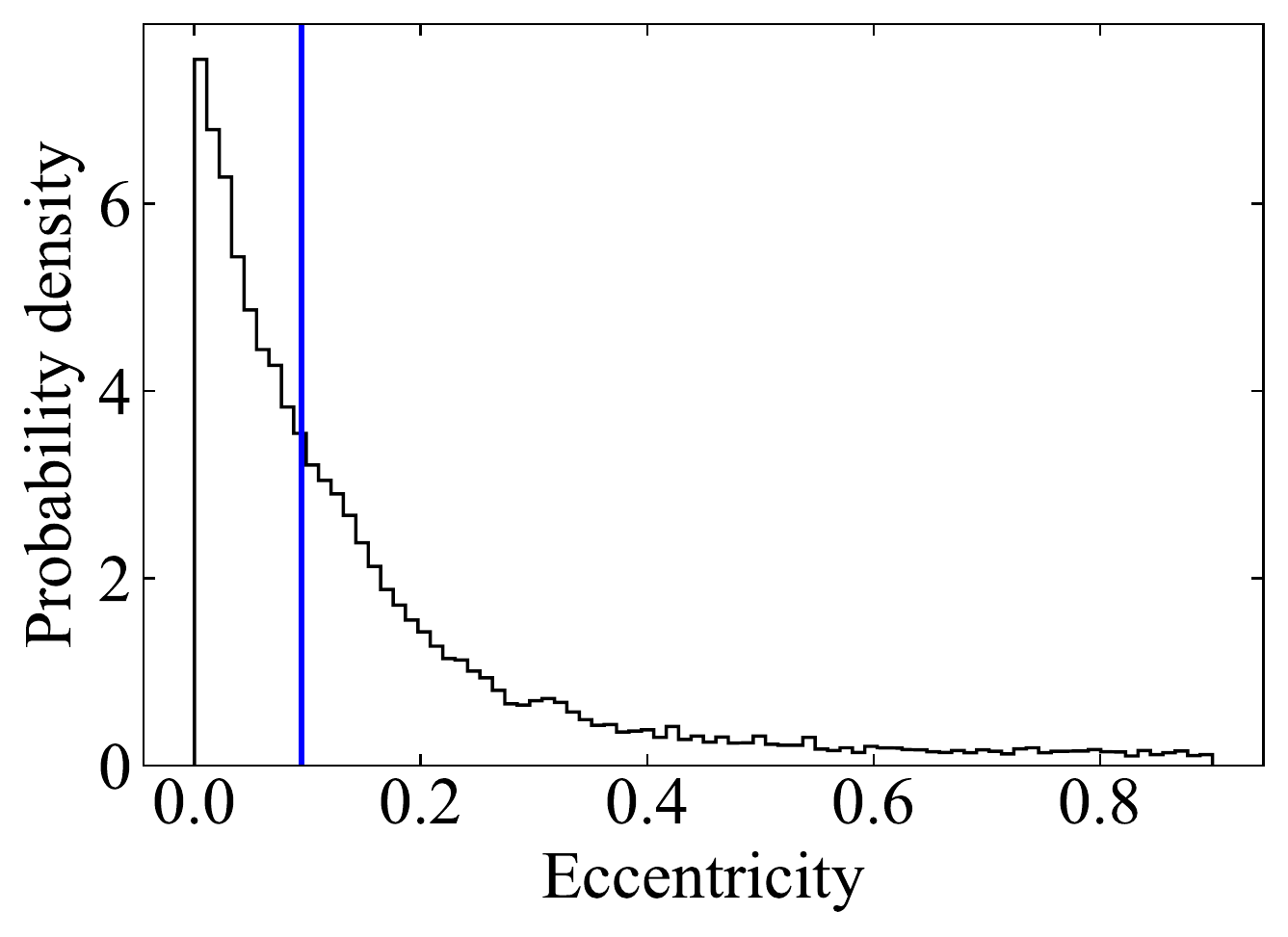}
         \includegraphics[width=0.245\linewidth]{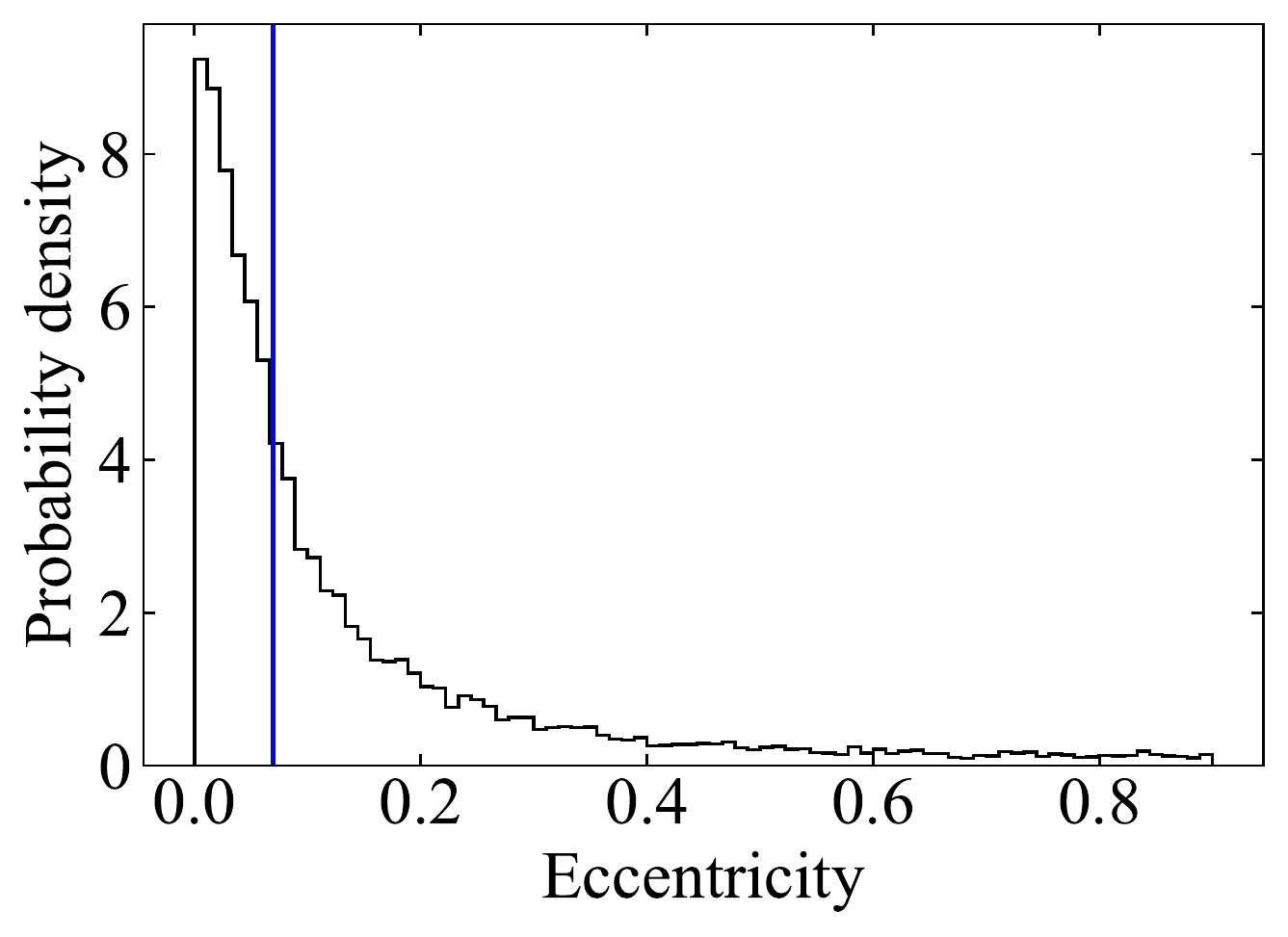}
         \includegraphics[width=0.245\linewidth]{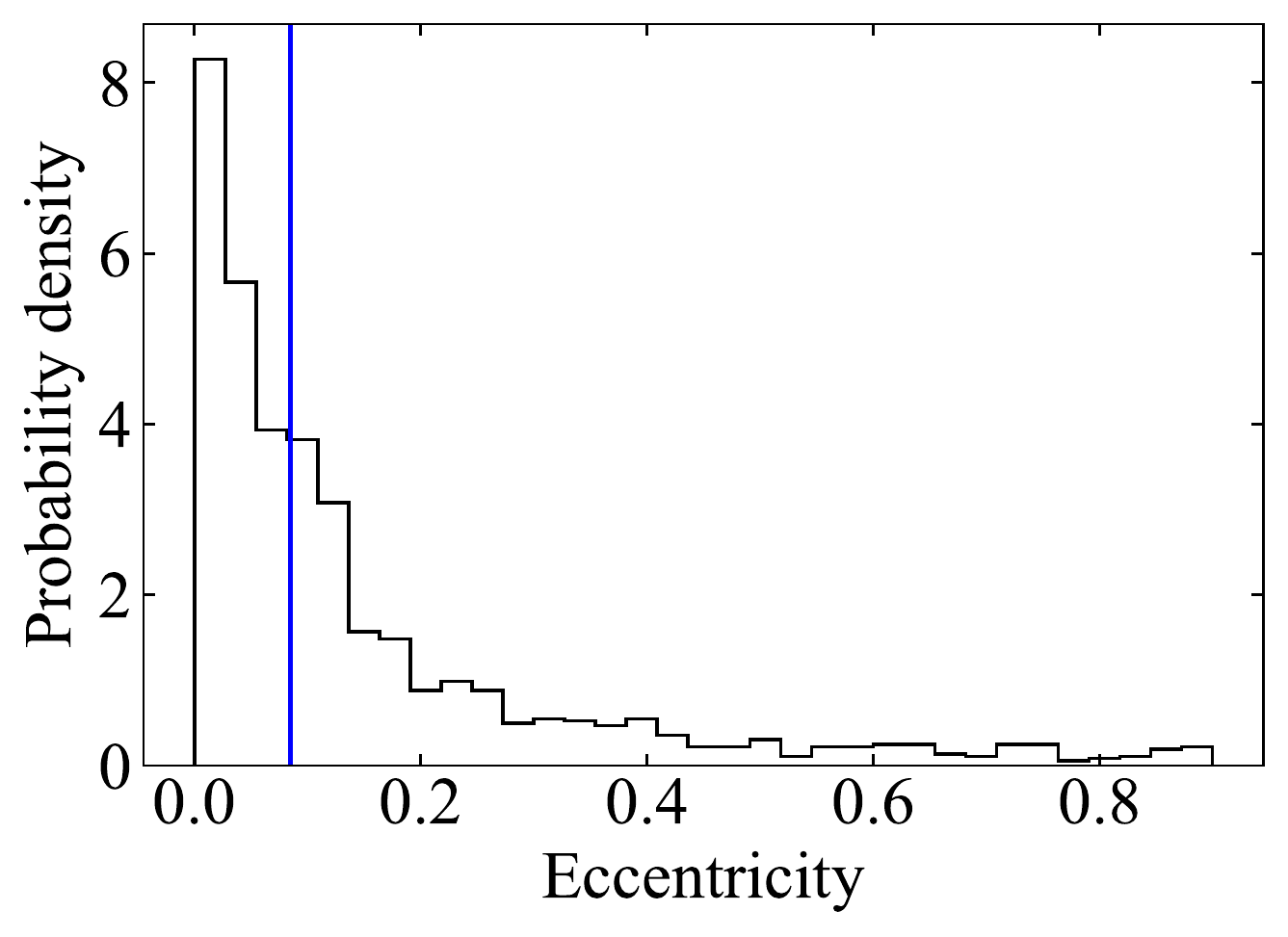}
         \includegraphics[width=0.245\linewidth]{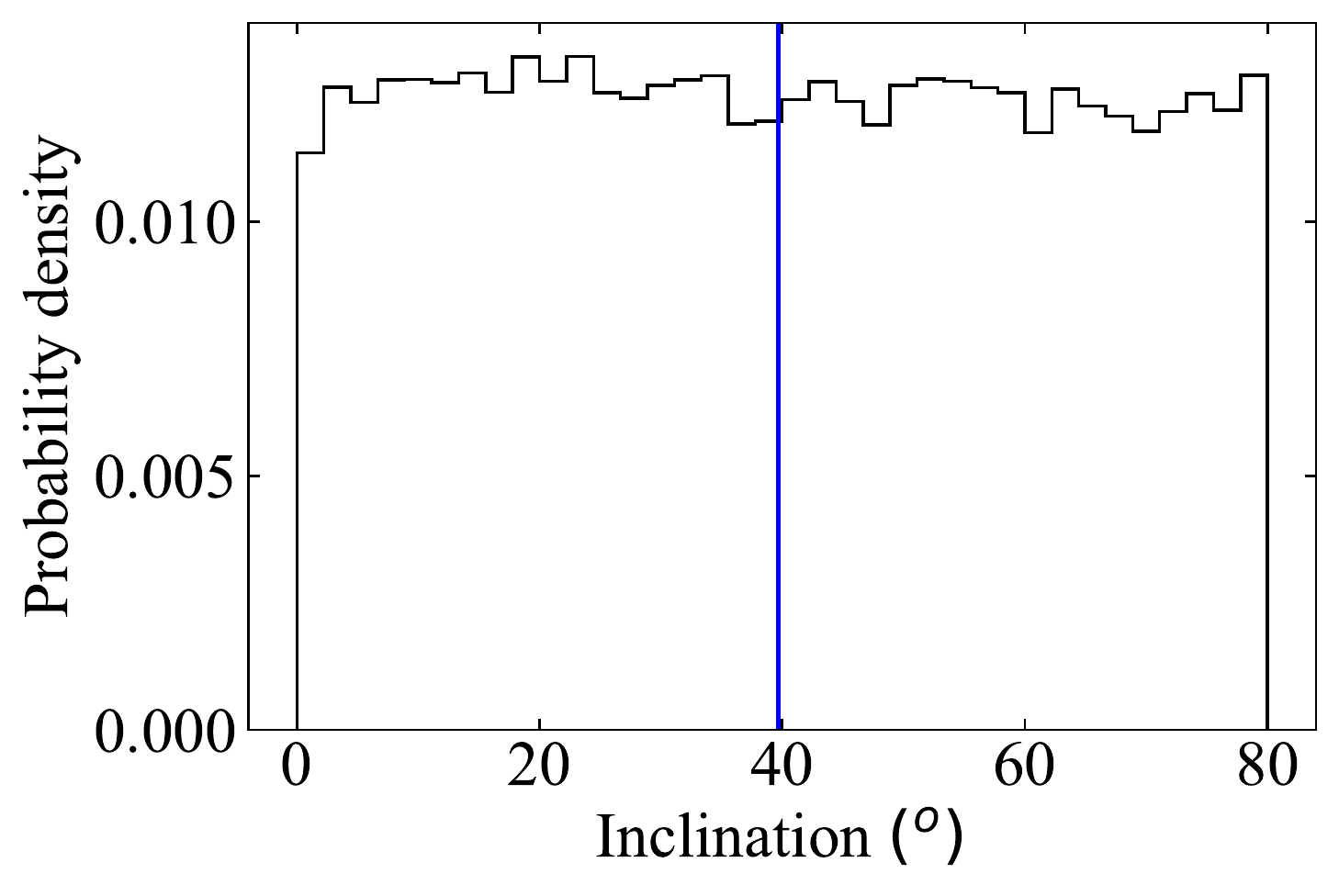}
         \includegraphics[width=0.245\linewidth]{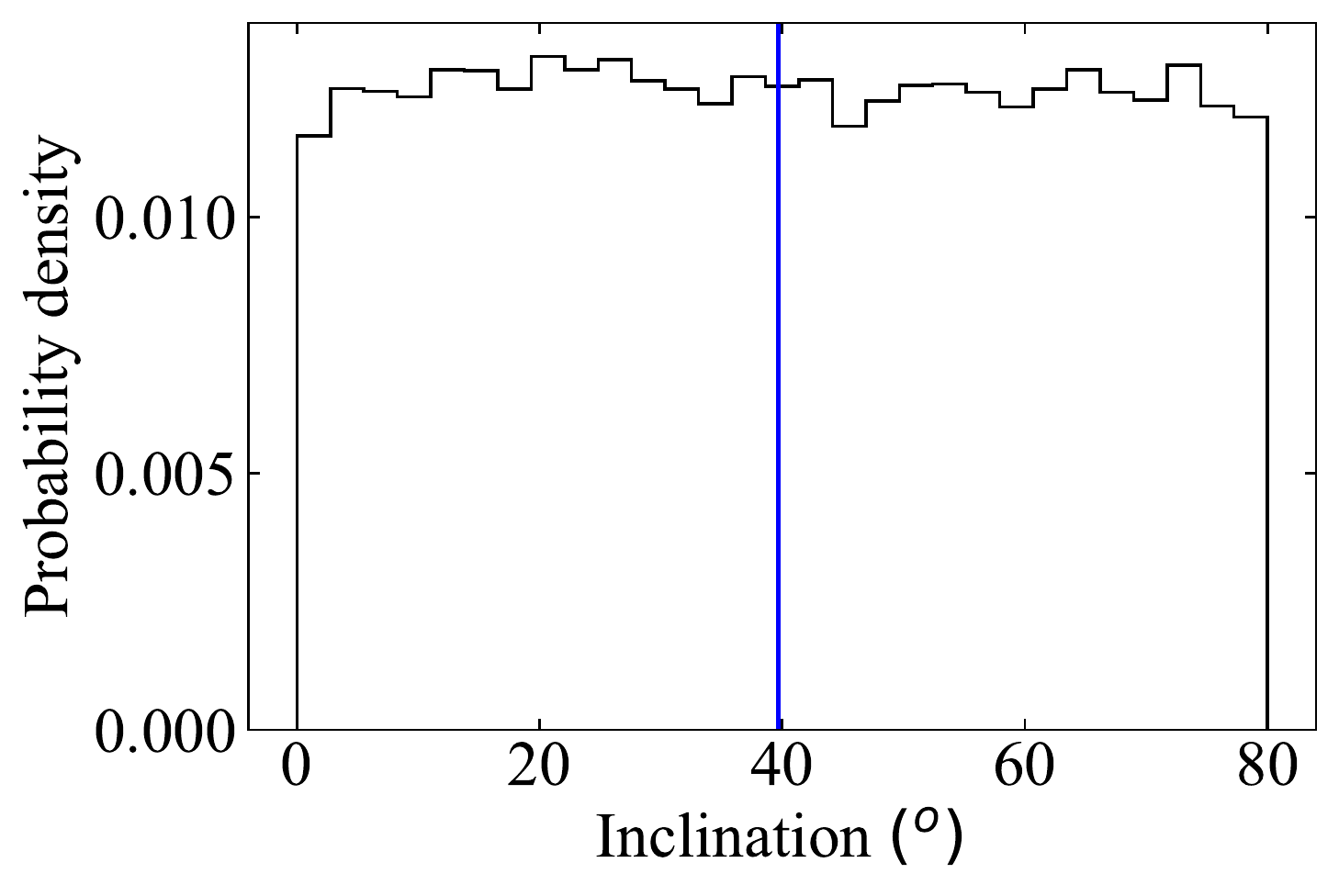}
         \includegraphics[width=0.245\linewidth]{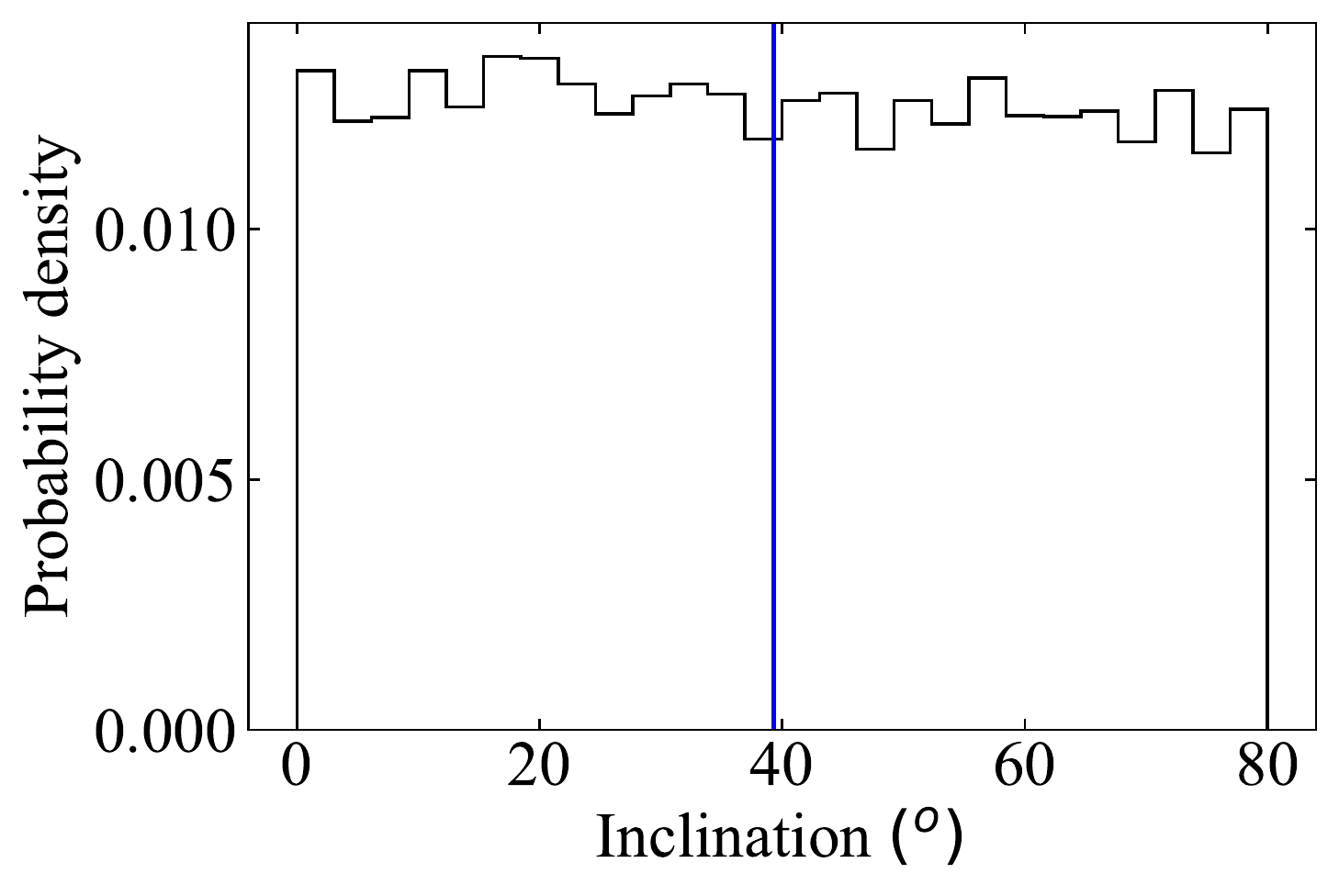}
         \includegraphics[width=0.245\linewidth]{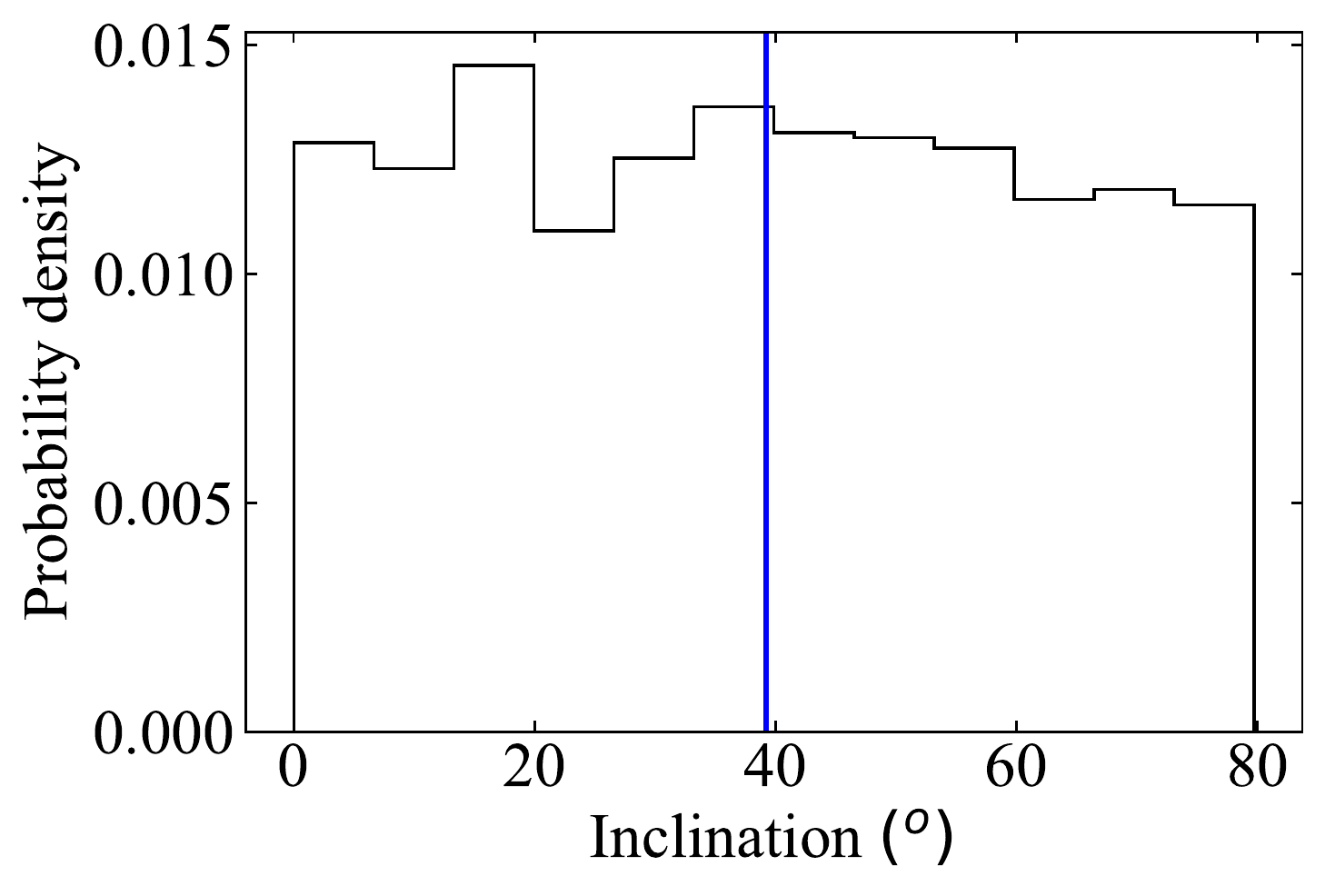}
         \includegraphics[width=0.245\linewidth]{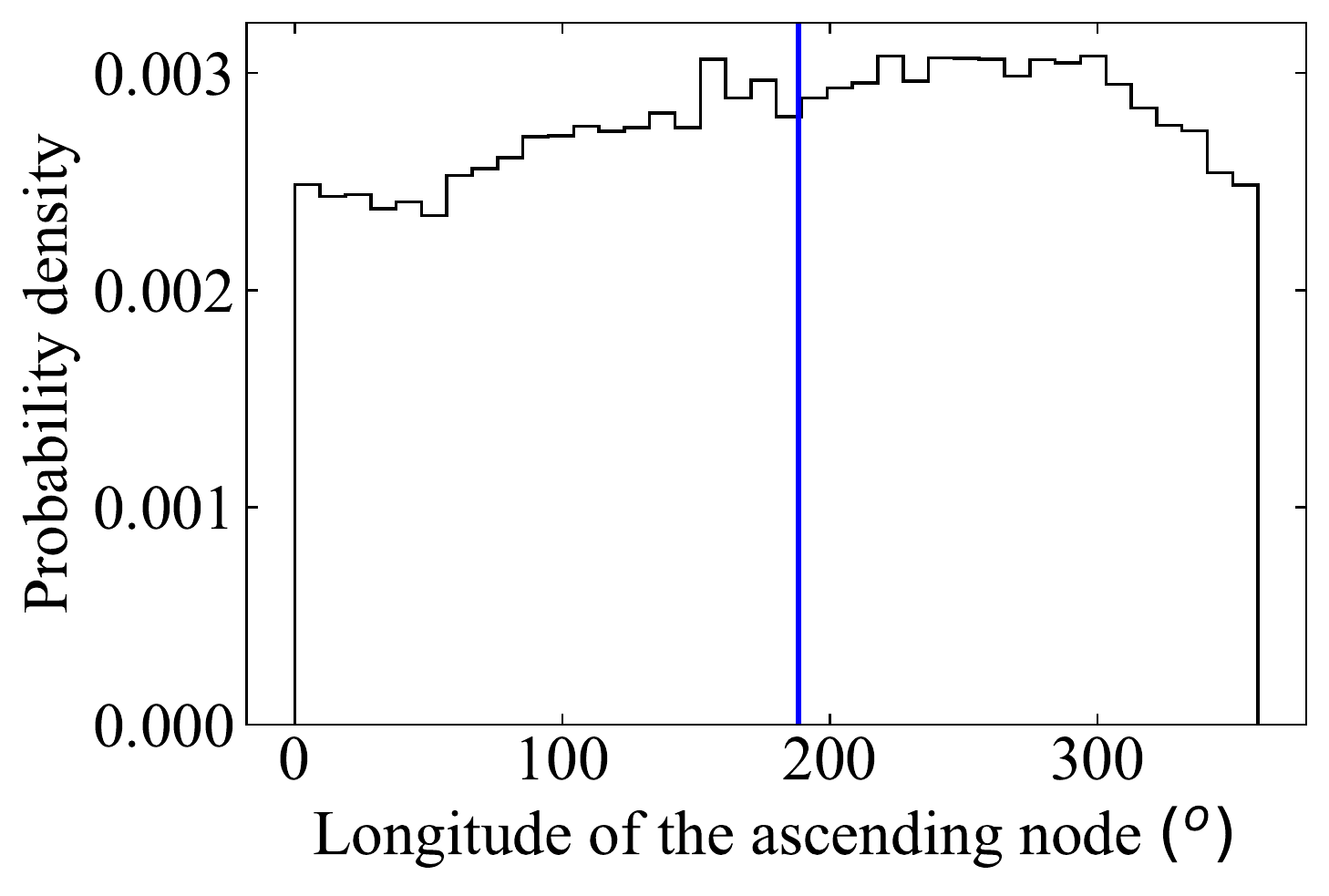}
         \includegraphics[width=0.245\linewidth]{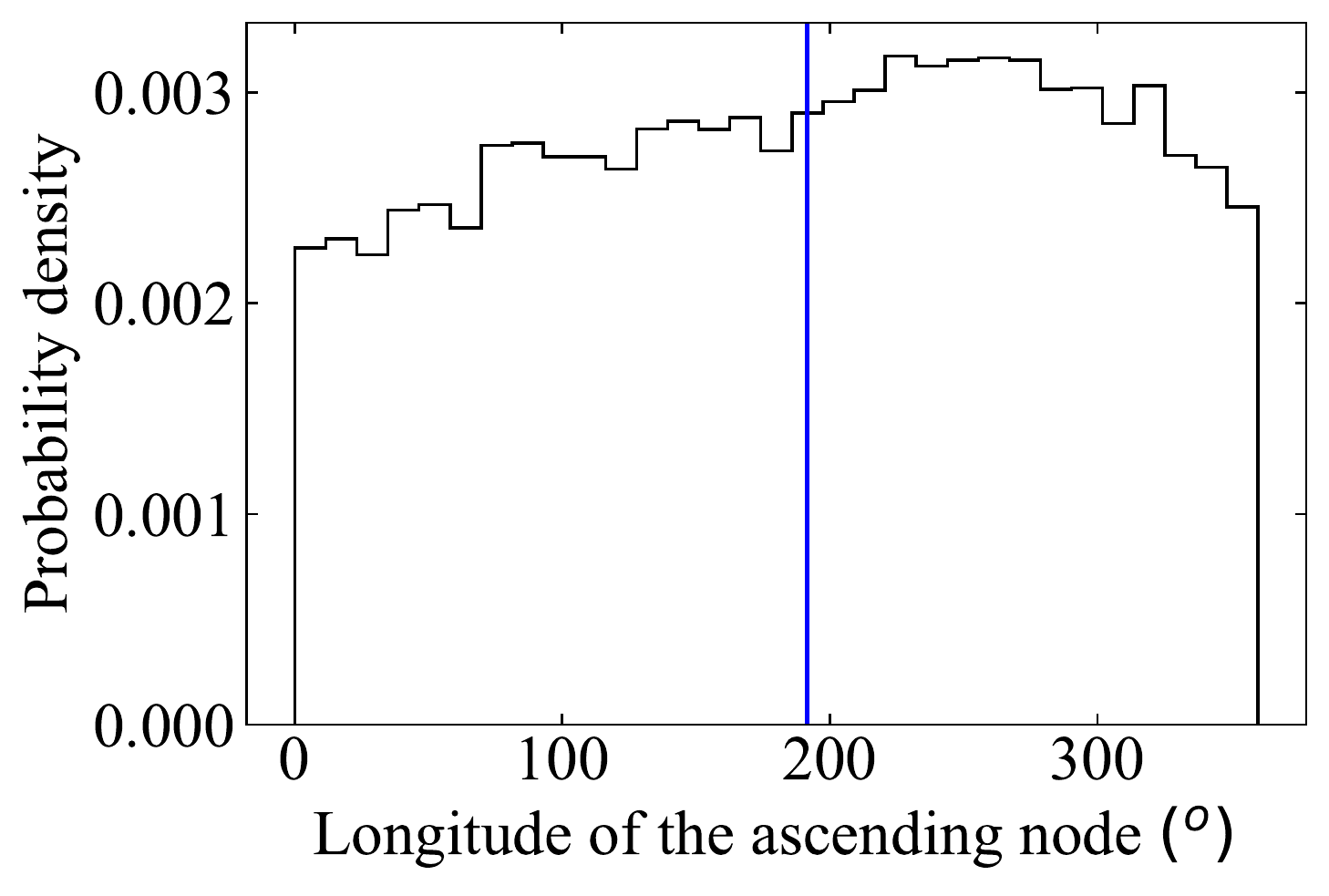}
         \includegraphics[width=0.245\linewidth]{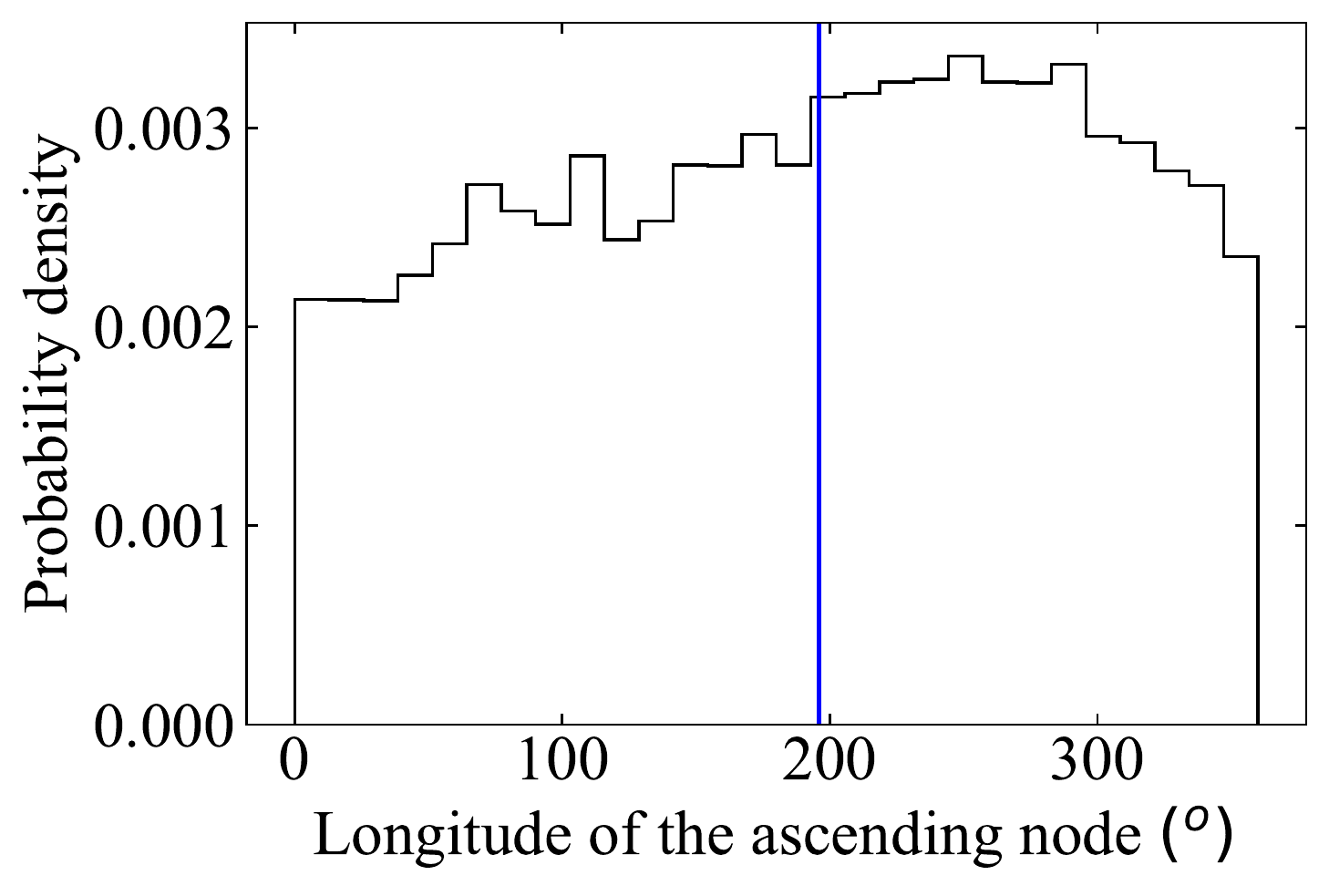}
         \includegraphics[width=0.245\linewidth]{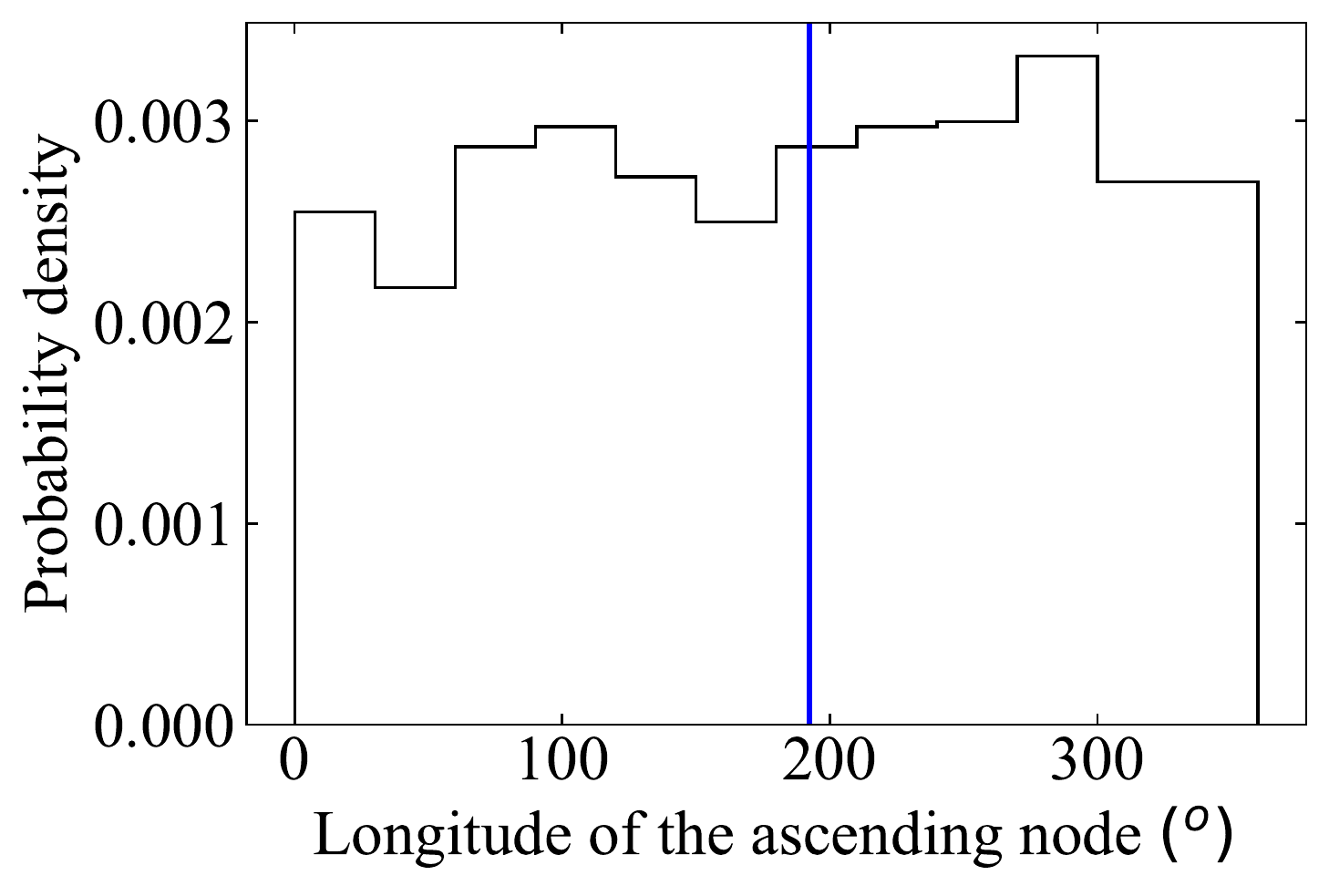}
         \includegraphics[width=0.245\linewidth]{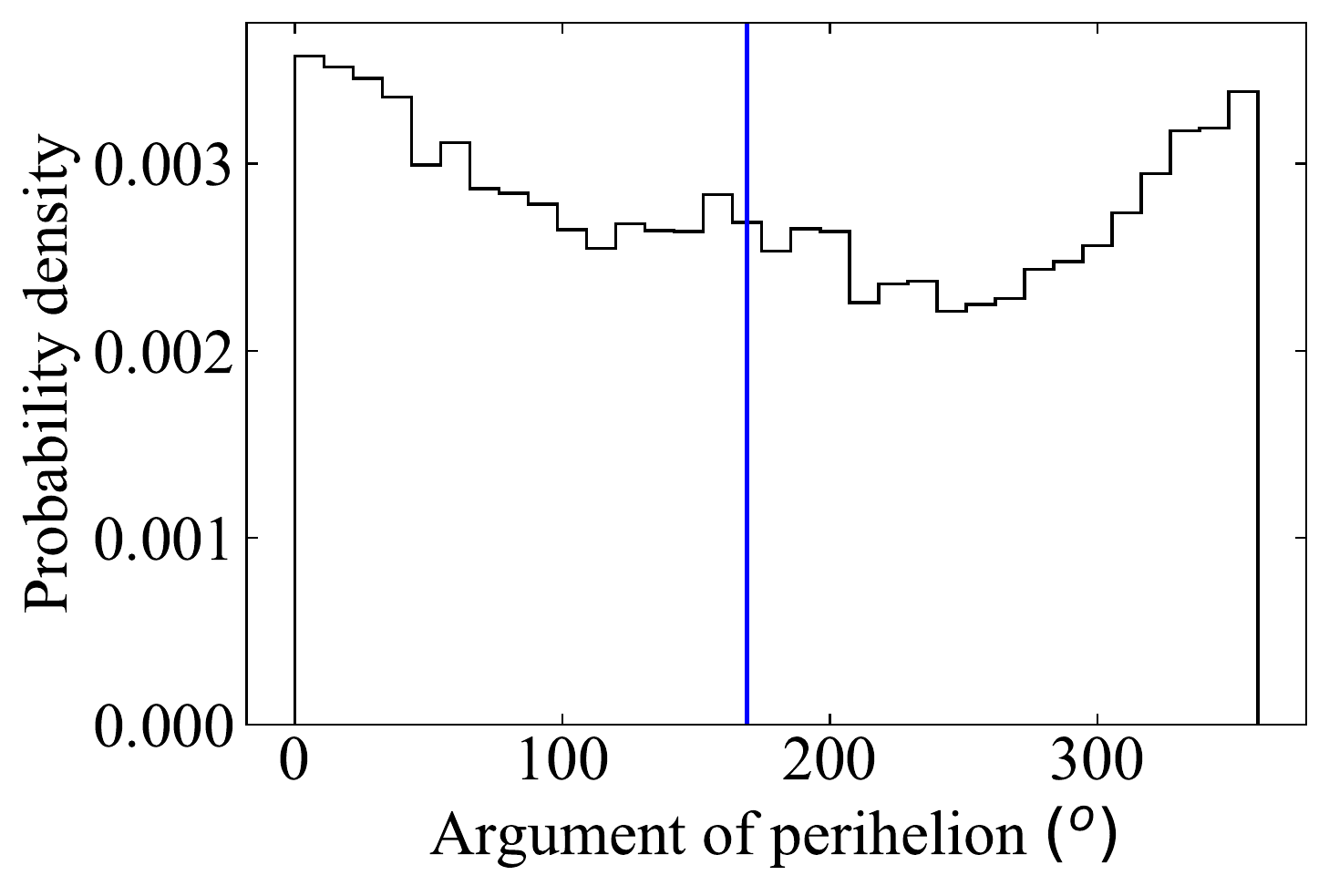}
         \includegraphics[width=0.245\linewidth]{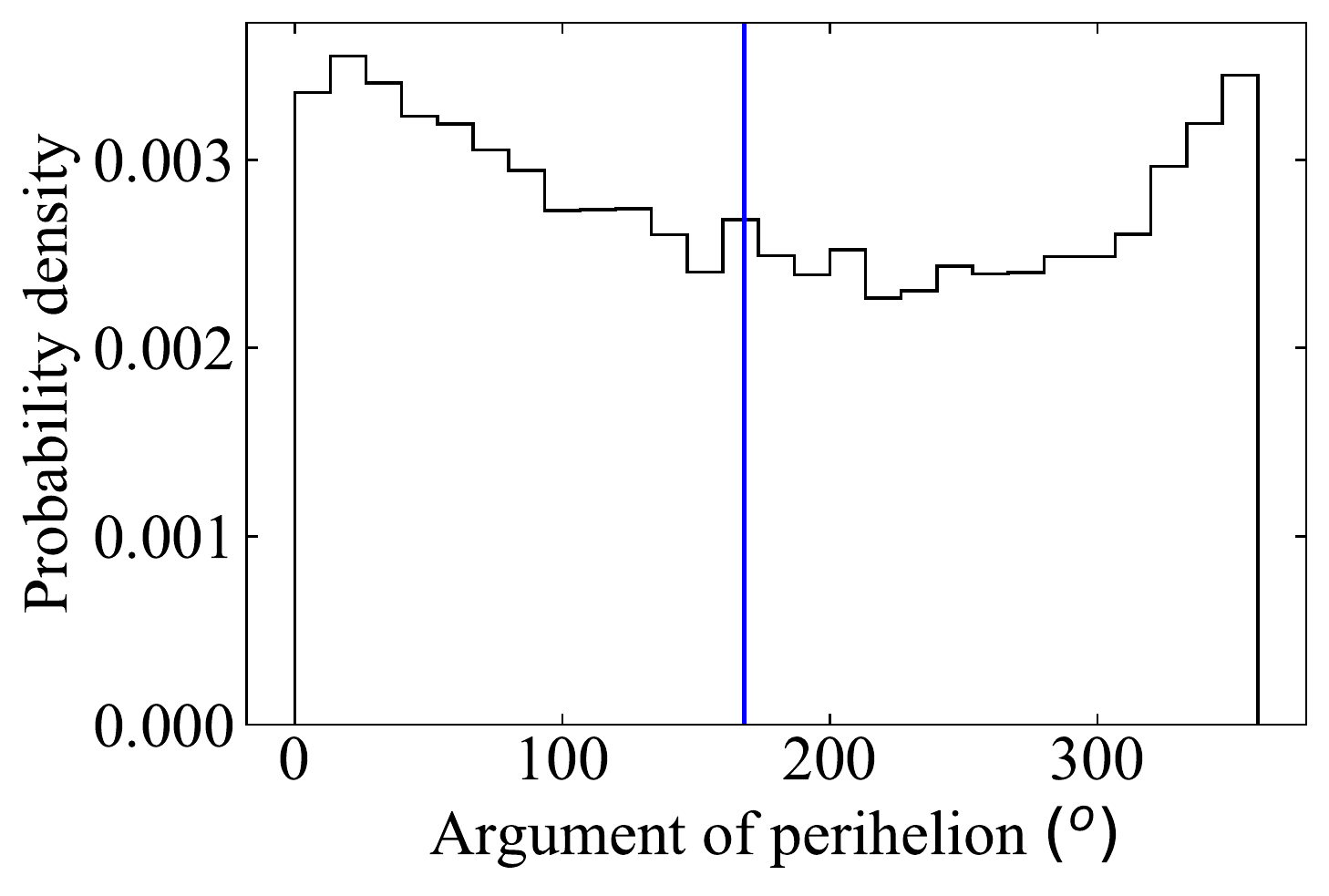}
         \includegraphics[width=0.245\linewidth]{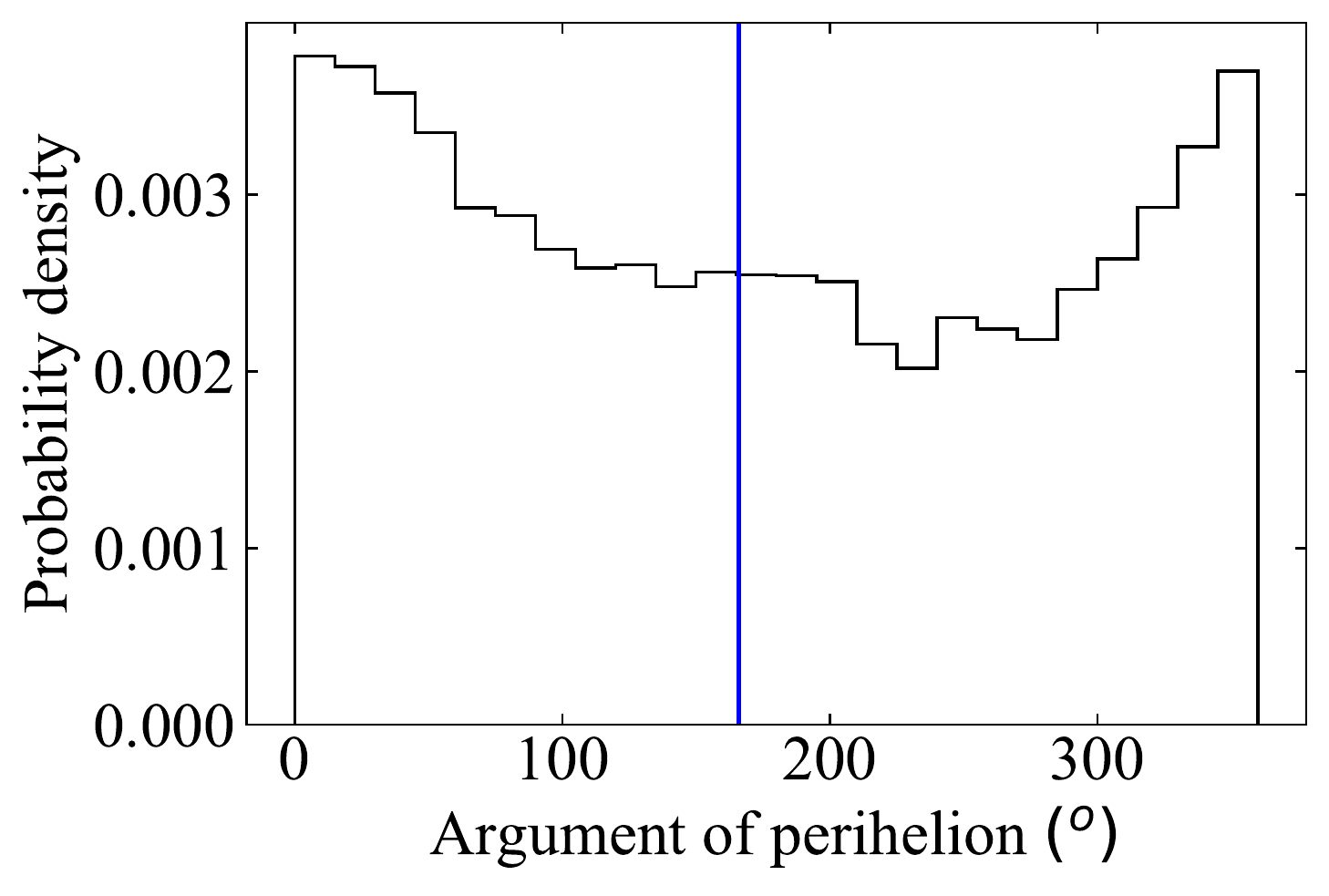}
         \includegraphics[width=0.245\linewidth]{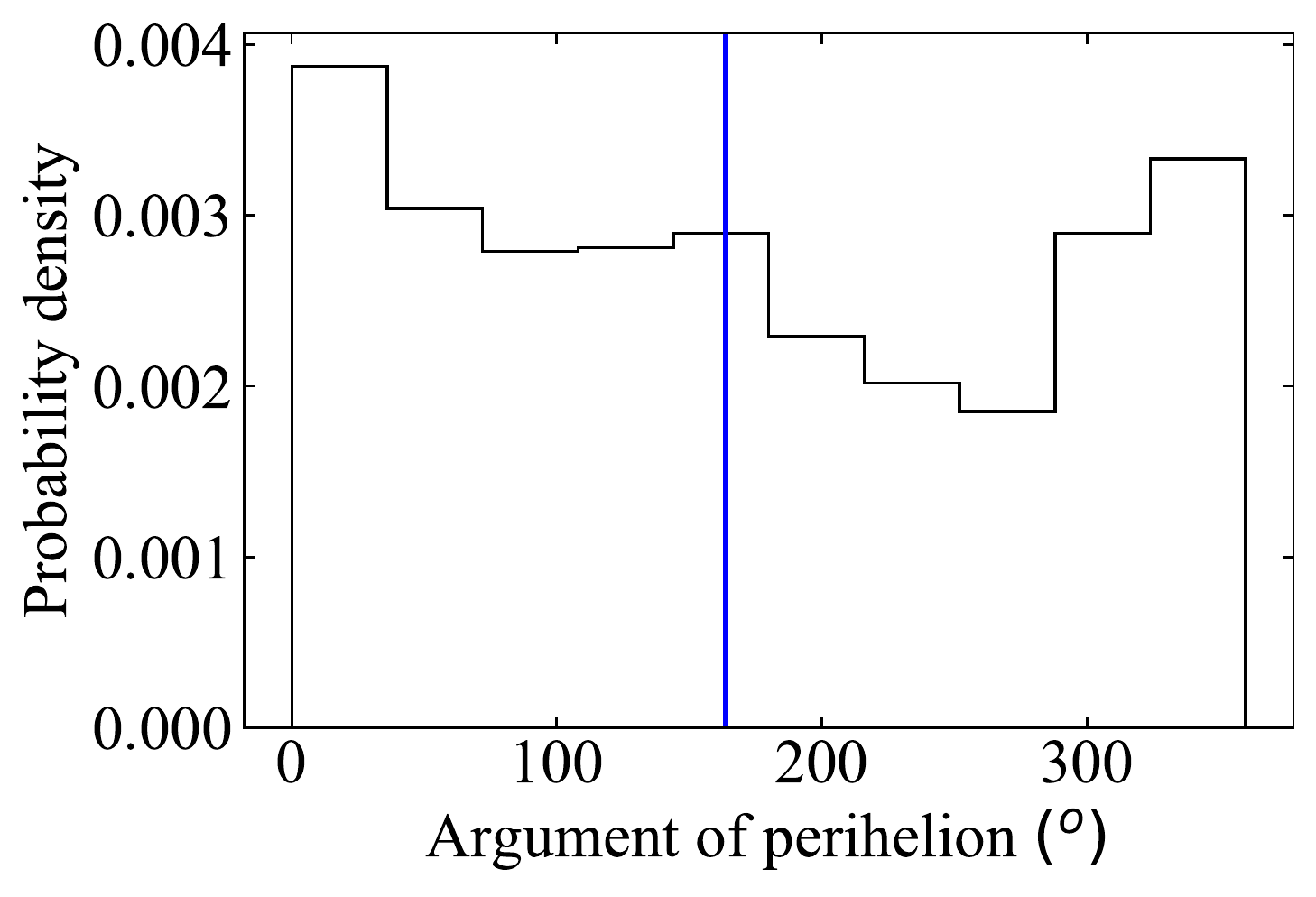}
         \caption{Heliocentric orbital elements of putative perturber planets. As Fig.~\ref{histograms} but using scrambled data as input.
                 }
         \label{histogramsCHK}
      \end{figure*}

         In order to verify that our results do not come from statistical artifacts, we randomly scramble the orbit parameters data used
         as input and repeat the uniform Monte Carlo random searches discussed in Sect.~\ref{Results} and Appendix~\ref{heliocentric}. In 
         these experiments, the set of synthetic ETNOs is such that, for example, the first object may have the value of $a$ of object \#5, 
         $e$ of \#27, $i$ of \#7, $\Omega$ of \#37, and $\omega$ of \#13. By randomly rearranging the values of the orbital elements of the 
         ETNOs, we retain the original distributions of the parameters, but destroy any possible correlations existing among them. The 
         results of these statistical significance tests are shown in Figs.~\ref{histogramsCHKB} and \ref{histogramsCHK}. The distributions 
         in  $i_{\rm p}$ and $\Omega_{\rm p}$ are flattened and no statistically significant perturbing orbits are produced.
%
%
%
%
%
%

      \section{Results at 200~AU\label{200au}}
\begin{figure}
        \centering
         \includegraphics[width=0.49\linewidth]{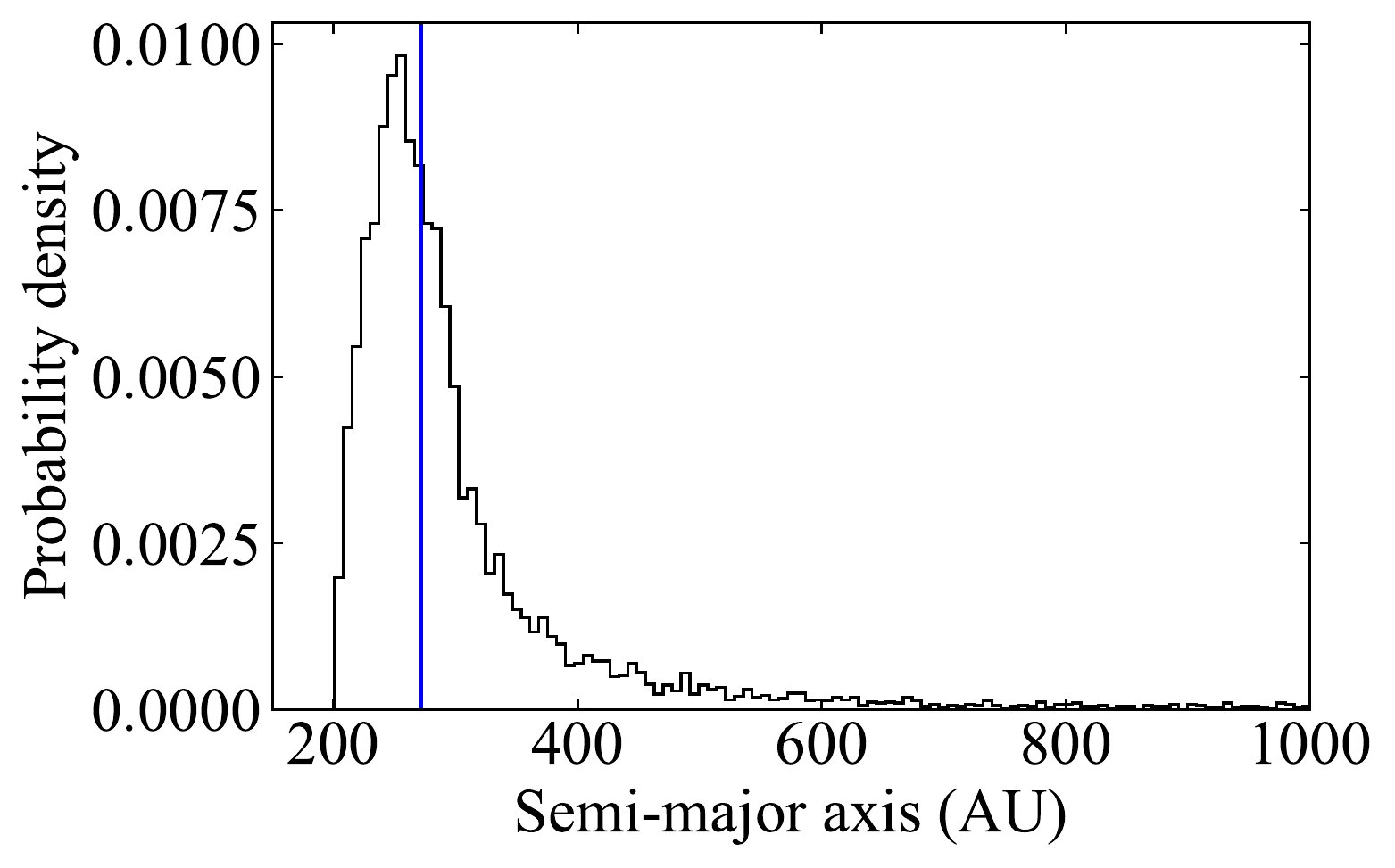}
         \includegraphics[width=0.49\linewidth]{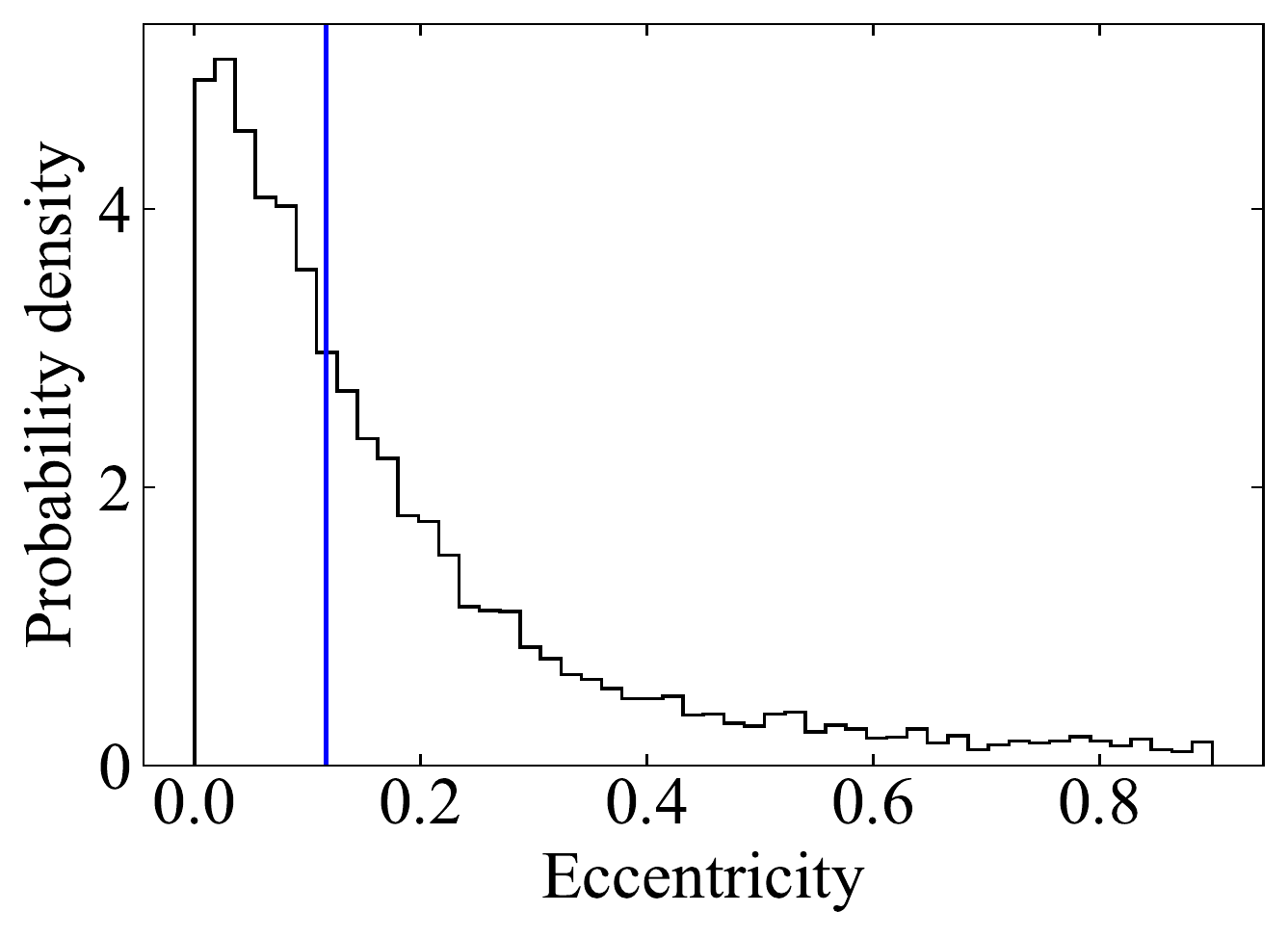}
         \includegraphics[width=0.49\linewidth]{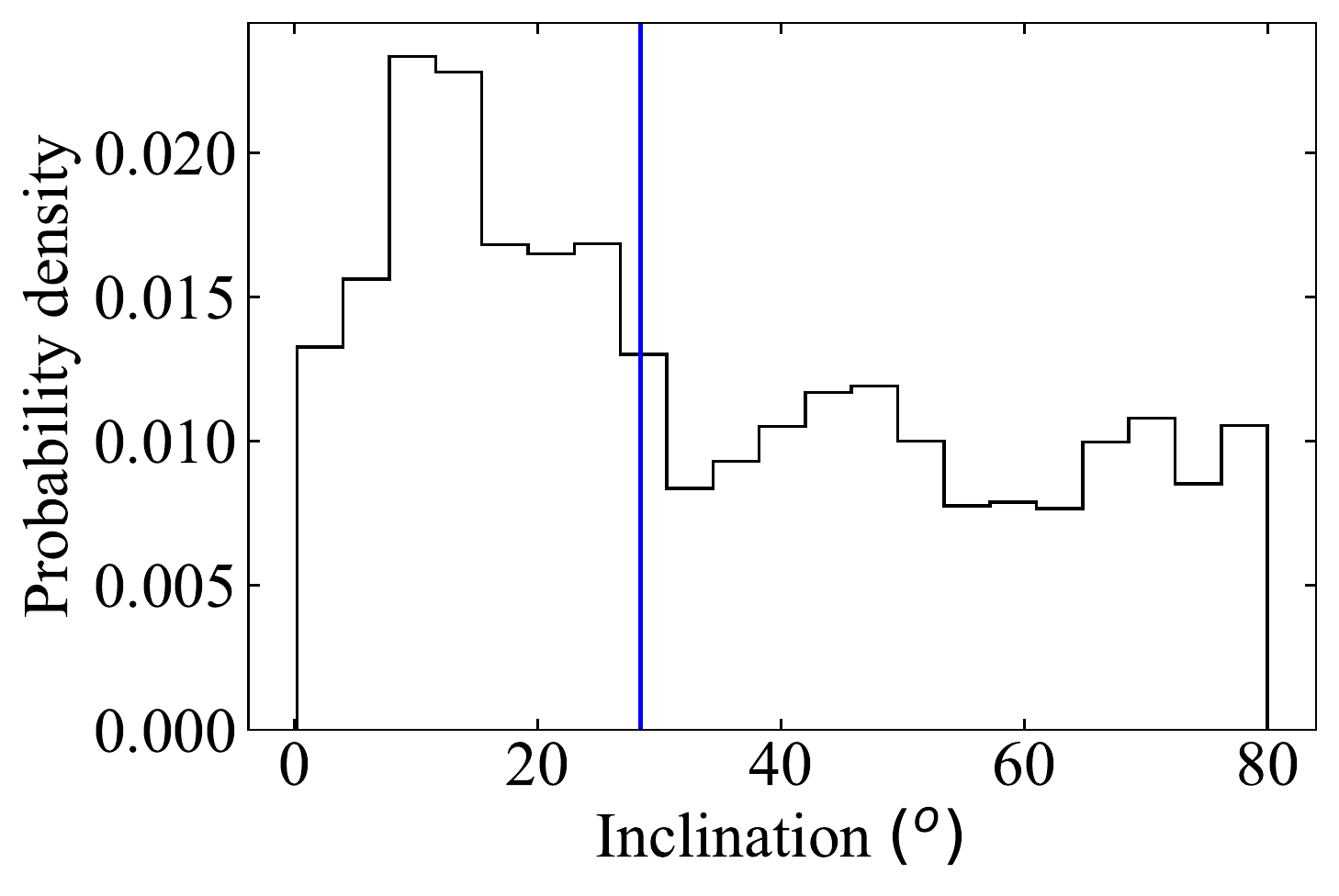}
         \includegraphics[width=0.49\linewidth]{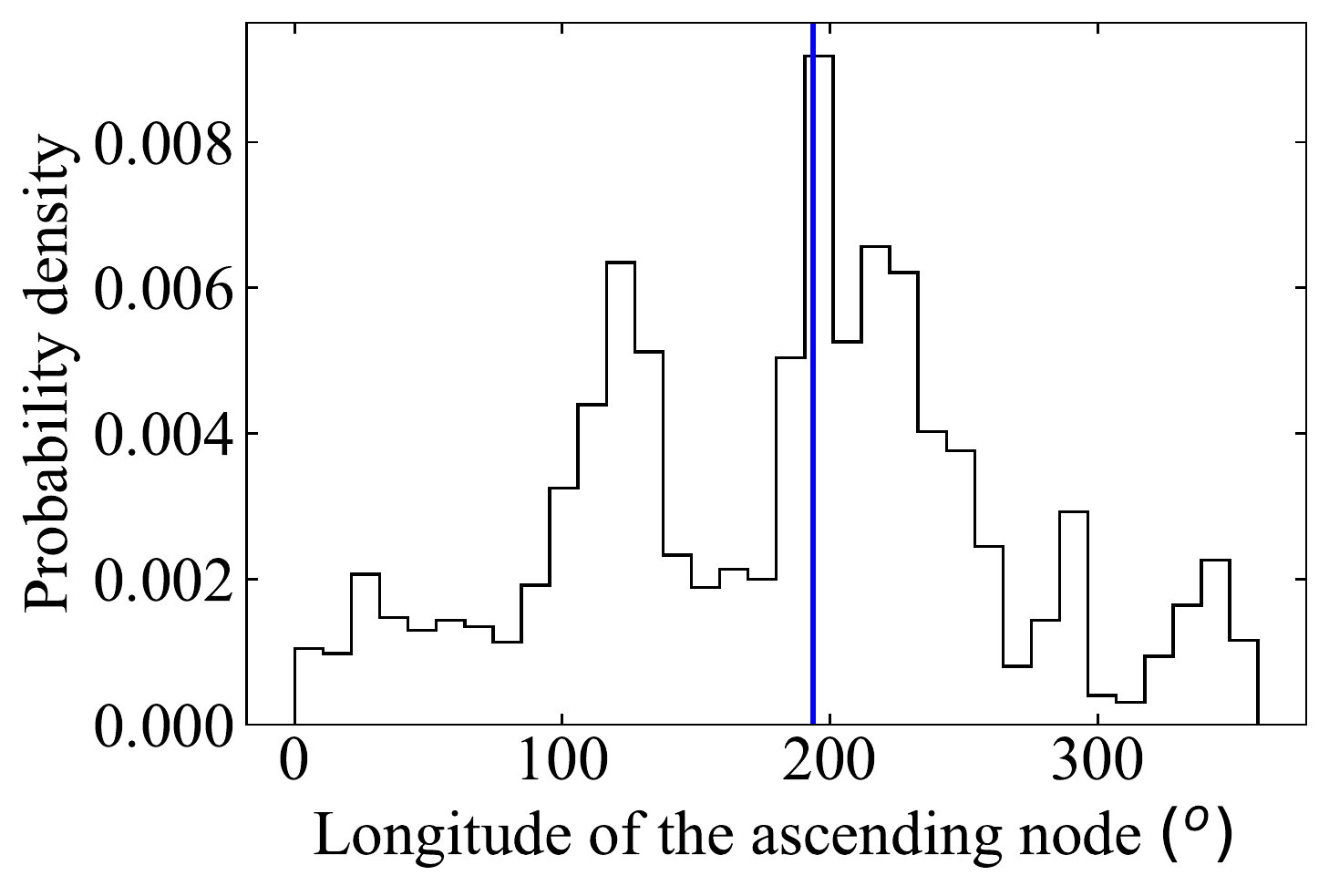}
         \includegraphics[width=0.49\linewidth]{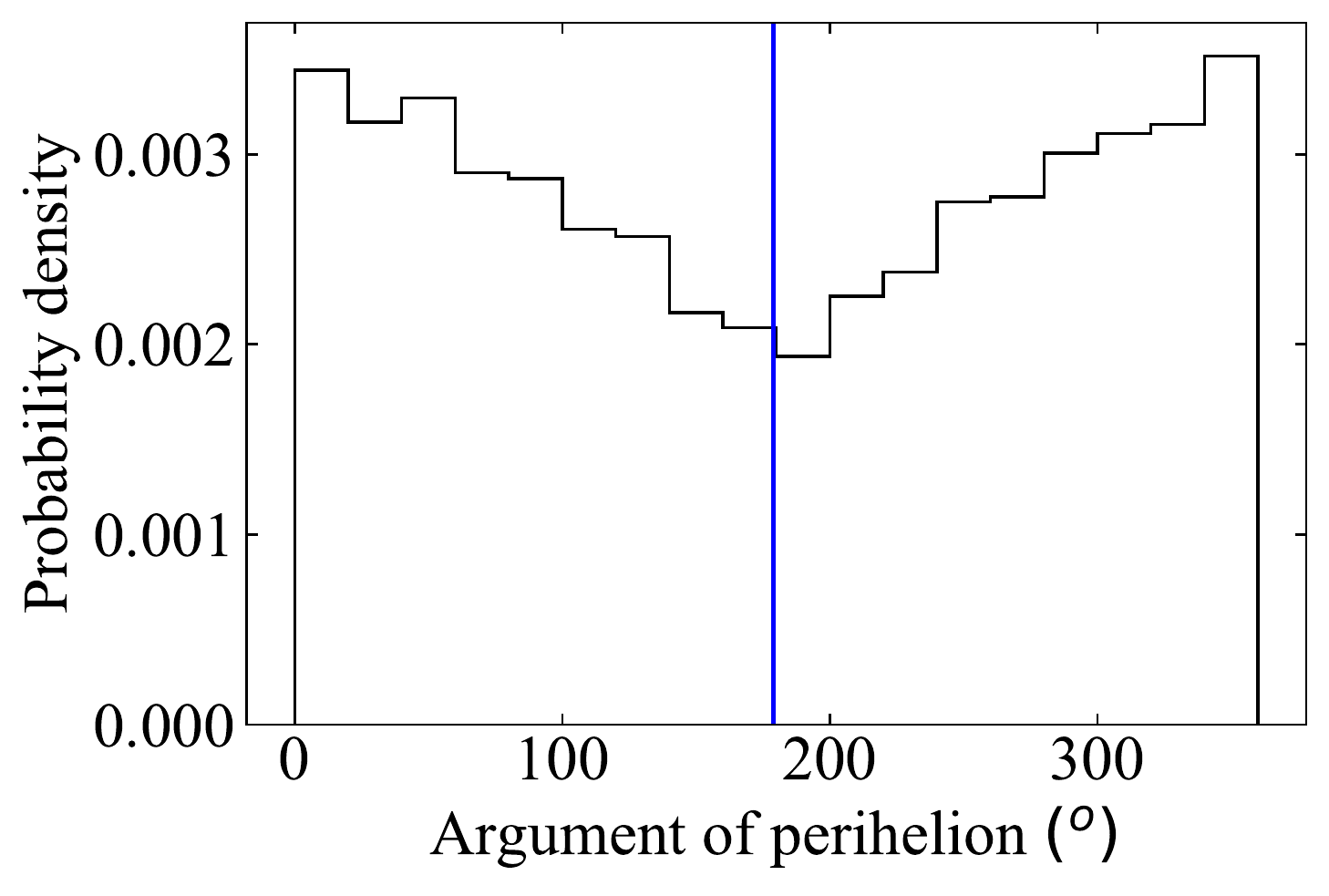}
         \caption{Barycentric orbital elements of putative perturber planets. As Fig.~\ref{histogramsB} but for $q_{\rm p}>200$~AU.
                 }
         \label{histograms200B}
      \end{figure}

         \citet{2020A&A...640A...6F} focused on testing for the presence of possible planets at 400, 500, 600, 650, 700, 750, and 800~AU 
         with masses of 5 or 10~$M_{\oplus}$. The existence of 5~$M_{\oplus}$ planets at 400 or 500~AU is strongly disfavored by their 
         results (see their Fig.~5, top panels). However, a hypothetical Earth-like planet at 200--400~AU from the Sun may still induce 
         significant gravitational effects if close encounters are possible, due to its relatively large value for the Hill radius (e.g., 2.2~AU if $a_{\rm p}=200$~AU, $e_{\rm p}=0.1$ and 1~$M_{\oplus}$). We repeated the analysis, imposing 
         $q_{\rm p}>200$~AU, and we obtained 8234 orbits with a number of potential close approaches in the range between 5--7. Here, we  count 
         how many synthetic ETNOs had at least one mutual nodal distance with the planet under 2~AU. The median values and 16th and 84th 
         percentiles (absolute maximum in parentheses) from the Monte Carlo random searches whose distributions are shown in 
         Fig.~\ref{histograms200B} are: $a_{\rm p}=272_{-38}^{+94}$~AU (258~AU), $e_{\rm p}=0.12_{-0.08}^{+0.21}$ (0.03), 
         $i_{\rm p}=28{\degr}_{-18\degr}^{+35\degr}$ (10{\degr}), $\Omega_{\rm p}=194{\degr}_{-90\degr}^{+58\degr}$ (196{\degr}), and 
         $\omega_{\rm p}=179{\degr}_{-130\degr}^{+132\degr}$ (350{\degr}). 
%
%
%
%
   \end{appendix}

\end{document}